\documentclass[12pt]{article}

\usepackage[top=20mm,bottom=30mm,left=20mm,right=20mm]{geometry}
\usepackage[utf8]{inputenc}
\usepackage[english]{babel}

\usepackage{amsmath,amssymb,latexsym,bm}
\usepackage[mathscr]{euscript}

\usepackage{xcolor}
\usepackage{amsmath,amssymb,latexsym}
\usepackage{cite}

\usepackage{doi}                    
\usepackage{hyperref}               
\hypersetup{
	colorlinks,
	linkcolor={black},
	citecolor={black},
	urlcolor={blue}
}
\usepackage{orcidlink}

\usepackage[T1]{fontenc}

\usepackage{slashed}

\usepackage{graphicx, float} 

\usepackage{caption}
\usepackage{enumitem}

\usepackage{placeins}
\def\clb{ }

\let\a=\alpha \let\b=\beta \let\g=\gamma \let\d=\delta \let\e=\epsilon
 \let\k=\kappa
\let\m=\mu \let\n=\nu \let\x=\xi \let\p=\pi 
\let\s=\sigma 

\let\w=\omega       \let\D=\Delta \let\Th=\Theta 
\let\X=\Xi  \let\S=\Sigma  \let\Y=\Psi

 \def\bd{\begin{document}} \def\ed{\end{document}}

\let\Br=\Bigr \let\Bl=\Bigl
\let\na=\nabla
\def\tU{{\widetilde U}}
\let\pa=\partial \let\ov=\overline
\def\ie{{\it i.e.\ }}
\newcommand{\be}{\begin{equation}}
\newcommand{\ee}{\end{equation}}
\newcommand{\cross }[1]{#1 \hspace{-0.45em} /}
\newcommand{\crosss}[1]{#1 \hspace{-0.6em} /}
\newcommand{\vect}[1]{\bm{\mathbf{#1}}}
\newcommand{\sumint}[0]{\int \hspace{-1.35em} \sum}
\def\ba{\begin{array}}
	\def\ea{\end{array}}

\def\bei{\begin{itemize}}
	\def\eei{\end{itemize}}

\def\ben{\begin{enumerate}}
	\def\een{\end{enumerate}}

\usepackage{tikz}
\newcommand*\circled[1]{\tikz[baseline=(char.base)]{
		\node[shape=circle,draw,inner sep=2pt] (char) {#1};}}

\def\ft#1#2{{\textstyle{{\scriptstyle #1}\over {\scriptstyle #2}}}}
\def\fft#1#2{{#1 \over #2}}
\def\F#1#2{{ F_{#1}^{(#2)} }}
\def\cF#1#2{{ {\cal F}_{#1}^{(#2)} }}

\def\R{{\bf R}}
\def\sst#1{{\scriptscriptstyle #1}}
\def\oneone{\rlap 1\mkern4mu{\rm l}}
\def\e7{E_{7(+7)}}
\def\td{\tilde}
\def\wtd{\widetilde}
\def\im{{\rm i}}
\def\bog{Bogomol'nyi\ }
\newcommand{\ho}[1]{$\, ^{#1}$}
\newcommand{\hoch}[1]{$\, ^{#1}$}
\newcommand{\bea}{\begin{eqnarray}}
\newcommand{\eea}{\end{eqnarray}}
\newcommand{\ra}{\rightarrow}
\newcommand{\lra}{\longrightarrow}
\newcommand{\Lra}{\Leftrightarrow}
\newcommand{\ap}{\alpha^\prime}
\newcommand{\bp}{\tilde \beta^\prime}
\newcommand{\cB}{{\cal B}}
\newcommand{\cO}{{\cal O}}
\newcommand{\vecx}{\vec{x}}
\newcommand{\vecy}{\vec{y}}
\newcommand{\vecp}{\vec{p}}
\newcommand{\vecq}{\vec{q}}
\newcommand{\tr}{{\rm tr} }
\newcommand{\Tr}{{\rm Tr} }
\newcommand{\NP}{Nucl. Phys. }
\newcommand{\mbf}[1]{\mathbf{#1}}
\newcommand{\mbfs}[1]{\boldsymbol{#1}}

\usepackage{relsize}
\newcommand{\sumintinline}[0]{\text{\large$\int$} \hspace{-1.15em} \sum}

\newcommand{\cL}{{\cal L}}
\newcommand{\cA}{{\cal A}}
\newcommand{\cT}{{\cal T}}
\newcommand{\cD}{{\cal D}}
\newcommand{\cH}{{\cal H}}

\def\th{\theta}

\def\sst#1{{\scriptscriptstyle #1}}
\def\0{{\sst{(0)}}}
\def\1{{\sst{(1)}}}
\def\2{{\sst{(2)}}}
\def\3{{\sst{(3)}}}
\def\4{{\sst{(4)}}}
\def\5{{\sst{(5)}}}
\def\6{{\sst{(6)}}}
\def\7{{\sst{(7)}}}
\def\8{{\sst{(8)}}}
\def\9{{\sst{(9)}}}
\def\p{{\sst{(p)}}}
\def\q{{\sst{(q)}}}

\def\ssa{{\sst{(\alpha)}}}
\def\ssb{{\sst{(\beta)}}}
\def\ssg{{\sst{(\gamma)}}}
\def\j{{\sst{(j)}}}

\def\ve{\varepsilon}
\def\vf{\varphi}
\def\F{\Phi}
\def\wg{\wedge}

\def\thb{\bar{\theta}}
\def\Thb{\bar{\Theta}}
\def\barp{\bar{p}}
\def\barq{\bar{q}}
\def\barc{\bar{c}}
\def\bard{\bar{d}}
\def\e{\epsilon}

\def \bi{\bibitem}
\def \la {\label}
\def\foot{\footnote}
\def \adss {$AdS_5 \times S^5$\ }
\newcommand{\rf}[1]{(\ref{#1})}
\def \ov {\over}
\def\Th{\Theta}
\def\vth{\vartheta}
\def\btheta{{\bar\theta}}
\def\ttheta{{{\tilde\theta}}}
\def\bttheta{{{\bar\ttheta}}}
\def\vth{\vartheta}

\def\ra{\rightarrow}
\def\N{{\cal N}}
\def\uM{\underline{M}}
\def\uA{\underline{A}}
\def\uN{\underline{N}}
\def\uP{\underline{P}}
\def\ua{\underline{a}}
\def\ub{\underline{b}}
\def\uc{\underline{c}}
\def\ud{\underline{d}}
\def\ue{\underline{e}}
\def\uf{\underline{f}}
\def\ui{\underline{i}}
\def\uj{\underline{j}}
\def\uk{\underline{k}}
\def\ual{\underline{\alpha}}
\def\ube{\underline{\beta}}
\def\um{\underline{m}}
\def\un{\underline{n}}
\def\up{\underline{p}}
\def\uq{\underline{q}}
\def\ur{\underline{r}}
\def\us{\underline{s}}

\def\umu{\underline{\mu}}
\def\unu{\underline{\nu}}
\def\uka{\underline{\k}}
\def\usi{\underline{\s}}
\def\urh{\underline{\r}}
\def\cc{\circ}
\def\eqv{\equiv}

\def\ni{\noindent}

\def\Ep{E^{{}^{(+)}}}
\def\Em{E^{{}^{(-)}}}

\def\Mp{M^{{}^{(+)}}}
\def\Mm{M^{{}^{(-)}}}

\def \ha{{1\ov 2}}

\def\r{\rho}

\def\Y{{\rm Y}}
\def\X{{\rm X}}
\def\tY{\tilde{\rm Y}}
\def\tX{\tilde{\rm X}}
\def\dY{\dot{\rm Y}}
\def\dX{\dot{\rm X}}

\def \J {\mathcal{J}}
\def \del {\partial}

\def\dF{\dot{F}}
\def\dG{\dot{G}}
\def\df{\dot{f}}
\def\dx{\dot{x}}
\def \E {{\cal E}}
\def \S {{\cal S}}
\def \J {{\cal J}}

\def\ms{\mathcal{S}}
\def\mj{\mathcal{J}}
\def\soj{\fr{\ms}{\mj}}
\def \R {{\bf R}}
\def \om {\omega}
\def \bE {\bar E}
\def \x {{\cal X}}

\def \bi{\bibitem}
\def \la {\label}
\def\foot{\footnote}
\def \adss {$AdS_5 \times S^5$\ }
\def \ov {\over}

\def \varpi {{\rm w}}

\def\thb{\bar{\theta}}
\def\Thb{\bar{\Theta}}
\def\mb{\bar{\m}}
\def\ab{\bar{\a}}
\def\zb{\bar{z}}
\def\psib{\bar{\psi}}
\def\barp{\bar{p}}
\def\barq{\bar{q}}
\def\barc{\bar{c}}
\def\bard{\bar{d}}
\def\e{\epsilon}
\def\wb{\bar{w}}
\def\Jb{\bar{J}}
\def\Nb{\bar{N}}
\def\Zb{\bar{Z}}
\def\pab{\bar{\pa}}

\def\At{\tilde{A}}
\def\Bt{\tilde{B}}
\def\Ct{\tilde{C}}
\def\Dt{\tilde{D}}
\def\Et{\tilde{E}}
\def\Ft{\tilde{F}}
\def\Gt{\tilde{G}}
\def\Ht{\tilde{H}}
\def\Mt{\tilde{M}}
\def\Rt{\tilde{R}}
\def\at{\tilde{a}}
\def\bt{\tilde{b}}
\def\ct{\tilde{c}}
\def\dt{\tilde{d}}
\def\et{\tilde{e}}
\def\ft{\tilde{f}}
\def\gt{\tilde{g}}
\def\mt{\tilde{\mu}}
\def\nt{\tilde{\nu}}

\def\asth{\hat{*}}
\def\phh{\hat{\phi}}

\def\bA{{\bf A}}

\def\ola{\overleftarrow}
\def\ora{\overrightarrow}
\def\alt{\tilde{\a}}

\def\eh{\hat{e}}
\def\eph{\hat{\e}}
\def\ph{\hat{p}}
\def\alh{\hat{\a}}
\def\beh{\hat{\b}}
\def\gah{\hat{\g}}
\def\Fh{\hat{F}}
\def\muh{\hat{\m}}
\def\nuh{\hat{\n}}
\def\thh{\hat{\th}}
\def\dh{\hat{d}}
\def\ih{\hat{i}}
\def\jh{\hat{j}}
\def\kh{\hat{k}}
\def\deh{\hat{\d}}
\def\wh{\hat{w}}
\def\Ah{\hat{A}}
\def\Ch{\hat{C}}
\def\Omh{\hat{\Omega}}

\def\xh{\hat{x}}

\def\ps{\rlap{\, /}\;\,p }
\def\ks{\rlap{\, /}\;\,k }

\def\gym{g_{YM}}

\def\adot{\dot{a}}
\def\bdot{\dot{b}}
\def\bpa{\bar{\pa}}

\def\pr{\prime}
\def\ssk{\medskip}
\def\bsk{\bigskip}

\def\t{\tau}
\def\cM{\mathcal{M}}
\def\S{\Sigma}

\def\N{\nabla}

\def\cR{\mathcal{R}}
\def\cL{\mathcal{L}}

\def\hb{\hbar}
\def\an{\hat{a}}
\def\ac{\hat{a}^\dag}
\def\hp{\hat{p}}

\def\Ec{{\cal E}}

\usepackage{authblk}

\renewcommand\Affilfont{\small}

\begin{document}

\title{Field-theoretical description of the deuteron breakup in the clothed particle representation}
\date{ }

\author[1]{O.~Shebeko\orcidlink{0000-0003-3357-2286}}

\author[1]{A.~Arslanaliev\orcidlink{0000-0002-8667-9688}}

\author[1]{Y.~Kostylenko\orcidlink{0000-0003-3869-2885}}

\author[6,2]{V.~Chahar}

\author[2]{J.~Golak\orcidlink{0000-0002-5210-6910}}

\author[3,4]{H.~Kamada\orcidlink{0000-0001-6519-9645}}

\author[5]{W.~N.~Polyzou\orcidlink{0000-0001-9014-1250}}

\author[6,2]{D.~Ramírez\orcidlink{0000-0002-2932-1399}}

\author[2]{R.~Skibi\'nski\orcidlink{0000-0003-0806-4634}}

\author[2]{K.~Topolnicki\orcidlink{0000-0002-9312-1842}}

\author[2]{H.~Witała\orcidlink{0000-0001-5487-4035}}

\affil[1]{Akhiezer Institute for Theoretical Physics of NSC KIPT, Kharkiv, Ukraine}

\affil[2]{M. Smoluchowski Institute of Physics, Faculty of Physics, Astronomy and Applied Computer Science, Jagiellonian University,
 PL-30348 Kraków, Poland}

\affil[3]{Research Center for Nuclear Physics, Osaka University, Ibaraki 567-0047,
Japan}

\affil[4]{Department of Physics, Kyushu Institute of Technology, 1-1 Sensuicho, Tobata, Kitakyushu 804-8550, Japan}

\affil[5]{Department of Physics and Astronomy, The University of Iowa, Iowa City, Iowa 52242, USA}

\affil[6]{Doctoral School of Exact and Natural Sciences, Jagiellonian University,
PL-30348 Kraków, Poland}

\maketitle

\begin{abstract}
We present a field-theoretical description of the deuteron electrodisintegration reaction $d(e,e'p)n$ induced by unpolarized and polarized electrons. The approach combines the Leh\-mann--Symanzik--Zimmermann $in(out)$ formalism with the clothed particle representation in the instant form of relativistic dynamics, providing a fully relativistic and gauge-independent framework based on the Fock--Weyl criterion. Within the method of unitary clothing transformations, one and the same transformation that generates the relativistic nucleon--nucleon interaction (the Kharkiv potential) also induces a fresh family of electromagnetic current operators. As a result, one-body and two-body (meson-exchange) currents emerge on a common footing.
We compute differential cross sections and polarization observables with the inclusion of final-state interaction effects and meson-exchange current contributions and compare the results with Saclay and Jefferson Lab data as well as with earlier theoretical predictions. The role of relativistic ingredients (one- and two-body currents, Fermi-motion effects, etc.) and the interplay between them are analyzed in several kinematic regimes of the experiments at Saclay and Jefferson Lab. 
\end{abstract}

\section{Introduction}

Explorations of the deuteron electrodisintegration $d(e,e'p)n$ have the eventful history, starting from the first coincidence experiments that have been performed in: 
Stanford \cite{Croissiaux1962}, Orsay \cite{Bounin1964}, Kharkiv \cite{Antufev1974} and Saclay \cite{Saclay1981, Saclay1984}. We have seen that their torch has been picked up for the 80s and 90s with the MIT-Bates \cite{MITBates1992}, Mainz \cite{Mainz1998}, NIKHEF \cite{Steenhoven1994, Jager1993, Nikhef1999} experiments (see, e.g., our survey \cite{KotMelShe95} and Refs. therein) and continued nowadays at Jefferson Lab \cite{Ulmer2002, Egiyan2007, Boeglin2011, Hoiaalo2017, Yero2020}. 
Of course, the corresponding results are a source of inspiration for many theoretical studies. We do not intend to give any review of a lot of them. 
Nevertheless, we would like to remind of the non-relativistic treatment of the deuteron electrodisintegration by H. Arenh{\"o}vel and his colleagues \cite{Arenhovel1979} with inclusion of the final-state interaction (FSI) between the outgoing proton and neutron, and effects of meson-exchange currents (MECs). Note also the remarkable work \cite{Mosconi1993}. 
{\clb Furthermore, relativistic descriptions of this reaction have been extensively developed, including the systematic inclusion of relativistic corrections by Arenhövel et al. \cite{Arenhovel1997, Arenhovel2000},  the diagrammatic models by Laget \cite{LAGET2005}, the covariant description by Jeschonnek and Van Orden \cite{Jeschonnek2008}, and the studies by Sargsian and Boeglin \cite{Sargsian2001,Sargsian2010,BoeglinSargsian2024}.}

In addition, let us recall our calculations of the proton angular distributions and polarization observables in the  $d(e,e'p)n$ reaction with polarized beam and target \cite{KorMelShe88, KorMelShe90, MelShe93, KotMelShe95}. They have much in common with preceding works, including the same ingredients, i.e., finding the $n$-$p$ wave functions (WFs) in continuum for a given nucleon-nucleon (NN) potential, FSI and MEC effects.

Recent experiments in deuteron electrodisintegration at high momentum transfers, particularly at Jefferson Lab \cite{Yero2020, Hoiaalo2017, Boeglin2011, BoeglinSargsian2024}, have pushed deuteron studies into the relativistic regime, reaching missing neutron momenta $k_n$ up to $\sim 1.0$ ${GeV}/c$. Under such conditions, a non-relativistic description of the $np$ interaction could be insufficient, requiring a more sophisticated approach that incorporates relativistic invariance. Furthermore, it may be necessary to consider sub-nucleonic degrees of freedom.

In this research, a new field-theoretical approach extends the in(out) formalism by Lehmann, Symanzik and Zimmermann (LSZ) \cite{LSZ1955}, being combined with the clothed particle representation (CPR) \cite{GreShv, SheShi2001, KoCaSh2007}.
Under this framework, the nucleon-nucleon (NN) interaction, MEC, and boost operators are constructed in a consistent manner, ensuring relativistic invariance by satisfying the Poincaré algebra and requiring the current to transform as a true four-vector. 
We are working in the instant form of relativistic dynamics \cite{Dirac1949}, where only the total Hamiltonian $H$ and the boost operator $\mbf{B}$ carry interactions.
All operators act within a Fock space of clothed particles that incorporate their physical properties, such as observed mass and charge. The relativistic NN potential used in this work (Kharkiv potential) was originally obtained in \cite{SheDub2010}. Applications of this potential to describe deuteron properties are given in \cite{SheDub2012}, while its extension to the three-nucleon (3N) system can be found in \cite{KamSheArs17,FBS2021, Ars2022} and the recent modern exposition \cite{Kamada26}.
Following our recent work \cite{FBS2024}, we employ a fully relativistic framework and propose a way towards the gauge-independent (GI) treatment. The latter can be achieved by imposing the Fock-Weyl criterion (details can be  found in \cite{She14}).

The electromagnetic current operators used in this work are derived directly from the conserved Noether current of the underlying meson–nucleon field model (we include $\pi$, $\rho$, $\delta$, $\eta$, $\omega$ and $\sigma$ mesons) via its reformulation in terms of clothed particle operators. Within the method of unitary clothing transformation (UCT), the same transformation that generates the relativistic NN interaction operator also induces a new family of current operators. As a result, the one-body, two-body and other many-body current operators emerge on a common footing. 
In contrast to other approaches, where MECs are often constructed separately from a potential and supplemented with meson form factors, the present two-nucleon currents (MECs) are consistent with the NN potential and the Poincaré group generators. 
In the course of our studies, we encounter both isoscalar and isovector MECs. The former were first derived in our studies devoted to elastic electron–deuteron scattering \cite{FBS2024, YanThesis}, where only the isoscalar components contribute to the reaction amplitude. In the present work, however, we extend this construction to include the isovector two-nucleon currents as well. The latter becomes essential when describing the deuteron breakup, where both isospin mechanisms contribute to the observables.

This paper is organized as follows. In Section 2, we introduce the basic notations and definitions necessary for describing the deuteron breakup process. Section 3 is devoted to the underlying formalism of clothed particles, where we find the essential links between the CPR and $in(out)$ formalism in the quantum field theory (QFT) \cite{BjorkenDrellVol2}. This section also details the construction of the deuteron state and the moving $np$ state within the CPR, alongside the derivation of the MEC operators. It concludes with a discussion of the path towards a GI description of the reaction. Finally, in Sec. 4, we present numerical results and give prospects of our approach.

\section{Basic expressions and definitions}
\begin{figure}[t]
  \centering
  \includegraphics[scale=0.25]{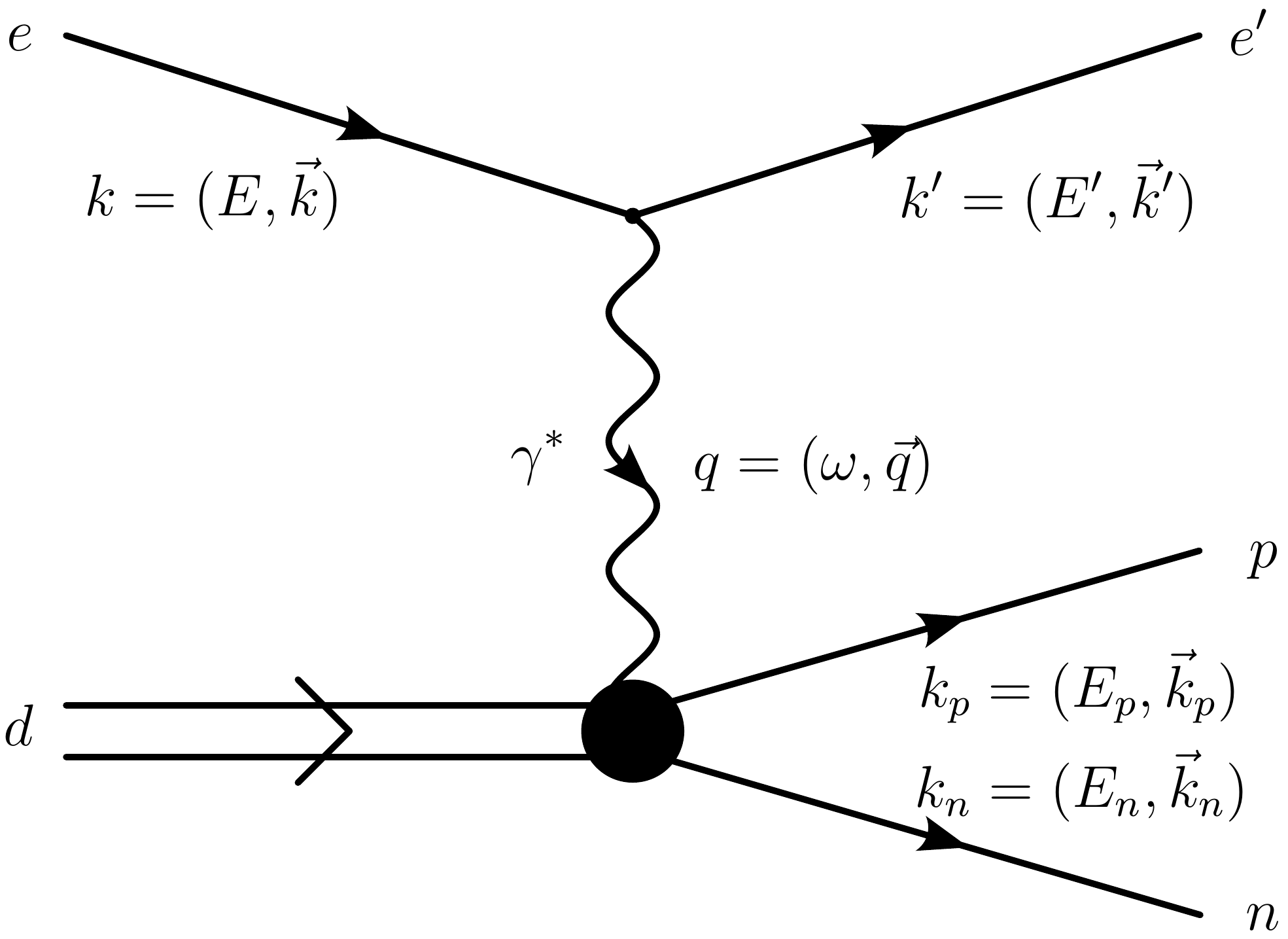}
  \caption{
	One-photon–exchange diagram.
  }
  \label{fig:01}
\end{figure}
	Within the one-photon–exchange approximation (OPEA) (see Fig.~\ref{fig:01}) the $d(e,e'p)n$ cross section for unpolarized particles in the laboratory frame can be written as 
	\be
	\sigma_0\equiv 
      \frac{d^3 \sigma}{d E^{\prime} d \Omega'_e d \Omega_p}
      =\frac{4\a^2E'^2R}{(Q^2)^2}l^{\m\n}W_{\m\n},
	\label{unpolarized_cross_section}
	\ee
	\be
	W_{\m\n}=\frac13\text{Tr}[\mathcal{F}_\m(\mbf{p}_0,\mbf{q})\mathcal{F}^\dagger_\n(\mbf{p}_0,\mbf{q})],
	\label{hadronic_tensor}
	\ee
	where neglecting the electron mass 
      $l_{\m\n}=(k'_\m k_\n + k'_\n k_\m - g_{\m\n}\, k'\cdot k)/2E'E$ is the leptonic tensor, $g_{\m\n}$ is the metric tensor, $k=(E,\mbf{k})$ and $k'=(E',\mbf{k}')$ are the four-momenta of the incident and outgoing electrons, $\mbf{q}=\mbf{k}-\mbf{k}'$ is the momentum transfer, $\vect{p}_0$ is the relative momentum of $np$-pair, $\alpha$ -- the fine-structure constant and $Q^2=\mbf{q}^2-\w^2$ with $\w$ is the energy transfer.
	
	The kinematic factor $R$ equals to
	\be
	      R=E_p|\vect{k}_p|\left[1-\frac{E_p}{E_n}\left(
            \frac{|\vect{q}|}{|\vect{k}_p|}
            \cos(\widehat{\mbf{q},\mbf{k}_p})-1\right)\right]^{-1},
	\ee
	where $E_p$ ($E_n$) is the energy of the knocked-out proton (recoil neutron) and $\mbf{k}_p$ is the proton momentum. Here $\mathcal{F}_\m(\mbf{p}_0,\mbf{q})$ denotes the matrix element
	\be\label{F_mu}
		\mathcal{F}^\m(\mbf{p}_0,\mbf{q})
		=
		\langle  \Psi^{(-)}_{\mbf{q},\mbf{p}_0SM_S}|J^\m(0)|\Psi_{1M_d}\rangle,
	\ee
	where $J^\m(0)=(J_0(0),\mbf{J}(0))$ is the electromagnetic (e.m.) Noether current density operator taken at the space-time point $x=(t,\mbf{x})=0$ sandwiched between the initial deuteron state $|\Psi_{1M_d}\rangle$ at rest and final $np$-pair state $|\Psi^{(-)}_{\mbf{q},\mbf{p}_0SM_S}\rangle$ with spin $S$, its projection $M_S$, total momentum $\mbf{q}$ and relative momentum $\mbf{p}_0$.
	Unlike
    \begin{equation}
        {F}_\mu (\vect{p}_0,\vect{q}) =
        \langle\Psi^{(-)}_{\vect{p}_0 SM_s}|\,J_\mu(\vect{q})|\Psi_{M_d}\rangle,
    \end{equation} 
    from Eq.~(3) of Ref.~\cite{KorMelShe90} we handle $\mathcal{F}^\mu(\vect{p}_0,\vect{q})$ defined by Eq.~\eqref{F_mu}, where, in general, the momenta $\vect{p}_0$ and $\vect{q}$ can not be separated.
\begin{figure}[t]
  \centering
  \includegraphics[scale=0.2]{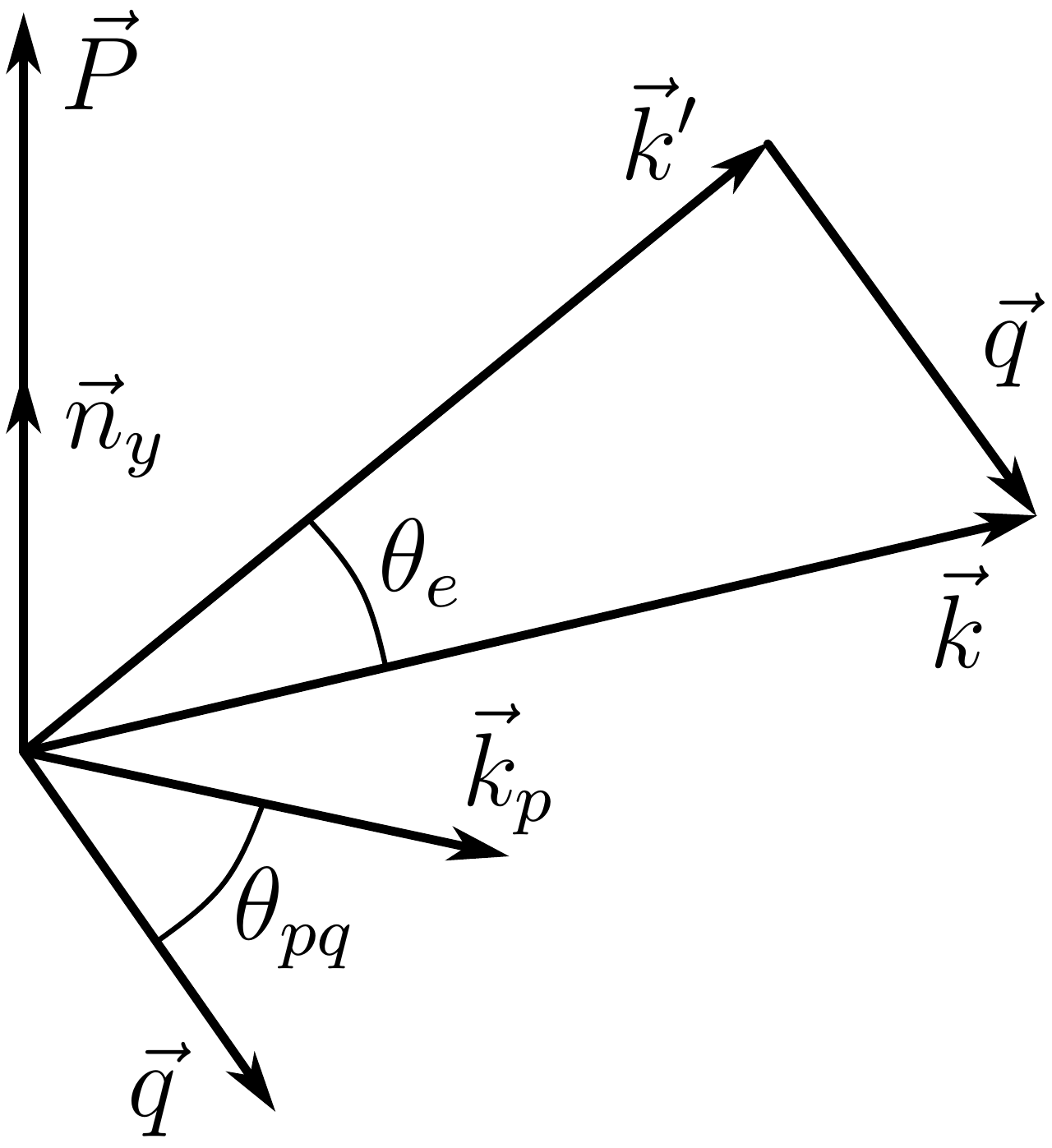}
  \caption{
	Proton polarization in the reaction $d(e,e'\vec{p})n$ for the coplanar geometry.
  }
  \label{fig:02}
\end{figure}
	Let us consider the coplanar case and introduce the orthonormal basis
	\be
	\mbf{n}_Z=\frac{\mbf{q}}{q}, \;\; \mbf{n}_Y=\frac{\mbf{k}\times\mbf{k}'}{|\mbf{k}\times\mbf{k}'|},\;\; \mbf{n}_X=\mbf{n}_Y\times\mbf{n}_Z,
	\label{basis}
	\ee
	assuming that $\mbf{n}_X$ and $\mbf{n}_Y$ form the reaction plane (Fig.~\ref{fig:02}). According to \cite{DeForest1983}, one has
	\be
	\label{cross_section_SF}
	\begin{split}
\frac{4\a^2E'^2}{(Q^2)^2}l^{\m\n}W_{\m\n} = \s_M&\left\{\xi^2W_C + \left(\frac12\xi+\eta\right)W_T+(\xi + \eta)W_S + \xi\sqrt{\xi+\eta}W_I\right\},\\
	&\xi = \frac{Q^2}{\mbf{q}^2},\;\;\;\; \eta=\tan^2\frac{\th_e}{2},
	\end{split}
	\ee
	where $\s_M$ is the Mott cross section and $\th_e$ is the electron emission angle. 
	
	The structure functions (SFs) $W_i$ ($i=C,T,S,I$) are bilinear combinations of the amplitude (\ref{F_mu}) and determined in terms of the hadronic tensor. The Coulomb function $W_C=W_{00}$ is determined by the longitudial current component, while $W_{T}=2W_{YY}$ and $W_{S}=W_{XX}-W_{YY}$ depend on the transverse current component only. The function $W_I=-W_{X0}-W_{0X}$ is the interference term.
	
	It is important to emphasize that, when deriving equation (\ref{cross_section_SF}), the continuity equation
	\be
		q_\m \mathcal{F}^\m(\mbf{p}_0,\mbf{q})=0
	\label{CEforF}
	\ee
	 was used to eliminate the matrix elements with longitudinal current component.
	 
	 In the coplanar case the proton polarization
	 \be
	\mbf{P}=\frac{\text{Tr}\{\mbfs{\s}(1)\mathcal{F}\mathcal{F}^\dagger\}}{\text{Sp}\mathcal{F}\mathcal{F}^\dagger}
	 \ee
	in the reaction $d(e,e'\vec{p})n$ with unpolarized electrons and deuteron is orthogonal to the reaction plane
	\be
	\mbf{P}=P\mbf{n}_Y.
	\label{inducedPa}
	\ee
	By using condition (\ref{CEforF}) one can show that
	\be
	\begin{split}
	&\hspace{10mm}\s_0P=\frac{4\a^2E'^2R}{(Q^2)^2}l^{\m\n}\S^Y_{\m\n}\\
	&= \s_M\left\{\xi^2\S_C + \left(\frac12\xi+\eta\right)\S_T+(\xi + \eta)\S_S + \xi\sqrt{\xi+\eta}\S_I\right\},
	\label{inducedPb}
	\end{split}
	\ee
	where the polarization SFs $\mbfs{\S}_i$ ($i=C,T,S,I$) are related to the componets of the polarization hadronic tensor
	\be
	\mbfs{\S}_{\m\n}=\frac13\text{Tr}\{\mbfs{\s}(1)\mathcal{F}_\m(\mbf{p}_0,\mbf{q})\mathcal{F}^\dagger_\n(\mbf{p}_0,\mbf{q})\}
	\ee
	are related by the same relations as those connecting $W_i$ with $W_{\m\n}$. Here $\mbfs{\s}(1)$ is the Pauli vector for the nucleon labelled by $1$. 
	In a theory that preserves both $P$- and $T$-invariance, 
	one can show  that the proton polarization $\vect{P}$ vanishes when the FSI contributions are neglected \cite{GoldWat,KorMelShe90}. 
	
	In the reaction $d(\vec{e},e'\vec{p})n$ with polarized electrons and unpolarized deuterons the proton polarization $\mbf{P}_\lambda$ has a component $\mbf{P}'$ related to the electron helicity $\lambda$
	\be
	\mbf{P}_\lambda=\mbf{P}+\lambda\mbf{P}',
	\label{polarizationLambda}
	\ee
	where $\mbf{P}$ is the polarization vector (\ref{inducedPa}) in case of unpolarized electrons.
	For the coplanar kinematics (\ref{basis}) the polarization-transfer vector lies in the plane and its components are given by
	\be
	\s_0P'_{X,Z}=\s_MR\sqrt{\eta}\left\{\xi\S'^{X,Z}_I+\sqrt{\eta+\xi}\S'^{X, Z}_T\right\}, \;\; P'_Y=0,
	\label{inducedXZ}
	\ee
	where the interference $\mbfs{\S}'_I$ and the transverse $\mbfs{\S}'_T$ SFs are given by
	\be
	\mbfs{\S}'_I = i\left[\mbfs{\S}_{0Y}-\mbfs{\S}_{Y0}\right], \;\; \mbfs{\S}'_T = i\left[\mbfs{\S}_{YX}-\mbfs{\S}_{XY}\right].
	\label{transver_P_SF}
	\ee
	Thus, the transferred polarization $\mbf{P}'$ depends on reaction amplitudes that do not appear in the so-called induced polarization $\mbf{P}$. In contrast to the latter, the transferred polarization does not vanish when FSI is neglected. This circumstance can be used to extract additional information on the electromagnetic properties of a bound nucleon (e.g., the neutron) under conditions where FSI and MEC effects can be neglected.

	\section{Underlying formalism}
	
	At the beginning we allow ourselves to recall a few details concerning the $in(out)$ formalism in QFT \cite{BjorkenDrellVol2}.
	We do it, keeping in mind that this topic is not popular amongst the nuclear and few-body physics community.

	\subsection{Reaction amplitude in the \textit{in(out)} formalism}
	In this subsection we will show how within the field-theoretical description the amplitude for an electron scattering off a hadron system in the OPEA can be reduced to the expression 
	\be
	\mathcal{T}_{if}=\frac{em_e}{\sqrt{EE'}}\d(p'+k'-p-k)\bar{u}_e(k')\g_\m u_e(k)\frac{\langle h'|J^\m_h(0)|h\rangle}{q^2},
	\label{amp_in_OPEA}
	\ee
	where $m_e$ is the mass of physical electron, $u_e(k)\equiv u_e(k\sigma)$ electron's Dirac spinor with 4-momentum $k$ and polarization $\sigma$ and $J^\m_h(0)$ is the hadronic current density operator at the space-time point $x=(t, \mbf{x})=(0, \vect{0})$, 
	that is sandwiched between the initial $|h\rangle$ and final $|h'\rangle$ hadronic states.

	In this context, it seems pertinent to address a field model for interacting particles, viz., let us consider the Lagrangian density
	\begin{eqnarray}\label{chpt3:A.1}
	\mathcal{L} = \mathcal{L}_e + \mathcal{L}_h + \mathcal{L}_{\gamma} + \mathcal{L}_{\gamma e} + \mathcal{L}_{\gamma h}
	\end{eqnarray}
	for interacting electrons, hadrons and photons, where after Schwinger we have
	\begin{eqnarray}
	\mathcal{L}_e &=& \frac{i}{4} [\bar{\psi}_e, \gamma^{\mu} \partial_{\mu} \psi_e] - \frac{i}{4} [\partial _{\mu} \bar{\psi}_e \gamma^{\mu}, \psi_e] - \frac{m_0^e}{2}[\bar{\psi}_e, \psi_e] ,
	\\
	\mathcal{L}_h &=& \mathcal{L}_N + \mathcal{L}_M + \mathcal{L}_{\mathcal{M}N} ,
	\\
	\mathcal{L}_N &=&  \frac{i}{4} [\bar{\psi}, \gamma^{\mu} \partial_{\mu} \psi] - \frac{i}{4} [\partial _{\mu} \bar{\psi} \gamma^{\mu}, \psi] - \frac{{m}_0}{2}[\bar{\psi}, \psi].
	\end{eqnarray}
	Here the density $\mathcal{L}_M$ consists  of the pionic (pseudoscalar) part
	\begin{eqnarray}
	\mathcal{L}_M &=&  \frac{1}{2} [ \partial_{\mu} \mbfs{\varphi}_{ps} \partial ^{\mu} \mbfs{\varphi}_{ps} - \mu_0^2 \mbfs{\varphi}_{ps}^2]
	\end{eqnarray}
	and heavier meson contributions, while
	$\mathcal{L}_{\mathcal{M}N} = \mathcal{L}_{s} + \mathcal{L}_{ps} + \mathcal{L}_{v} +\cdots$ is the density of the meson-nucleon interaction.
      Here $s$, $ps$ and $v$ stand for scalar, pseudoscalar and vector couplings.
	We use notation $m_0^e$, $m_0$ for the electron, nucleon bare (trial) masses, respectively, and $e_0$ for bare electron and proton charge.
	
	Electromagnetic interactions with the lepton current $J_e^{\mu} = \frac{e_0}{2} [\bar{\psi}_e, \gamma^{\mu} \psi_e]$ and the hadron current $J_h = J_N + J_M$, where 
	$J^{\mu}_N = \frac{e_0}{2} [\bar{\psi}, \gamma^{\mu} \frac{1 + \tau^z}{2} \psi]$ are given by
	\begin{eqnarray}
	\mathcal{L}_{\gamma e} &=&  J_e^{\mu} A_{\mu} = J_e A 
	\\
	\label{chpt3:A.8}
	\mathcal{L}_{\gamma h} &=&  J_h A
	\end{eqnarray}
	
	At last, the Lagrangian density for the radiation field after St\"uckelberg is defined as
	\begin{eqnarray}
	\mathcal{L}_{\gamma } &=&  -\frac{1}{4} F_{\mu \nu} F^{\mu \nu} - \frac{1}{2} (\partial A)^2 + \frac{\lambda^2}{2} A^2
	\end{eqnarray}
	with infinitesimally small photon mass $\lambda$. $\partial A \equiv \partial_{\mu} A^{\mu}$ is the divergence of the e.m. field vector $A = (A^0(x), A^1(x), A^2(x), A^3(x))$.

	Note also that these currents originate from the so-called minimal substitution $\partial ^{\mu} \to \partial ^{\mu} + ie_0 A^{\mu}$. Introduction of non-minimal currents is a separate problem, but their inclusion does not violate the gauge invariance of the theory.
	
	The corresponding Euler-Lagrange equations
	\begin{eqnarray}
	(i \slashed{\partial} - m_0) \psi_e &=& e_0\slashed{A} \psi_e ,
	\label{chpt3:eq:A.9}
	\\
	\bar{\psi}_e (-i \slashed{\partial} - m_0)  &=& e_0  \bar{\psi}_e\slashed{A}  ,
	\label{chpt3:eq:A.9a}
	\\
	(i \slashed{\partial} - {m}^N_0) \psi &=& \frac{\partial \mathcal{L}_{\mathcal{M}N}}{\partial \bar{\psi}} + e_0 \slashed{A} \frac{1 + \tau^z}{2}\psi,
	\\
	\bar{\psi} (-i \slashed{\partial} - {m}^N_0)  &=& \frac{\partial \mathcal{L}_{\mathcal{M}N}}{\partial \psi} + e_0 \frac{1 + \tau^z}{2} \bar{\psi} \slashed{A} ,
	\\
	\Box A^{\mu}(x) &=& J_e^{\mu}(x) + J_h^{\mu}(x) \equiv {J}^{\mu}(x) ,
	\label{chpt3:eq:A.11}
	\end{eqnarray}
	where we introduced the symbol $\slashed{O} = \gamma_{\mu}O^{\mu}$ for any vector $O$ and $\frac{1 + \tau^z}{2}$ is the proton isospin projection operator.
	
	Using these equations we would like to address the reduction technique within the $in(out)$ formalism, in particular, the so-called asymptotic LSZ approach (see, e.g., monographs by Schweber (1961) \cite{Schweber}
	, Goldberger and Watson (1964) \cite{GoldWat}, 
	Bjorken and Drell (1962) \cite{BjorkenDrellVol2}). 
	In this context, relying upon the previous experience (see 
	Ref.\cite{SheISHEPP2002}
	and Appendix~B in PhD thesis~\cite{YanThesis}), 
	one can show that the $S$-matrix for an electron scattering off a hadron system $h$
	\begin{eqnarray}
	e + h \to e + h'
	\end{eqnarray}
	with the conservation law
	$k + p = k'+p'$
	is given by
	\begin{eqnarray}
	\langle f | S | i \rangle = \langle e'h'; out | eh; in \rangle = 2 \pi i \frac{m_e}{\sqrt{EE'}}e_0 \bar{u}_e(k') \gamma_{\mu} u_e(k) \frac{\langle h' | {J}^{\mu}(0) | h \rangle}{q^2} \delta(p'+k'-p-k) \label{eq:A.13}
	\end{eqnarray}
	where $m_e$ is the mass of the physical electron, $q^2 = (k'-k)^2 = (p'-p)^2$ is the 4-momentum transferred, $u_e(k) \equiv u_e(k \sigma)$ is the electron's Dirac spinor with 4-momentum $k$ and polarization $\sigma$. The matrix element $\langle h' | {J}^{\mu}(0) | h \rangle$ can be replaced by $\langle h' | J^{\mu}_h(0) | h \rangle$ since, first, $J^{\mu}_e(0) = e_0 : \bar{\psi}_e(0) \gamma ^{\mu} \psi(0) :$ 
	with the symbol $:~:$ for the normal ordering of operators,
	and, second,
	\begin{eqnarray}
	\langle h' | :\bar{\psi}_e(0) \gamma^{\mu} \psi(0) : | h \rangle = 0
	\end{eqnarray}
	with accuracy up to terms of the $e^1$ order. 
	In addition, assuming that $e_0 - e \propto O(e^3)$ we can resort to the OPEA
	\begin{eqnarray}\label{in_out_amplitude}
            \langle e'h'; out | eh; in \rangle = 2 \pi i \frac{m_e}{\sqrt{EE'}}e \bar{u}_e(k') \gamma_{\mu} u_e(k) \frac{\langle h' | J_h^{\mu}(0) | h \rangle}{q^2} \delta(p'+k'-p-k),
	\end{eqnarray}
	with the matrix element $\langle h' | J_h^{\mu}(0) | h \rangle$ between the strong Hamiltonian ($H_S = H_N + H_M + H_{MN}$ in our case) eigenstates.
	
	In this respect, the UCT method \cite{SheShi2000,SheShi2001, KoCaSh2003} has an advantage allowing us to build up new interaction operators \cite{SheDub2010} and solve the respective eigenvalue problem non-perturbatively \cite{DubShe13}.
	In the CPR all ingredients of matrix elements  
	\be
	\langle\Psi^{(-)}_{\mbf{q}, \mbf{p}_0 SM_S}|J_h^\m(0)|\Psi_{M_d}\rangle,
	\ee
	namely, the hadron states and current operator, can be consistently found within the field-theoretical calculations. This is achieved, in particular, via certain links between the $in(out)$ and clothed particle states.

	\subsection{Links between the clothed-particle representation and \textit{in}(\textit{out}) formalism}
	\label{subsec:links_with_in_out}

	Recall that along with the one-clothed-particle states 
	there are other one-particle states which are the $H$ eigenvectors as well.
	In Ref.~\cite{She2004} there was presented the equivalence theorem that connects the definitions of the $S$-matrix within the $in(out)$ formalism by LSZ and the CPR approach.
	It has turned out that the $S$ operators that are determined by the time evolution from a distant past to a distant future, for the two decompositions $H=H(\{\alpha\})=H_F+H_I$ and $H=K(\{\alpha_c\})=K_F+K_I$, being sandwiched between the bare states and the corresponding clothed ones are equal to each other. Such a coincidence becomes possible owing to certain isomorphism between the $\{\alpha_c\}$ algebra and the $\{\alpha\}$ algebra once the UCT obey the asymptotic condition 
	\begin{equation}\label{chpt2:equiv_condition}
	\lim\limits_{t \rightarrow \pm \infty} W_D(t)=1,
	\end{equation}
	where the UCT in the Dirac (D) picture $W_D(t) = e^{iH_F t} W e^{-iH_F t}$ and the limit is implied in the strong sense.
	
	As well-known, when evaluating the $S$--matrix in the Heisenberg (H) picture,
	\begin{equation}
	S_{if} = \langle f;out \mid  i;in \rangle
	\end{equation}
	one has to deal with the $in(out)$ states (see, e.g., \cite{GoldWat})
	\be
	|\mbf{k}_1\cdots\mbf{k}_2;in(out)\rangle \equiv a_{in(out)}^\dagger(\mbf{k}_1)\cdots a_{in(out)}^\dagger(\mbf{k}_n)|\Omega\rangle
	\ee
	in particular, one-particle state
	\be
	|\mbf{k};in(out)\rangle \equiv a_{in(out)}^\dagger(\mbf{k})|\Omega\rangle,
	\ee
	where $ \left| \Omega \right\rangle $ is the physical vacuum.
	The creation (destruction) $in(out)$ operators $a^\dagger_{in(out)}$ ($a_{in(out)}$) meet the same canonical commutation relations that clothed ones $a_c^\dagger$ ($a_c$) do. By definition, the $in(out)$ states are eigenvectors of the energy-momentum operator $P^\mu=(H,\mathbf{P})$
	\begin{equation}\label{inout_eigen}
	P^\mu| \mathbf k_1 \cdots \mathbf k_n;in(out)\rangle=(k_1^\mu+\cdots+k_n^\mu)| \mathbf k_1 \cdots \mathbf k_n;in(out)\rangle,
	\end{equation}
	with $k^\mu=(E_{\mathbf{k}},\mathbf{k})$.
	
	In Appendix A one can find (see also Sec.~4 in~\cite{She2004} and in~\cite{YanThesis}) the derivation of important relations between states in the CPR and $in(out)$ formalism for the one particle
	\be
	|\mbf{k};in(out)\rangle \equiv a_{in(out)}^\dagger(\mbf{k})|\Omega\rangle=a_c^\dagger(\mbf{k})|\Omega\rangle
	\label{oneparticlerelation}
	\ee
	and two particles
	\begin{align}
		&|\mbf{k}_1,\mbf{k}_2;in\rangle \equiv a_{in}^\dagger(\mbf{k}_1)a_{in}^\dagger(\mbf{k}_2)|\Omega\rangle=\Omega_{c}^{(+)}a_{c}^\dagger(\mbf{k}_1)a_{c}^\dagger(\mbf{k}_2)\Omega\rangle,
		\label{twoparticlerelation_1}
		\\
		&|\mbf{k}_1,\mbf{k}_2;out\rangle \equiv a_{out}^\dagger(\mbf{k}_1)a_{out}^\dagger(\mbf{k}_2)|\Omega\rangle=\Omega_{c}^{(-)}a_{c}^\dagger(\mbf{k}_1)a_{c}^\dagger(\mbf{k}_2)\Omega\rangle,
		\label{twoparticlerelation_2}
	\end{align}
	where $\Omega_{c}^{(\pm)}$ denotes the  M{\o}ller operators
	\begin{equation}\label{Moller_def}
		\Omega_{c}^{(\pm)} \equiv \lim\limits_{t\to\mp\infty}
		\exp({iKt})\exp({-iK_Ft}).
	\end{equation}
	These relations hold under the condition (\ref{chpt2:equiv_condition}). To some extent, relation (\ref{oneparticlerelation}) does not seem unexpected, since both one-particle clothed and $in(out)$ states 
	are the $H$ eigenvectors. Of course, it does not mean that $a_{in(out)}(\mbf{k})=a_c(\mbf{k})$!

	Furthermore, one should keep in mind the following recipe of practical calculations with
	\begin{equation}
	\Omega^{ ( \pm ) }_{c} \mid E;c \rangle = \Omega^{ ( \pm ) }_{c}(E)\mid E;c \rangle
	= \pm i \lim_{\epsilon \to + 0} \epsilon
	G(E \pm \imath \epsilon ) \mid E;c \rangle,
	\label{chpt2:eq_2.16}
	\end{equation}
	where we have introduced the notation $G(z) = (z - K)^{-1}$ for the Hamiltonian resolvent.
	In its turn, the resolvent can be expressed through the corresponding $T$ or $R$
	operator to make a path to known methods. 
	Just such an approach has been used in papers \cite{SheDub2010, DubShe13} when studying the properties of two-nucleon systems.
	Evidently, the appearance of these operators reflects the inclusion of the initial (final) state interaction effects into entrance (exit) reaction channels.

	\subsection{The deuteron and \textit{np}-pair states in the CPR}
	\label{subsec:deuteron_and_np_states}

	Working in the two-nucleon sector of the Fock space, the initial and final states in the center of momentum system (c.m.s.) can be represented as
\be
  |\Psi_{M_d}\rangle =
  \int d\mbf{p}'\, |\mbf{p}'S'M'_S\rangle
  \langle \mbf{p}'S'M'_S| \Psi_{M_d}\rangle
  \equiv
  \int d\mbf{p}'\, \Psi_{M_d}(\mbf{p}'S'M'_S) |\mbf{p}'S'M'_S\rangle
  ,
  \label{DState}
\ee
\be
  |\Psi^{(-)}_{\mbf{0},\mbf{p} SM_S}\rangle
  \equiv 
  |\Psi^{(-)}_{\mbf{p} SM_S}\rangle 
  =
  \int d\mbf{p}'\,
  |\mbf{p}' S'M'_S\rangle
  \langle \mbf{p}' S'M'_S|\Psi^{(-)}_{\mbf{p}\, SM_S}\rangle
  \equiv
  \int d\mbf{p}'\,
  \Psi^{(-)}_{\mbf{p}\, SM_S}\left(\mbf{p}' S'M'_S\right) |\mbf{p}' S'M'_S\rangle,
\ee
where
\be \label{SM_S_basis}
  |\mbf{p}SM _S\rangle = 
  \sum_{\m_1\m_2}(\tfrac12\,\m_1\, \tfrac12\,\m_2 | S\,M_S)
  b_c^\dagger(\mbf{p}\mu_1)b_c^\dagger(-\mbf{p}\mu_2)|\Omega\rangle.
\ee
These states are the $P=(H,\vect{P})$ eigenvectors, viz.,
\be\label{Eq_for_Psi_pm}
  P|\Psi_{M_d}\rangle = (m_d,\vect{0})|\Psi_{M_d}\rangle,
  ~~~
  P|\Psi^{(-)}_{\mbf{p} SM_S}\rangle = (2E_{\vect{p}},\vect{0})|\Psi^{(-)}_{\mbf{p} SM_S}\rangle,
\ee
with $E_{\vect{p}}=\sqrt{\vect{p}^2 + m^2}$, where $\vect{p}$ is the c.m.s. relative momentum and $m$ ($m_d$) is the physical nucleon (deuteron) mass.

Further, accordingly Appendix B of Ref.~\cite{SheDub2010} the basis vectors \eqref{SM_S_basis}
are the common eigenstates of the operators $\mbf{S}_{\textrm{ferm}}^2$ and ${S}_{\textrm{ferm}}^z$:
\be
\mbf{S}_{\textrm{ferm}}^2 |\mbf{p}SM_S\rangle = S(S+1) |\mbf{p}SM_S\rangle,
~~~
{S}_{\textrm{ferm}}^z |\mbf{p}SM_S\rangle = M_S |\mbf{p}SM_S\rangle.
\ee
Then the $np$-pair states  are also $\mbf{S}_{\textrm{ferm}}^2$ eigenvectors
\be
\mbf{S}_{\textrm{ferm}}^2 |\Psi^{(-)}_{\mbf{p} S M_S}\rangle 
=\sum_{M'_S}\int d\mbf{p}'\, \Psi^{(-)}_{\mbf{p}SM_S}(\mbf{p}'SM'_S)
\mbf{S}_{\textrm{ferm}}^2 |\mbf{p}'SM'_S\rangle
= S(S+1) |\Psi^{(-)}_{\mbf{p} S M_S}\rangle.
\ee
The latter follows from the fact that our Hamiltonian is diagonal with respect to the spin quantum numbers
\be
\langle\mbf{p}'S'M'_S| H |\mbf{p}SM_S\rangle =
\d_{S'S}
\langle\mbf{p}'SM'_S| H |\mbf{p}SM_S\rangle.
\ee

The $\mathbb{C}$-number coefficient $\Psi_{M_d}(\mbf{p}SM_S)$ can be written in terms of the S- and D-components of the deuteron WF in momentum space
\be 
  \Psi_{M_d}(\mbf{p}1M_S)=\sum_{l=0,2}
  (lm_l1M_S|1M_d)u_l(|\mbf{p}|)Y_{lm_l}(\hat{\mbf{p}}).
\ee
They have been already calculated in 
Refs.~\cite{She13, Ars2022}.

In a moving frame the corresponding $np$-pair state can be found via the link
\be\label{boost_link}
  |\Psi^{(-)}_{\mbf{q},\mbf{p}_0 SM_S}\rangle
  =
  e^{ -i \vect{\beta} \cdot \mbf{B}}
  |\Psi^{(-)}_{\mbf{p} SM_S}\rangle
  .
\ee
This relation is realized with the boost operator $\mbf{B}=\mbf{B}_F+\mbf{B}_I$, 
where its free part $\mbf{B}_F$ is determined via the Belinfante ansatz $\mbf{B}_F = - \int \mbf{x} K_F(\mbf{x}) d\mbf{x}$ and $K_F(\mbf{x})$ is the free Hamiltonian density.
Perhaps, one should note that the required $P|\Psi^{(-)}_{\mbf{q},\mbf{p}_0 SM_S}\rangle = (E_{np},\vect{q})|\Psi^{(-)}_{\mbf{q},\mbf{p} SM_S}\rangle$ follows from the property of the energy-momentum operator to be four-vector, viz.,
\begin{equation}\label{P_is_4_vector}
  e^{ i \vect{\beta} \cdot \mbf{B}} P^\mu e^{ -i \vect{\beta} \cdot \mbf{B}}
  =
  P^\nu L^{\mu}_{\;\nu}(\vect{\beta}),
\end{equation}
\begin{equation}
  L^{\mu}_{\;\nu}(\vect{\beta})=
  \begin{pmatrix}
    u_0 & \vdots & u_j \\
    \cdots & \cdots & \cdots\\
    -u^i & \vdots & \delta^i_j - \frac{u^iu_j}{1+u_0}
    \end{pmatrix}
\end{equation}
with the Lorentz transformation $(2E_{\vect{p}},\vect{0}) \Rightarrow (E_{np}=E_{\vect{k}_p}+E_{\vect{k}_n},\vect{q})=P$ and 4-velocity vector 
\begin{equation}
  u^\mu=(u_0,\vect{u})=(\textrm{ch}\beta, \vect{n}\,\textrm{sh}\beta) =
  \frac{P^\mu}{\sqrt{E_{np}^2 - \vect{q}^2}}
  .
\end{equation}
Relative momenta in c.m.s. $\vect{p} $ is expressed through $\vect{q}$ and $\vect{p}_0$ in lab system as follows
\begin{equation}
      \vect{p} = 
      \vect{p}_0 + \frac{\vect{q}}{2} + \left(E_{\vect{p}_0 + \frac{\vect{q}}{2}} + \frac{\vect{u} \cdot(\vect{p}_0 + \frac{\vect{q}}{2})}{1+u_0}\right)\vect{u}.
\end{equation}

In order to meet Eq.~\eqref{P_is_4_vector} it is sufficient to meet the Lie-Poincaré commutation relations
\begin{equation}
  [H,\vect{B}] = i \vect{P}, ~~
  [P^i,B^j]= i \delta^{ij} H.
\end{equation}
To apply Eq.~\eqref{boost_link}, one can employ the following operator relation
\begin{equation}\label{BF_BI_separation}
    e^{-i \vect \beta (\mbf B_F + \mbf{B}_I)}=
    Z( e^{-i \vect \beta \mbf B_F}\mbf{B}_Ie^{ i \vect \beta \mbf B_F})
    e^{-i \vect \beta \mbf B_F}  
    ,
\end{equation}
where the operator $Z$ is a measure of non-commutativity of the operators $\mbf B_F$ and $\mbf{B}_I$. 
In this context, let us consider the relation 
\begin{equation}\label{Z_derivation_start}
e^{\tau(a+b)} = Z(\tau)e^{\tau a}
\end{equation}
for arbitrary operators $a$ and $b$ (in our case $a=-i \vect \beta \cdot \mbf B_F$ and $b=-i \vect \beta \cdot \mbf B_I$), where the operator  $Z(\tau)$ meets the equation 
\begin{equation}\label{Z_diff_eq}
\frac{\partial Z(\tau)}{\partial \tau} = Z(\tau)\, e^{\tau a} b e^{-\tau a}.
\end{equation}
Recall that the similarity transformation in this equation is the Campbell--Hausdorff series
\begin{equation}
e^{\tau a} b e^{-\tau a} = \sum\limits_{n=0}^{\infty} \frac{\tau^n}{n!}\big[ \underbrace{a[a \cdots [a}_{n} , b]] \cdots \big].
\end{equation}
There are different ways on solving the equations similar to \eqref{Z_diff_eq}. First of all we keep in mind those based on the Baker--Hausdorff and Magnus theorems \cite{Wei1963, Magnus54}.

Taking into account $U_F(L)=\exp(-i \vect \beta \mbf B_F)$, the separation \eqref{BF_BI_separation} allows us to use the property
\begin{equation}\label{chpt3:b_U_transf_property}
  e^{ -i \vect \beta \mbf B_F} b^\dagger(\vect{p}\mu) e^{ i \vect \beta \mbf B_F} = 
  D^{[1/2]}_{\mu\bar\mu}(W(L,p))b^\dagger(L\vect{p}\bar\mu),
\end{equation}
in order to get
\begin{multline}\label{chpt3:qM_state_boost}
    |\Psi^{(-)}_{\mbf{q},\mbf{p}_0 SM_S}\rangle 
    = \int 
    {d \mbf{p}'}
    \,
    \Psi^{(-)}_{\mbf{p}\, SM_S}\left(\mbf{p}' \bar\mu_1 \bar\mu_2\right)
    D^{[1/2]}_{\bar\mu_1 \mu'_1}(W(L,p'))
    D^{[1/2]}_{\bar\mu_2 \mu'_2}(W(L,p'_-))
    \\\times
    Z
    ( e^{-i \vect \beta \mbf B_F}\mbf{B}_Ie^{ i \vect \beta \mbf B_F})
    b_c^\dagger(L\vect{p}' \mu'_1)b_c^\dagger(L\vect{p}'_- \mu'_2) | \Omega \rangle,
\end{multline}
where the $D$-function depends on the Wigner rotation $W(L,p)$ and we introduced 3-vectors 
\begin{equation}
      (L\vect{p}')^i \equiv L^i_\mu p'^\mu,
      \quad
      (L\vect{p}'_-)^i \equiv L^i_\mu  p'^\mu_-
\end{equation}
with $p'=(E_{\vect{p}'}, \vect{p}')$ and $p'_-=(E_{\vect{p}'}, -\vect{p}')$.
We will confine ourselves to calculations with $Z = 1$, {\clb i.e., we neglect the interaction term in the total boost operator}.

Now, let us address the partial wave decomposition of the $np$-pair final state in the c.m.s.
\begin{equation}
  \langle \Psi^{(-)}_{\mbf{p}SM_ST0}| \equiv \langle \Psi^{(-)}_{\mbf{p}SM_S}| \langle T 0| ,
\end{equation}
i.e.,
\begin{multline}
  \langle \Psi^{(-)}_{\mbf{p}SM_ST0}|\mbf{p}'S'M_{S'}T'0\rangle = \d_{S'S}\d_{T'T}\sum_{J=0}^{J_{\textrm{max}}}\sum_{M_J=-J}^{J} 
  Y_{lm}(\hat{\mbf{p}})(lmSM_S|JM_J)
  \\\times
  \Psi^{\a(-)}_{pll'}(p')(l'm'SM_{S'}|JM_J)Y^*_{l'm'}(\hat{\mbf{p}'}),
\end{multline}
where we restore the indices for the total isospin $T$ and its projection $M_T=0$, $J_{\textrm{max}}$ is the maximum value of the total angular momentum in the $np$-pair partial-wave scattering states $\alpha = \{J,S,T\}$ and summation over $l$, $l'$ as well as $m$ and $m'$ is implied. 
Here we use the basis vectors
\be
|pJ(lS)M_JTM_T\rangle = \int d\hat{\mbf{p}}\,Y_{lm_l}(\hat{\mbf{p}})|\mbf{p}SM_STM_T\rangle (lm_lSM_S|JM_J)
\label{chpt5::JSTbasis}
\ee
and vice versa
\be
|\mbf{p}SM_STM_T\rangle = (lm_lSM_S|JM_J)\,Y^*_{lm_l}(\hat{\mbf{p}})|pJ(lS)M_JTM_T\rangle,
\label{chpt5::JSTbasisReverse}
\ee
with the unit vector $\hat{\mbf{p}}=\mbf{p}/p$. Details of the construction and physical properties of this basis can be found in Appendix B of Ref.~\cite{SheDub2010}.

The partial waves $\Psi^{\a(\pm)}_{pll'}(p')$ are related by
\be
\Psi^{\a(\pm)}_{pll'}(p')=\sum_{l''}{O}^{\a(\pm)}_{ll''}(p)\varphi^{\a}_{pl''l'}(p')
,
\label{chpt5::coupledLink}
\ee
with the WFs $\varphi^{\a}_{pl''l'}(p')$, which have the asymptotics of standing waves.
The latter meet the equation
\begin{equation}\label{standing_waves_eq}
  \varphi^{\a}_{pl'l}(p') = \frac{1}{p'^2}\d(p'-p)\d_{l'l}+P\frac{1}{E_{p}-E_{p'}}R_{l'l}^\a(p',p),
\end{equation}
where elements of the $R$-matrix satisfies its own Lippman-Schwinger type equations.
For coupled partial waves ($l',l=J\pm1$) the matrix ${O}^{\a(\pm)}(p)$ is parametrized as in Ref.~\cite{BlattBiedenharn1952},
\be
\hat{O}^{\a(\pm)}(p)=\begin{pmatrix}\D_-^{\a(\pm)}\cos^2\ve_\a+\D_+^{\a(\pm)}\sin^2\ve_\a&\frac12(\D_-^{\a(\pm)}-\D_+^{\a(\pm)})\sin2\ve_\a\\
\frac12(\D_-^{\a(\pm)}-\D_+^{\a(\pm)})\sin2\ve_\a&\D_-^{\a(\pm)}\sin^2\ve_\a+\D_+^{\a(\pm)}\cos^2\ve_\a\end{pmatrix},
\label{chpt5::Omatrix}
\ee
where $\ve_\a$ is the mixing parameter, while for uncoupled partial waves ($l'=l=J$) we have ${O}^{\a(\pm)}(p) = \D^{\a(\pm)}$.
The quantities $\D^{\a(\pm)}_\pm$ ($\D^{\a(\pm)}$) can be expressed through the corresponding phase shifts $\d^\a_\pm=\d^\a_{J\pm1}$ ($\d^\a_J$): $\D^{\a(\pm)}=\exp(\pm i\d^\a)\cos\d^\a$.

In order to get expressions for the partial states of standing waves, we will employ the so-called matrix inversion method \cite{BrowJackson1969}. Let us consider the typical integral of the generalized functions
\begin{equation}
  I = \int_{0}^{\infty} \varphi(q) h(q) q^2 dq
\end{equation}
with { $h(q)$ from the class of functions on which the quadrature calculation scheme works.} 
Using Eq.~\eqref{standing_waves_eq} one can show
\begin{equation}\label{typical_integral_final_form}
  I = h(N+1) - \sum_{j=1}^{N+1} \Omega_j p_j^2 R(j,N+1) h(j)
  = \sum_{j=1}^{N+1} B(j)h(j),
\end{equation}
\begin{equation}
  B(j) \equiv \delta_{j,N+1} - \Omega_j p_j^2 R(j,N+1)
\end{equation}
so
\begin{equation}
  \varphi(p') = \sum_{j=1}^{N+1} B(j)\frac{\d(p'-p_j)}{p'^2}.
\end{equation}
This idea can be applied to all partial standing waves:
\be
\varphi^\a_{pll'}(p')=\sum_{j=1}^{N+1}B^\a_{pll'}(j)\frac{\d(p'-p_j)}{p_j^2},
\label{chpt5:phiMIM}
\ee
where coefficients $B^\a_{pll'}(j)$ are the solutions of the set of linear algebraic equations approximately equivalent to the underlying Lippman-Schwinger type equation for the $n$-$p$ scattering problem in c.m. system. Here $N$ is the number of Gaussian nodes on the interval $[-1,1]$, $p_j$ are the corresponding grid points and $p_{N+1}=p$ (see details in Ref.~\cite{KorSheKIPTPreprint1977}). 
It is important to stress that the representation \eqref{chpt5:phiMIM} should be meant in an integral sense.
Previously this method was successfully applied for the treatment of
the final–state interaction in \cite{KorMelShe90,LadShe2003,ArsEtAl2021}.
{\clb In Appendix C one can find an additional test of our $np$-pair WF calculations.}

	\subsection{The current density operator in the CPR without partial wave expansions}

	Our matrix elements \eqref{F_mu} with the states \eqref{DState} and \eqref{chpt3:qM_state_boost} are reduced to the expectation value with respect to the physical vacuum $|\Omega\rangle$
\be\label{expec}
	\Big\langle L\vect{p}'\, \mu'_1, L\vect{p}'_-\, \mu'_2 \Big| J^\mu(0) \Big| \mbf{p}{\m}_1,-\mbf{p}{\m}_2 \Big\rangle,
\ee
where $| \mbf{p}_1\m_1,\mbf{p}_2\m_2 \rangle \equiv b^\dagger_c(\mbf{p}_1{\m}_1)b^\dagger_c(\mbf{p}_2{\m}_2)|\Omega\rangle$.

The Noether current operator $J(0)$ in the CPR can be found by using the Campbell-Hausdorff formula \cite{SheShi2001,SheDub2010, FBS2024}
\be
J^\mu(0)=e^{R}J^\mu_c(0)e^{-R}=J_c^\mu(0)+[R,J_c^\mu(0)]+\frac12[R,[R,J_c^\mu(0)]] + ...\,,
\label{JUCT}
\ee
where $J_c^\mu(0)$ is the primary Noether current in which ``bare'' operators $\{\alpha\}$ are replaced by their clothed counterparts $\{\alpha_c\}$ and $R$ is the generator of the clothing transformation $W=\exp(R)$. 
Remind that the Noether current density $J^\mu(\vect{x})= J_{N}^\mu(\vect x) + J_{\mathcal{M}}^\mu(\vect x)$ in the Schrödinger (S) picture consists of the nucleon part
\begin{equation}\label{chpt3:J_N_def}
    J_{N}^\mu(\vect x)= e :\bar\psi(\vect x) \gamma^\mu \frac{1 + \tau^z}{2} \psi(\vect x):,
    ~\textrm{where}~
    \psi=
    \begin{pmatrix}
            \psi_{1/2} \\
            \psi_{-1/2}
    \end{pmatrix},
\end{equation}
and mesonic one.
For the latter, we have the following definitions
\begin{align}
      \label{chpt3:J_ps_def}
      J_{ps}^\alpha(\vect x)= e:[\vect\varphi_{ps}(\vect x) \times \partial^\alpha\vect\varphi_{ps}(\vect x)]^z:,
      \\ 
      \label{chpt3:J_s_def}
      J_{s}^\alpha(\vect x)= e:[\vect\varphi_{s}(\vect x) \times \partial^\alpha\vect\varphi_{s}(\vect x)]^z:,
      \\
      \label{chpt3:J_v_def}
      J_{v}^\alpha(\vect x)= e:[\vect\varphi_{v}^{\alpha\beta}(\vect x) \times \vect\varphi_{v\,\beta}(\vect x)]^z:,
\end{align}
with the isovectors $\vect{\varphi}$.

Decomposition (\ref{JUCT}) involves one-body, two-body and more complicated interaction currents that are often called the meson exchange currents (MECs), if one uses the terminology which is adopted in the theory of e.m. interactions with nuclei.

In Eq.~(\ref{expec}) the current density operator $J^\mu(0)$ is sandwiched between the clothed two-nucleon states. As shown in Ref.~\cite{FBS2024} for such expectation values the current $J^\mu(0)$ will contribute only with its one-body $J_{[1]}$ and two-body $J_{[2]}$ parts 
\be
\langle \mbf{p}'_1\m'_1,\mbf{p}'_2\m'_2|J^\mu(0)|\mbf{p}_1\m_1,\mbf{p}_2\m_2\rangle = \langle \mbf{p}'_1\m'_1,\mbf{p}'_2\m'_2|
(J^{\mu}_{[1]} + J^{\mu}_{[2]})
|\mbf{p}_1\m_1,\mbf{p}_2\m_2\rangle,
\label{1n2n}
\ee
which, in general, can be written as 
\be\label{one-body_current}
J_{[1]}^\m = \sum_{\mu'\mu}\int d\mbf{p}'d\mbf{p}\,F^\m(\mbf{p}'\mu',\mbf{p}\mu)b_c^\dagger(\mbf{p}'\mu')b_c(\mbf{p}\mu),
\ee
\be
J_{[2]}^\m = \sum_{\m'_1\m'_2,\m_1\m_2}\int d\mbf{p}'_1d\mbf{p}'_2d\mbf{p}_1d\mbf{p}_2 F^\m_{MEC}(\mbf{p}'_1\m'_1,\mbf{p}'_2\m'_2;\mbf{p}_1\m_1,\mbf{p}_2\m_2)b_c^\dagger(\mbf{p}'_1\m'_1)b^\dagger_c(\mbf{p}'_2\m'_2)b_c(\mbf{p}_1\m_1)b_c(\mbf{p}_2\m_2).
\ee

The \mbox{$\mathbb{C}$-number} functions $F^\mu$ and $F_{MEC}^\mu$ can be found from the series \eqref{JUCT} by separating the respective $b_c^\dagger b_c$ and $b_c^\dagger b_c^\dagger b_c b_c$ contributions
\begin{equation}\label{chpt3:J1N_J2N_series}
      \begin{split}
            &J_{[1]}= J_{c}(0)_{b_c^\dagger b_c} + [R,J_{c}(0)]_{b_c^\dagger b_c} + \frac{1}{2} [R,[R,J_{c}(0)]]_{b_c^\dagger b_c} + \cdots,
            \\
            &J_{[2]}= J_{c}(0)_{b_c^\dagger b_c^\dagger b_c b_c} + [R,J_{c}(0)]_{b_c^\dagger b_c^\dagger b_c b_c} + \frac{1}{2} [R,[R,J_{c}(0)]]_{b_c^\dagger b_c^\dagger b_c b_c} + \cdots.
      \end{split}
\end{equation}
It is important to stress that the many-nucleon currents introduced in such a way do not depend on the choice of states between which we calculate the amplitude. In particular, it means that the one-nucleon matrix elements 
\begin{equation}
      \langle \vect{p}'\mu' | {J}_h(0)| \vect{p}\mu \rangle = 
	  \langle \vect{p}'\mu' | J_{[1]} | \vect{p}\mu \rangle = F(\vect{p}'\mu',\vect{p}\mu),
\end{equation}
where $| \vect{p}\mu \rangle = b_c^\dag(\vect{p}\mu) | \Omega \rangle$ and the operator $J_{[1]}$ is the same as in Eq.~\eqref{1n2n}.

For the one-nucleon current we have an alternative way to find the function $F(\vect{p}'\mu',\vect{p}\mu)$ than from the series \eqref{chpt3:J1N_J2N_series}.
It follows from the observation that the primary Noether current density, being sandwiched between the one-nucleon states, yields the usual on-mass-shell expression
\be\label{F_one-body}
F^\m(\mbf{p}'\mu',\mbf{p}\mu)=\bar{u}(\mbf{p}'\mu')\left[F_1[(p'-p)^2]\g^\m+i\s^{\m\n}\frac{(p'-p)_\n}{2m}F_2[(p'-p)^2]\right]u(\mbf{p}\mu)
\ee
in terms of the Dirac ($F_{1}$) and Pauli ($F_{2}$) nucleon FFs (see, for example, p.~242 in Ref.~\cite{BjorkenDrell}).
That fact allows us to use the form \eqref{F_one-body} to determine the \mbox{$\mathbb{C}$-number} coefficient in \eqref{one-body_current} instead of evaluating the infinite number of contributions in the series \eqref{chpt3:J1N_J2N_series}. It is impossible for the two-nucleon contribution that can be interpreted as a meson exchange current, so the determination of $F_{MEC}$ requires sequential evaluation of the $[R,\cdots [R,J_c(0)]\cdots ]$ commutators. The total hadronic current density $J_c$ composed of the meson $J_{\mathcal{M},c}$ and nucleon $J_{N,c}$ currents so the two-nucleon meson exchange current density also composed of the two parts
\begin{equation}
      J_{[2]} = e^{R}\left(J_{\mathcal{M},c}(0)+J_{N,c}(0)\right)e^{-R}\rvert_{b_c^\dagger b_c^\dagger b_c b_c}
      \equiv J_{\mathcal{M}CC} + J_{\mathcal{M}NN}
\end{equation}
with the so-called mesonic $J_{\mathcal{M}CC}$ and seagull $J_{\mathcal{M}NN}$ currents.

In Refs.~\cite{YanThesis, FBS2024} using the recursive technique for computing the multiple commutators from Ref.~\cite{KoCaSh2007} we considered the first non-vanishing contribution to the two-nucleon current from the series \eqref{chpt3:J1N_J2N_series} that stem from the commutator $\frac12[R,[R,J_c^\mu(0)]]$. Thus, we have found $F^\m_{MEC}$ for the model of interacting clothed nucleons and $\pi$, $\rho$, $\omega$, $\eta$, $\delta$, $\sigma$ mesons. Explicit expressions for the MEC operators can be found in Appendix B.

At this point, let us discuss the distinctive features of our MECs in comparison with several other models proposed in the literature. 
In Sec.~\ref{sec:results_and_discussion} our numerical results are compared with those of Ref.~\cite{KorMelShe90}, where the MECs of Maize and Kim \cite{maize1984} were used. 
Their mesonic current and seagull current (called the ``pair current'' in Ref.~\cite{maize1984}) are given by
\begin{equation}\label{maize_kim_CC}
      \mathbf{j}_{\pi C C}
      =
      \frac{i g_\pi^2}{2m^2(2 \pi)^3} 
      F_\pi\left(Q^2\right) \left[\boldsymbol{\tau}_1 \times \boldsymbol{\tau}_2\right]_z\left(\boldsymbol{\sigma}_1 \cdot \mathbf{k}_1\right)\left(\boldsymbol{\sigma}_2 \cdot \mathbf{k}_2\right) 
      \frac{\mathbf{k}_2-\mathbf{k}_1}{\mathbf{k}_2^2-\mathbf{k}_1^2} 
      \left[\frac{f_{\pi N N}^2\left(\mathbf{k}_1\right)}{\mathbf{k}_1^2+m_\pi^2}
      -\frac{f_{\pi N N}^2\left(\mathbf{k}_2\right)}{\mathbf{k}_2^2+m_\pi^2}\right],
\end{equation}
and 
\begin{equation}\label{maize_kim_NN}
      \mathbf{j}_{\pi N N}
      =
      \frac{i g_\pi^2}{4 m^2 (2 \pi)^3}
      F_1^V\left(Q^2\right)\left[\boldsymbol{\tau}_1 \times \boldsymbol{\tau}_2\right]_z
      \left[\frac{\left(\boldsymbol{\sigma}_2 \cdot \mathbf{k}_2\right) \boldsymbol{\sigma}_1}{\mathbf{k}_2^2+m_\pi^2} f_{\pi N N}^2\left(\mathbf{k}_2^2\right)\right. 
      \left.-\frac{\left(\boldsymbol{\sigma}_1 \cdot \mathbf{k}_1\right) \boldsymbol{\sigma}_2}{\mathbf{k}_1^2+m_\pi^2} f_{\pi N N}^2\left(\mathbf{k}_1^2\right)\right]
\end{equation}
with momenta $\mathbf{k}_1=\mathbf{p}_1-\mathbf{p}_1^{\prime}, \mathbf{k}_2=\mathbf{p}_2-\mathbf{p}_2^{\prime}$, $\pi NN$ vertex cutoff function
\begin{equation}\label{nrl_cutoff_def}
      f_{\pi N N}\left(\mathbf{k}^2\right)=\frac{\Lambda_\pi^2-m_\pi^2}{\Lambda_\pi^2+\mathbf{k}^2},
\end{equation}
e.m. pion FF $F_\pi$ and isovector nucleon FF $F_1^V = F_1^p - F_1^n$.
Being derived from Feynman diagrams, these expressions represent the matrix elements defined only on the energy shell. In contrast, in our approach we have no need to ``invent'' any analytical extensions to the off-energy-shell region.
For comparison with Ref.~\cite{maize1984}, we consider the non-relativistic reduction of the two-nucleon matrix elements of the pion-induced currents \eqref{FpiCC} and \eqref{FpiNN}:
\begin{multline}\label{JpiCC_nrl}
      \langle 1',2'|\vect{J}_{\pi C C}|1,2 \rangle 
      \approx
      \frac{i g_\pi^2}{2 m^2 (2 \pi)^6} \left[\boldsymbol{\tau}_1 \times \boldsymbol{\tau}_2\right]_z\left(\boldsymbol{\sigma}_1 \cdot \mathbf{k}_1\right)\left(\boldsymbol{\sigma}_2 \cdot \mathbf{k}_2\right) 
      \frac{\mathbf{k}_2-\mathbf{k}_1}{\mathbf{k}_2^2-\mathbf{k}_1^2}
      \left[\frac{f_{\pi N N}\left(\mathbf{k}_1^2\right) f_{\pi N N}\left(\mathbf{k}_2^2\right)}{\mathbf{k}_1^2+m_\pi^2}
      \right.
      \\
      \left.
      -\frac{f_{\pi N N}\left(\mathbf{k}_1^2\right) f_{\pi N N}\left(\mathbf{k}_2^2\right)}{\mathbf{k}_2^2+m_\pi^2}\right]
      ,
\end{multline}
\begin{equation}\label{JpiNN_nrl}
      \langle 1',2'|\vect{J}_{\pi N N}|1,2 \rangle 
      \approx
      \frac{i g_\pi^2}{4 m^2(2 \pi)^6}
      \left[\boldsymbol{\tau}_1 \times \boldsymbol{\tau}_2\right]_z 
      \left[
      \frac{\left(\boldsymbol{\sigma}_2 \cdot \mathbf{k}_2\right) \boldsymbol{\sigma}_1}{\mathbf{k}_2^2+m_\pi^2} f_{\pi N N}^2\left(\mathbf{k}_2^2\right)
      -
      \frac{\left(\boldsymbol{\sigma}_1 \cdot \mathbf{k}_1\right) \boldsymbol{\sigma}_2}{\mathbf{k}_1^2+m_\pi^2} f_{\pi N N}^2\left(\mathbf{k}_1^2\right)
      \right]
      .
\end{equation}
Note that the cutoff function \eqref{cutoff_def} reduces, in the non-relativistic limit, to \eqref{nrl_cutoff_def}, whereas the time components $\langle 1',2'|{J}^{0}_{\pi C C,\,\pi N N}|1,2 \rangle$ are smaller by two orders in $|\vect{p}|/m$ than the spatial ones.
The expressions \eqref{JpiCC_nrl}--\eqref{JpiNN_nrl} differ from the Maize--Kim currents in two aspects. First, they do not contain the phenomenological e.m. pion $F_\pi$ and isovector nucleon $F_1^V$ FFs. Second, the cutoff dependence in the mesonic current is different: Eq.~\eqref{JpiCC_nrl} contains the product $f_{\pi NN}(\mathbf{k}_1^2)f_{\pi NN}(\mathbf{k}_2^2)$, whereas Eq.~\eqref{maize_kim_CC} contains $f_{\pi NN}^2(\mathbf{k}_1^2)$ and $f_{\pi NN}^2(\mathbf{k}_2^2)$. Therefore, in the non-relativistic limit our pion-exchange MECs reduce, up to an overall normalization factor, to the Maize--Kim currents if one sets $F_\pi=F_1^V=1$ and takes the cutoff parameter $\Lambda_\pi\to\infty$.

One may also compare currents \eqref{JpiCC_nrl}--\eqref{JpiNN_nrl} with those from Ref.~\cite{Arenhovel1997} (see Eqs.~(A12) and (A17) therein). The only difference is the presence of the isovector FF $F_{1}^V$ in the latter. 
The authors of Ref.~\cite{Arenhovel1997} also provide  relativistic corrections to these currents, but a detailed comparison with those terms is less straightforward. In addition to the mesonic and seagull two-body currents (called exchange and contact currents, respectively, in Ref.~\cite{Arenhovel1997}), they include $\Delta$-isobar currents and the so-called dissociation currents associated with the $\gamma \pi \rho$ and $\gamma \pi \omega$ vertices. Such mechanisms are introduced there in a non-minimal way and are absent from our primary Lagrangian. We plan to consider such extensions of the considered field model in future research.

It is also important to emphasize that the currents \eqref{maize_kim_CC}--\eqref{maize_kim_NN} are purely isovector, whereas our MECs contain both isovector and isoscalar contributions. For the pseudoscalar-meson exchange, the isoscalar currents are suppressed in the non-relativistic limit $\langle 1',2'|\vect{J}^{is}_{\pi N N}|1,2 \rangle \approx 0$, while
the scalar- and vector-meson exchange generate nonvanishing isoscalar currents. 
From Eq.~\eqref{FdeltaNN} we get
\begin{equation}\label{FsigmaNN_is_nrl}
      \langle 1',2'|\vect{J}^{is}_{\sigma N N}|1,2 \rangle
      \approx
    \frac{e g_\sigma^2 }{4 m^2 (2 \pi)^6} 
    (1 + \tau_{1}^z)
    \frac{f_{\sigma NN}^2\left(\vect{k}_2^2\right)}{\vect{k}_2^2 + m_\sigma^2}
    \left\{
      (\vect{p}'_1 + \vect{p}_1)
      -
      i[\vect{\sigma}_{1} \times (\vect{k}_1 + \vect{k}_2)]
    \right\} 
    + \left(1 \leftrightarrow 2\right)
    .
\end{equation}
The corresponding result for the vector boson ($\omega$ meson) exchange looks as
\begin{multline}\label{FomegaNN_is_nrl}
      \langle 1',2'|\vect{J}^{is}_{\omega N N}|1,2 \rangle
      \approx
      \frac{ e g_\omega^2 }{4 m^2 (2 \pi)^6} 
      (1 + \tau_{1}^z)
      \frac{f_{\omega NN}^2\left(\vect{k}_2^2\right)}{\vect{k}_2^2 + m_\omega^2} 
    \bigg\{
      (\vect{p}'_2 + \vect{p}_2)
      \\
      -
      \left(1 + \frac{f_\omega}{g_\omega}
      \right)i\left[(
      \vect{\sigma}_{1}
      + \vect{\sigma}_{2}
      ) \times\vect{k}_2\right]
    \bigg\}
    + \left(1 \leftrightarrow 2\right)
    .
\end{multline}
Here we have also assumed $4m^2 \gg m_{\sigma,\omega}^2$.
For the mesons with isospin $T=1$ ($\delta$, $\sigma(1)$ and $\rho$), one should replace $1 + \tau_{1}^z \mapsto \vect\tau_{1} \cdot \vect\tau_{2} + \tau_{2}^z$. 
The expressions \eqref{FsigmaNN_is_nrl} and \eqref{FomegaNN_is_nrl} have much in common with the isoscalar currents derived in Refs. \cite{Ichii1987, Blunden1986}. For example, compare them with exchange currents in Eqs. (A.1a)-(A.1b) of Ref. \cite{Ichii1987}:
\begin{equation}\label{ichii_sigmaNN}
      \vect{j}_{\sigma NN}^{is} =
      \frac{e g_\sigma^2 }{4 m^2} (1 + \tau_{1}^z) \frac{1}{\vect{k}_2^2 + m_\sigma^2}
      \left\{
      (\vect{p}'_1 + \vect{p}_1) 
      -
      i[\vect{\sigma}_{1} \times (\vect{k}_1 + \vect{k}_2)] 
      \right\}
      + \left(1 \leftrightarrow 2\right),
\end{equation}
\begin{equation}\label{ichii_omegaNN}
      \vect{j}_{\omega NN}^{is} =
      \frac{e g_\omega^2 }{4 m^2} (1 + \tau_{1}^z) \frac{1}{\vect{k}_2^2 + m_\omega^2}
      \bigg\{
            (\vect{p}'_2 + \vect{p}_2)
            -
            \left(1 + \frac{f_\omega}{g_\omega} \right)i\left[(
            \vect{\sigma}_{1}
            + \vect{\sigma}_{2}
            ) \times \vect{k}_2 \right]
      \bigg\}
      + \left(1 \leftrightarrow 2\right).
\end{equation}
Once again, these expressions coincide with \eqref{FsigmaNN_is_nrl}--\eqref{FomegaNN_is_nrl}, up to an overall normalization factor, in the limit $\Lambda_\pi \to \infty$.

To summarize, the main features of our MECs are:
\begin{itemize}
      \item They are consistent with the nuclear Hamiltonian in the CPR.
      \item Instead of a common procedure of extracting MECs from the Feynman
      diagrams and extrapolating them to the off-shell region, we derive our MECs just from the given conserved nucleon and meson Noether currents. 
      \item The MEC operators do not depend on the choice of states between which we calculate the amplitude that considerably expands the range of applications.
      \item They contain both isovector and isoscalar contributions. The latter are often omitted in other approaches, although they are essential, for example, in elastic $ed$ scattering, where isovector currents do not contribute.
      \item Our approach produces relativistic expressions for the MECs, without any non-relativistic reduction. This allows us to use them in a broad range of energies and momentum transfers.
\end{itemize}

\subsection{Details of computations}

Using the commutation relations \eqref{bb_commutator} 
and 
the antisymmetric property of the deuteron WF,
we get
\begin{multline}\label{MEFinal}
	\mathcal{F}^\m(\mbf{p}_0,\mbf{q})=2\sum \int d\mbf{p}'d\mbf{p} 
    \,
      \Psi^{(-)*}_{\mbf{p}_0 SM_S}(\mbf{p}'\bar\mu_1\bar\mu_2)
	D^{[1/2]*}_{\bar\mu_1 \mu'_1}(W(L,p'))
    D^{[1/2]*}_{\bar\mu_2 \mu'_2}(W(L,p'_-))
	\\
	\bigg\{J^\m\left(L\vect{p}'\mu'_1, L\vect{p}'_-\mu'_2,\mbf{p}{\m}_1,-\mbf{p}{\m}_2\right)
	+J^\m\left(L\vect{p}'_-\mu'_2,L\vect{p}'\mu'_1,-\mbf{p}{\m}_2,\mbf{p}{\m}_1\right)\bigg\}\Psi_{M_d}(\mbf{p}{\m}_1{\m}_2),
\end{multline}
where the \mbox{$\mathbb{C}$-number} function $J^\m(\mbf{p}'_1\mu'_1,\mbf{p}'_2\mu'_2;\mbf{p}_1\mu_1,\mbf{p}_2\mu_2)$ consists of the one-body and two-body terms
\be
J^\mu(\mbf{p}'_1\mu'_1,\mbf{p}'_2\mu'_2;\mbf{p}_1\mu_1,\mbf{p}_2\mu_2)=J^\mu_{[1]}(\mbf{p}'_1\mu'_1,\mbf{p}'_2\mu'_2;\mbf{p}_1\mu_1,\mbf{p}_2\mu_2)+J^\mu_{[2]}(\mbf{p}'_1\mu'_1,\mbf{p}'_2\mu'_2;\mbf{p}_1\mu_1,\mbf{p}_2\mu_2)
\label{JME}
\ee
with
\be
J^\mu_{[1]}(\mbf{p}'_1\mu'_1,\mbf{p}'_2\mu'_2;\mbf{p}_1\mu_1,\mbf{p}_2\mu_2)=\delta(\mbf{p}'_2-\mbf{p}_2)\delta_{\m'_2\m_2}F^\m(\mbf{p}'_1\m'_1,\mbf{p}_1\m_1),
\label{1bdyJME}
\ee
\be
J^\mu_{[2]}(\mbf{p}'_1\mu'_1,\mbf{p}'_2\mu'_2;\mbf{p}_1\mu_1,\mbf{p}_2\mu_2)=-F^\mu_{MEC}(\mbf{p}'_1\mu'_1,\mbf{p}'_2\mu'_2;\mbf{p}_1\mu_1,\mbf{p}_2\mu_2).
\label{2bdyJME}
\ee
Here after $\mbf{p}_0$ is the $np$-pair relative momentum in c.m.s.

Further, it would be helpful to show the spin dependence of the quantities \eqref{JME}. Relying upon previous experience  \cite{KorMelShe90} it can be done as follows
\begin{multline}\label{spin_tensor_structure}
  J^\m\left(Lp'\mu'_1, Lp'_-\mu'_2,\mbf{p}\mu_1,-\mbf{p}\mu_2\right)
  \equiv 
  \langle \m'_1\m'_2| {J}(\mbf{p}',\mbf{p},\mbf{q}) |\m_1\m_2 \rangle
  \\
  =
  \sum_{k_1k_2,k\kappa}
  \langle \mu'_1 \mu'_2|\{\s_{k_1}(1)\otimes\s_{k_2}(2)\}_{k\k}|\mu_1\mu_2\rangle
  J^{k_1k_2,\mu}_{k\k}(\mbf{p}',\mbf{p},\mbf{q}), 
\end{multline}
with the irreducible tensor product $\{\s_{k_1}\otimes\s_{k_2}\}_{k}$ of rank $k$ and the spin-tensors $\s_{00}=\mathbb{I}$, $\s_{1\k}=(\mbfs{\s})_\k$.
Henceforth, $(\vect{a})_{\k=0,\pm1} = a_{0,\pm1}$ denotes the cyclic components for any 3-vector $\vect{a}$:
\begin{equation}
  a_{\pm1} = \mp \frac{\sqrt{2}}{2}(a_x \pm i a_y),
  ~~~
  a_0 = a_z,
  ~~
  a^{\k}= (-1)^\k a_{-\k}.
\end{equation}

The components of the spatial tensor $J^{k_1k_2,\m}_{k\k}$ can be easily found by taking the traces
\be
\begin{split}
    &J^{00,\m}_{00}(\mbf{p}', \mbf{p},\mbf{q}) = \frac14 \textrm{Tr}\left( {J}(\mbf{p}',\mbf{p},\mbf{q})\right),
    \\
    &J^{10,\m}_{1\k} (\mbf{p}',\mbf{p},\mbf{q}) = \frac14 \textrm{Tr}\left( {J}(\mbf{p}',\mbf{p},\mbf{q}) \sigma^{\k}(1)\right),
    \\
    &J^{01,\m}_{1\k} (\mbf{p}',\mbf{p},\mbf{q}) = \frac14 \textrm{Tr}\left( {J}(\mbf{p}',\mbf{p},\mbf{q}) \sigma^{\k}(2)\right),
    \\
    &J^{11,\m}_{1\k} (\mbf{p}',\mbf{p},\mbf{q}) = 
    \frac14 \textrm{Tr}\left( {J}(\mbf{p}',\mbf{p},\mbf{q})
    \sum_{\k_1\k_2} \sigma^{\k_1}(1) \sigma^{\k_2}(2) (1\,\k_1\,1\,\k_2|k\,\k) \right),
\end{split}
\ee
where $\textrm{Tr}\left(O\right) \equiv \sum_{\m_1\m_2} \langle \m_1\m_2| O |\m_1\m_2 \rangle$.

\bigskip

Substituting Eq.~\eqref{spin_tensor_structure} into Eq.~\eqref{MEFinal} we obtain the six-dimensional overlapping integrals
\be\label{FSIintegral}
	S^{\alpha}_{ll'm'LM_L}(\mbf{p}_0,\mbf{q})=\int d\mbf{p}d\mbf{p}'\psi^{\a}_{p_0ll'}(p')Y^*_{l'm'}(\hat{\mbf{p}}')J\left(\mbf{p}',\mbf{p},\mbf{q}\right)u_{L}\left(p\right)Y_{LM_L}\left(\hat{\mbf{p}}\right),
\ee
so the one-body current contributes via the 3D integrals
\be\label{1bdyIntegrals}
	S^{[1]\,\alpha}_{ll'm'LM_L}(\mbf{p}_0,\mbf{q})=\int d\mbf{p}' \psi^{\a}_{p_0ll'}(p')Y^*_{l'm'}(\hat{\mbf{p}}')
	J\left(\vect{p}'-\tfrac12\vect{q},\mbf{q}\right)
	u_{L}\left(\left|\vect{p}'-\tfrac12\vect{q}\right|\right)Y_{LM_L}\left(\widehat{\vect{p}'-\tfrac12\vect{q}}\right).
\ee
Their measure is decreasing owing to the
delta functions in the formula \eqref{chpt5:phiMIM},
while the remaining angular integrals in \eqref{1bdyIntegrals}
can be expressed through the quantities
\begin{equation}\label{chpt3:KorMelShe_int_def}
    I_{\lambda m,LM}(p,\vect{q}) = \int d\hat{\vect{p}} Y_{\lambda m}(\hat{\vect{p}})\, Y_{LM}(\widehat{\vect{p}-\tfrac12\vect{q}}) u_{L}(|\vect{p}-\tfrac12\vect{q}|)
	.
\end{equation}
In its turn, 
the Wigner–Eckart theorem
allows us to express them through the reduced quantities $a_L^{\lambda \lambda'}(p,q)$
\begin{multline}\label{chpt3:KorMelShe_int_res}
    I_{\lambda m,LM}(p,\vect{q}) =\sum_{\lambda'} \sqrt{\frac{(2L+1)(2\lambda+1)}{4 \pi (2\lambda'+1)}} (\lambda m LM|\lambda' \, m +M)
    (\lambda 0\,L 0|\lambda' 0)\, Y_{\lambda'  m+M}(\hat{\vect{q}}) a_L^{\lambda \lambda'}(p,q)
\end{multline}
(see details in Ref.~\cite{KorMelShe90}).
That gives us analytic results for the integrals \eqref{1bdyIntegrals} for a given deuteron WFs model.

We could reduce the integrals \eqref{FSIintegral} to the two-dimensional ones via the same technique,
but here we prefer to proceed more straightforwardly
by addressing the Korobov integration method \cite{Korobov1963}. This efficient method was successfully applied in studying the two-body $\,^3He$ breakup by linearly polarized photons \cite{KotMelShe95}.

	\subsection{Towards gauge-independent description}
	\label{sec:GI}
	
	According to Eq.(\ref{amp_in_OPEA}), the amplitude of the process $e+i\to e'+f$ within the OPEA, is expressed as
	\be
	T_{if}=\varepsilon_\m(e',e)\langle f; \mbf{P}_f|J^\m_h(0)|i;\mbf{P}_i\rangle
	\label{em_amp}
	\ee
	with the virtual photon polarization vector $\varepsilon_\mu (e',e) = \bar u_{e}(k^\prime) \gamma_\mu u_e(k)$. The initial $|h\rangle=|i; \mbf{P}_i\rangle$ and final $|h'\rangle=|f;\mbf{P}_f\rangle$ hadronic states are the eigenvectors of the operator $P^\m=(H_h,\mathbf{P})$
	\be
	\begin{split}
	&H_h|i;\mbf{P}_i\rangle= E_i|i;\mbf{P}_i\rangle,\;\;\;\;\;\; \mbf{P}|i;\mbf{P}_i\rangle=\mbf{P}_i|i;\mbf{P}_i\rangle\\
	&H_h|f;\mbf{P}_f\rangle= E_f|f;\mbf{P}_f\rangle,\;\;\mbf{P}|f;\mbf{P}_f\rangle=\mbf{P}_f|f;\mbf{P}_f\rangle
	\end{split}
	\ee
	where $H_h$ is a model Hamiltonian for the system of interacting meson and nucleons. 
	
	The current density operator $J^\m_h(\mbf{x})\equiv J^\m(t=0,\mbf{x})$ meets the continuity equation (CE)
	\begin{equation}\label{continuity_equation_coord}
	i \,\textrm{div} \mbf{J}(\mbf{x})= [H_h,\rho_h(\mbf{x})]
	\end{equation}
	or in more compact form
	\begin{equation}\label{GIP_compact}
	[P_\mu, J^\mu(0)]= 0.
	\end{equation}
	It is the $e^1$-order consequence of the gauge invariance principle that can be formulated via the Fock-Weyl criterion (for details see Ref. \cite{She2014}).
	Of great interest is the gauge independence condition
	\begin{equation}\label{chpt3:GI}
	q_\mu \langle f; \mathbf{P}_i+\mathbf{q} | J^\mu(0) |i; \mathbf{P}_i \rangle =0
	\end{equation}
	that follows directly from the CE being sandwiched between the exact $P^\m$ eigenstates. 
	
	It becomes clear that the amplitude (\ref{em_amp}) is GI being equal to zero after the replacement ${\varepsilon_\mu (e',e) \to q_\mu}$, only if, first, Eq.~\eqref{GIP_compact} fulfills and, second, the states $| f\rangle$, $| i \rangle$ are the exact $H_h$ eigenstates. The latter should be kept in mind in practical calculations, when one has to handle approximate states of a hadronic system.

	In our consideration, to provide the GI of calculations, we would like to make use of the extension of the Siegert theorem (cf. Ref.~\cite{FriFal1986}) proposed in \cite{She1989, LevShe1993}. This way allows us to express the amplitude of interest in explicitly GI way
	through the Fourier transforms of electric ($\mbf{E}(\mbf{q})$) and magnetic ($\mbf{H}(\mbf{q})$) field strengths
	\begin{equation}\label{amp_in_GI_form}
	\begin{split}
	&T_{if}=\mbf{E}(\mbf{q})\mbf{D}_{if}(\mbf{q})+\mbf{H}(\mbf{q})\mbf{M}_{if}(\mbf{q}),\\
	\mbf{E}(\mbf{q})=&i(\w\mbfs{\ve}(\mbf{q})-\mbf{q}\ve_0(\mbf{q})),\;\;\mbf{H}(\mbf{q})=i\mbf{q}\times\mbfs{\ve}(\mbf{q}),
	\end{split}
	\end{equation}
	with $\mbf{D}_{if}(\mbf{q})$ and $\mbf{M}_{if}(\mbf{q})$ being matrix elements of generalized electric and magnetic dipole moments of the hadronic system containing the information on the nuclear dynamics
	\be
	\mbf{D}_{if}(\mbf{q})=i\w^{-1}\int_0^1d\lambda\nabla_{\lambda\mbf{q}}\left[(E_f(\lambda\mbf{q})-E_i)\langle f; \mbf{P}_i+\lambda\mbf{q}|\r(0)|i; \mbf{P}_i\rangle\right],
	\ee
	\be
	\mbf{M}_{if}(\mbf{q})=i\int_0^1\lambda d\lambda\left[\nabla_{\lambda\mbf{q}}\times\langle f; \mbf{P}_i+\lambda\mbf{q}|\mbf{J}(0)|i; \mbf{P}_i\rangle\right]
	\ee
	These equations have been derived using the GI condition (\ref{GIP_compact}), the property of translational invariance
	\begin{equation}
	J^\mu(\mbf{x}) = e^{-i \mbf{x} \cdot \mbf{P}} J^\mu(\mbf{0}) e^{i \mbf{x} \cdot \mbf{P}}
	\end{equation}
	and the formula \cite{Foldy1953}
	\begin{equation}
	\mbf{a} e^{i \mbf{b} \cdot \mbf{c}} =
	\int_{0}^{1}d\lambda \left(
	\nabla_{\mbf{c}}((\mbf{a} \cdot \mbf{c}) e^{i \lambda \mbf{b} \cdot\mbf{c}}) + i \lambda [\mbf{c}\times[\mbf{a}\times\mbf{b}]]e^{i \lambda \mbf{b} \cdot \mbf{c}}
	\right)
	\end{equation}
	with arbitrary vectors $\mbf{a}$, $\mbf{b}$ and $\mbf{c}$.

	Within this approach the matrix element (\ref{F_mu}) now expressed in terms of $\mbf{D}_{if}(\mbf{q})$ and $\mbf{M}_{if}(\mbf{q})$
	\be
	\mathcal{F}^0_{if}=-i\mbf{q}\cdot \mbf{D}_{if}(\mbf{q}), \;\; 
      \mbf{\mathcal{F}}_{if}=-i\w\mbf{D}+i\mbf{q}\times\mbf{M}_{if}(\mbf{q}),
	\label{F_in_GIform}
	\ee
	and satisfy the continuity equation (\ref{CEforF}) by construction. From the practical point of view Eq.~(\ref{F_in_GIform}) gives an opportunity to study separately the influence of the electric and magnetic components on the respective process. In our opinion, this is an attractive feature of such approach.  

      Up to now, we have not paid  attention to another important property of the current density operator, viz., to be four-vector
      \begin{equation}\label{J_is_4_vector}
		e^{ i \vect{\beta} \cdot \mbf{B}}J^\mu(x) e^{ -i \vect{\beta} \cdot \mbf{B}}
		=
		J^\nu(Lx) L^{\mu}_{\;\nu}.
	\end{equation}

	In its turn, the latter leads to the 
	so-called local analogue of the Siegert theorem 
	\begin{equation}
		i \vect{J}(x=0)=[\vect{B},J^0(0)],
	\end{equation}
	that we encounter in Ref.~\cite{Shebeko1990} instead of
	Siegert's relation 
	\begin{equation}
		\vect{J}(\vect{q}=0) = i [\mathscr{H}, \vect{\mathscr{D}}]
	\end{equation}
	for integral quantities 
	\begin{equation}
	\vect{J}(\mbf{q}) \equiv 
	\int \vect{J}(\mbf{x}) e^{i\mbf{q}\cdot\mbf{x}} \, d\mbf{x},
	\quad
	\vect{\mathscr{D}} \equiv  \int  \mbf{x} 
	J^0(\mbf{x}) \, d\mbf{x},
	\quad
	\mathscr{H} \equiv  \int H(\mbf{x}) \, d\mbf{x}.
	\end{equation}
	This result justifies the title of the article 
      \cite{Shebeko1990}
      and allows us to restore the spatial part of the current $\vect{J}(0)$ from the time component $J^0(0)$. 
	In this context, the recipe for constructing the boost operators $\vect{B}$ in the CPR has been given in Ref.~\cite{FroShe2012}.

	We will come back to such GI and covariant calculations somewhere else.
\begin{figure}[t]
  \centering
  \includegraphics[width=\linewidth]{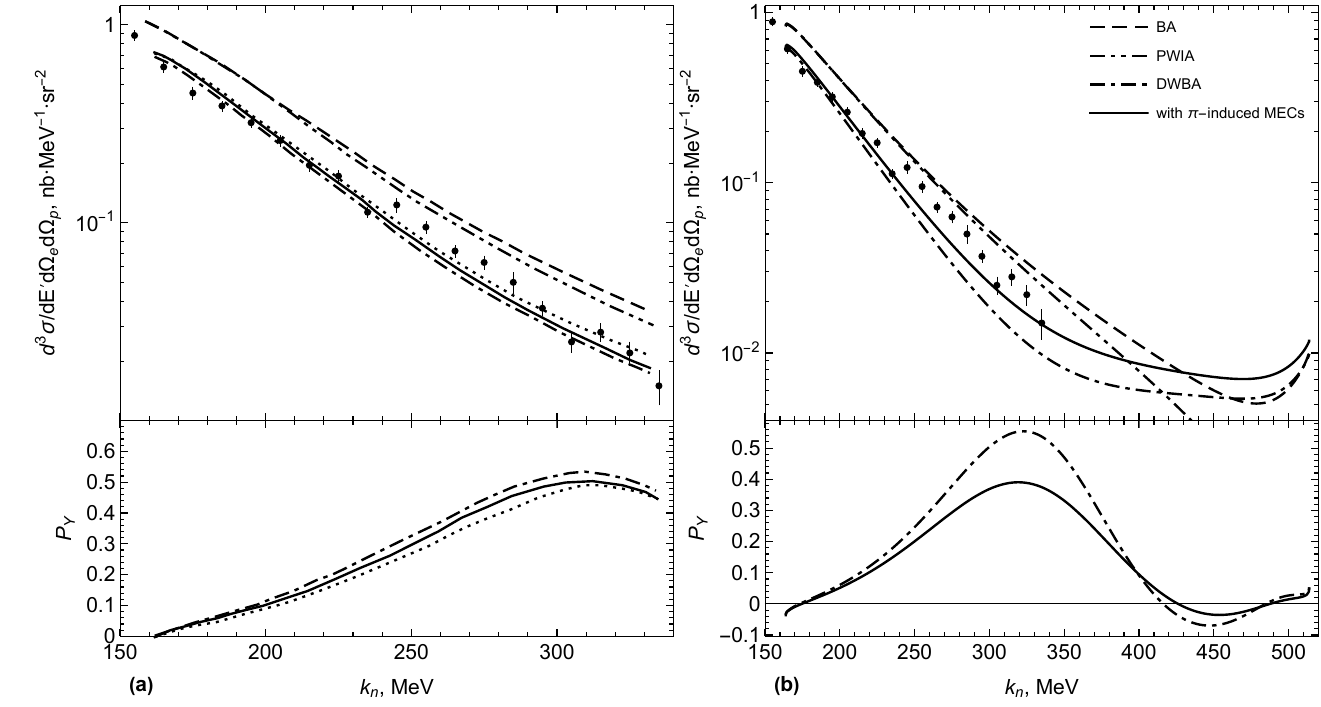}
  \caption{
  Differential cross section (upper part) and induced polarization (lower part) of outgoing protons  versus the neutron momentum for the Saclay kinematics \cite{Saclay1981}. The azimuth angle (reaction plane angle) is fixed $\varphi = 180^\circ$.
  \\
  Panel a: The results of Ref.~\cite{KorMelShe90} obtained within the non-relativistic approach with the Paris potential and pion-induced MECs by Maize and Kim \cite{maize1984}. 
  Here the following calculations are compared: PWIA (dash-double dotted curve), BA (dashed curve), DWBA (dash-dotted curve) and with inclusion of FSI effects and the MECs with $\Lambda = 4m_\pi$ ($\Lambda = \infty$) (solid (dotted) curve). 
  \\
  Panel b: Calculations made within the formalism of the present paper. Here, the solid curve corresponds to the calculation with the deuteron and $np$-pair WFs obtained within the CPR, as well as, pion-induced MECs.
  }
  \label{fig:03}
\end{figure}
\begin{figure}[t]
  \centering
  \includegraphics[width=\linewidth]{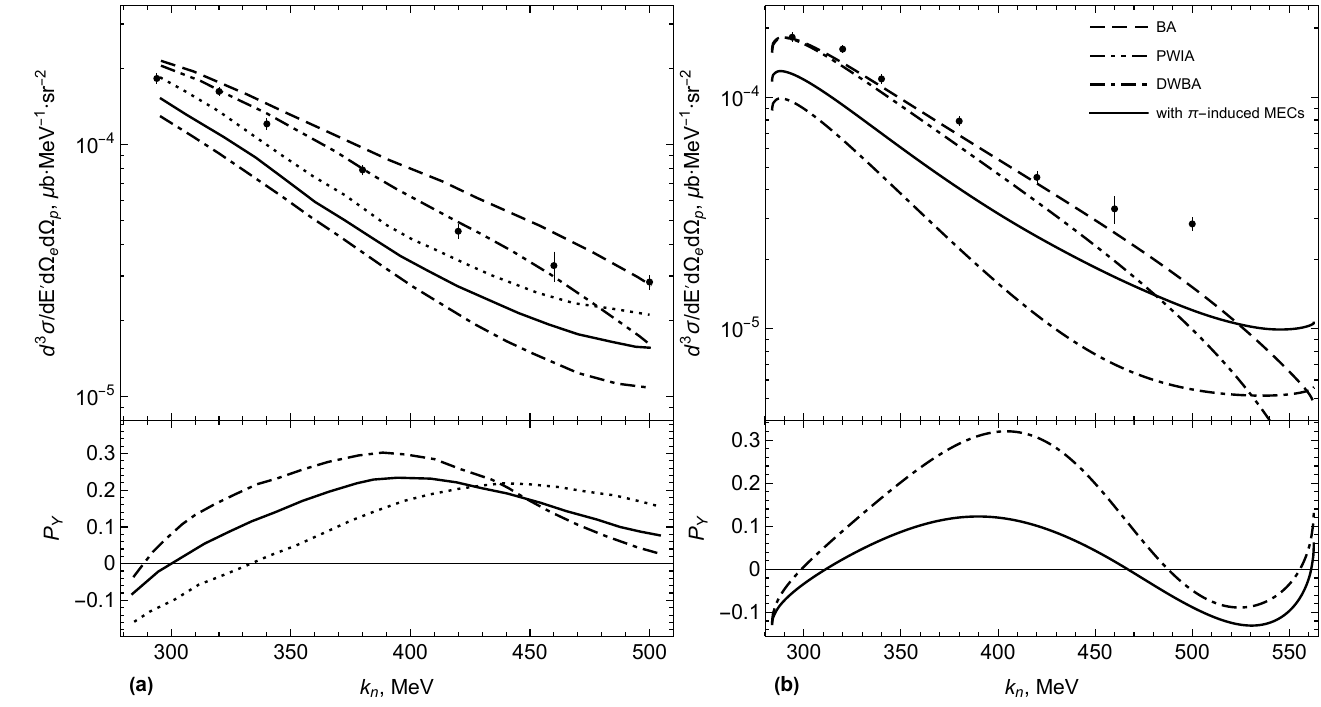}
  \caption{The same as in Fig.~\ref{fig:03} but for the Scalay kinematics \cite{Saclay1984}.}
  \label{fig:04}
\end{figure}
\begin{figure}[t]
  \centering
  \includegraphics[width=\linewidth]{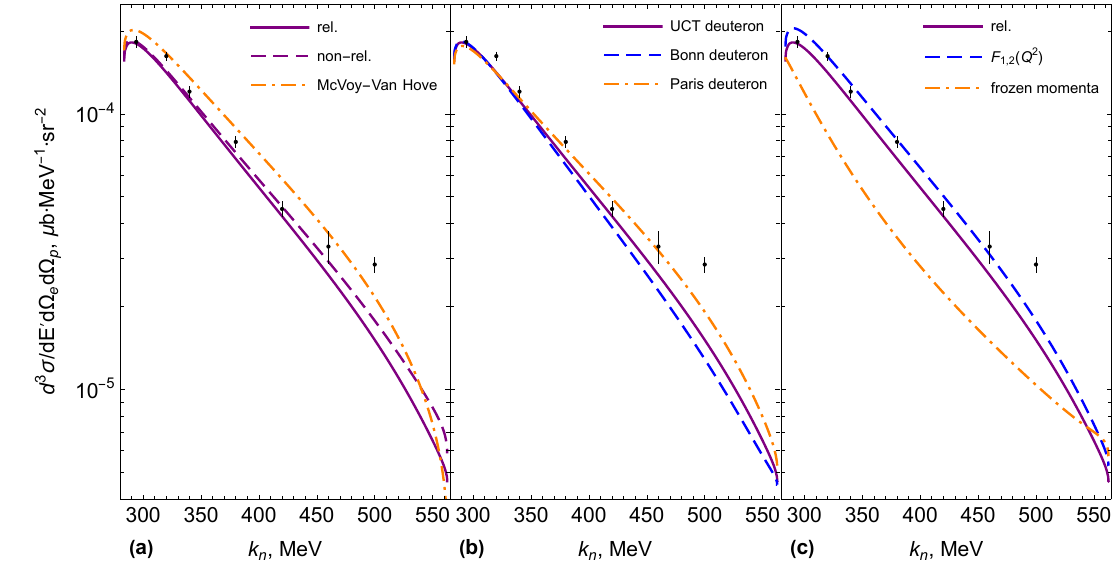}
  \caption{
	Plots of different BA calculations of the differential cross section versus the missing momentum for the Scalay kinematics \cite{Saclay1984}. 
	\\ 
	Panel a: calculations with different one-body currents. We use the deuteron WF from \cite{She13} for all calculations here.
	\\
	Panel b: calculations with relativistic one-body current but with different deuteron WFs. The purple solid curve -- with UCT deuteron WF components \cite{She13}, the blue dashed curve -- with the Bonn deuteron WFs \cite{Bonn89} and the orange dash-dotted one -- with the Paris deuteron WFs \cite{ParisPotential}.
	\\
	Panel c: demonstration of the role of the Fermi motion effects.
  }
  \label{fig:05}
\end{figure}
\begin{figure}[t]
  \centering
  \includegraphics[width=\linewidth]{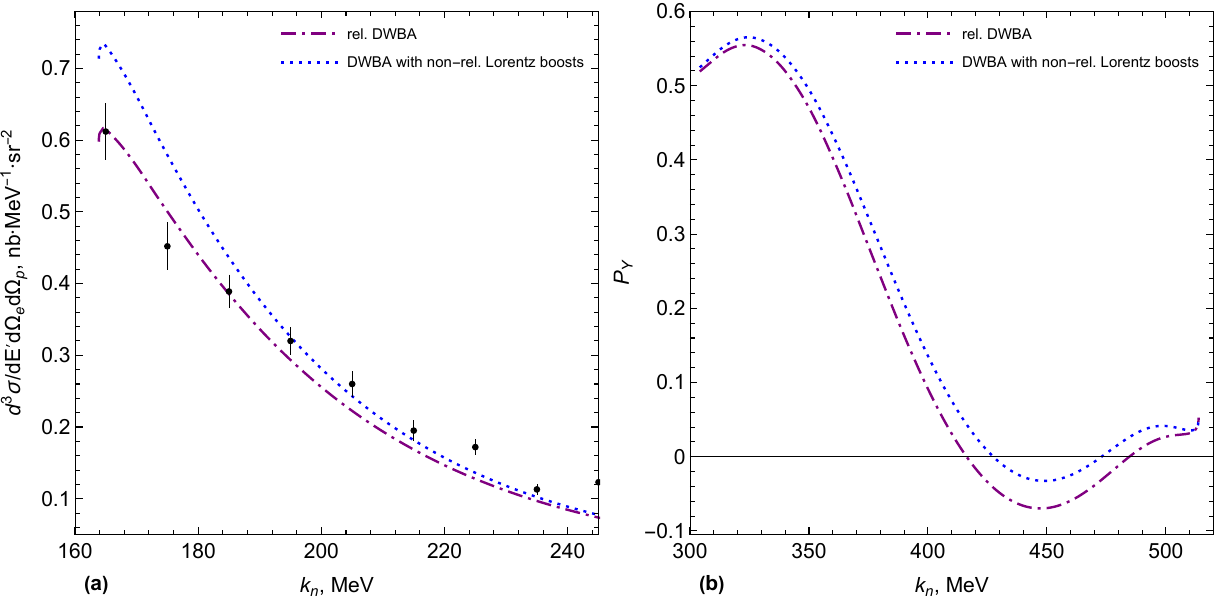}
  \caption{Demonstration of the role of the Lorentz boosts in the deuteron electrodisintegration observables for the Scalay kinematics \cite{Saclay1981}. The dash-dotted curves corresponds to the relativistic DWBA calculation, while the dotted curves -- with a non-relativistic approximation to the Lorentz boosts. (a): the differential cross section, (b): the induced polarization of knocked-out protons.}
  \label{fig:06}
\end{figure}
\begin{figure}[t]
  \centering
  \includegraphics[width=\textwidth]{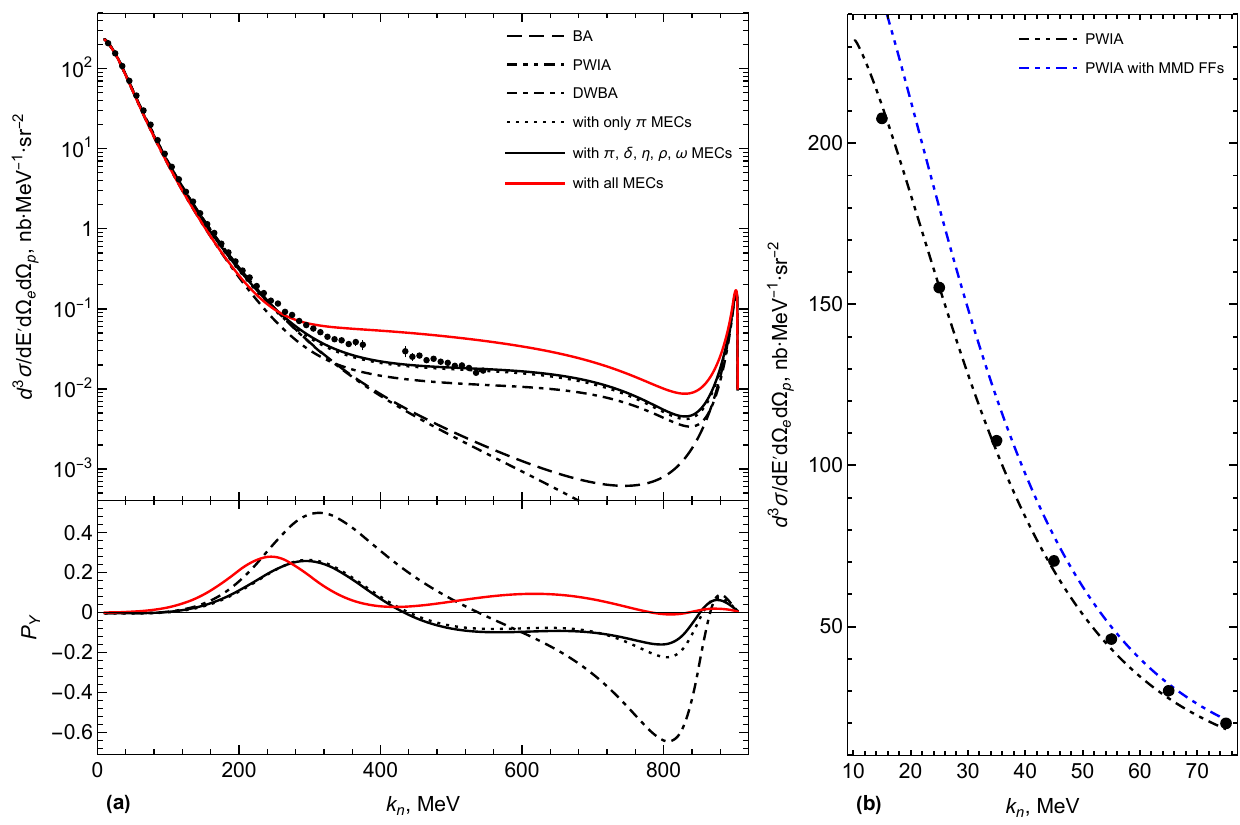}
  \caption{
	Differential cross section (upper part) and induced polarization (lower part) of knocked-out protons versus the neutron momentum for the kinematics of Ref.~\cite{Ulmer2002}. 
	The following calculations without MECs are compared: PWIA (dash-double dotted curve), BA (dashed curve), DWBA (dash-dotted curve). 
	As well as the calculations with inclusion of the FSI effects and the MECs: with the pion-induced MECs only (dotted), with MECs induced by the $\pi$, $\delta$, $\eta$, $\rho$, $\omega$ mesons (solid), and with full set of MECs (red solid). 
  }
  \label{fig:07}
\end{figure}
\begin{figure}[t]
  \centering
  \includegraphics[width=\linewidth]{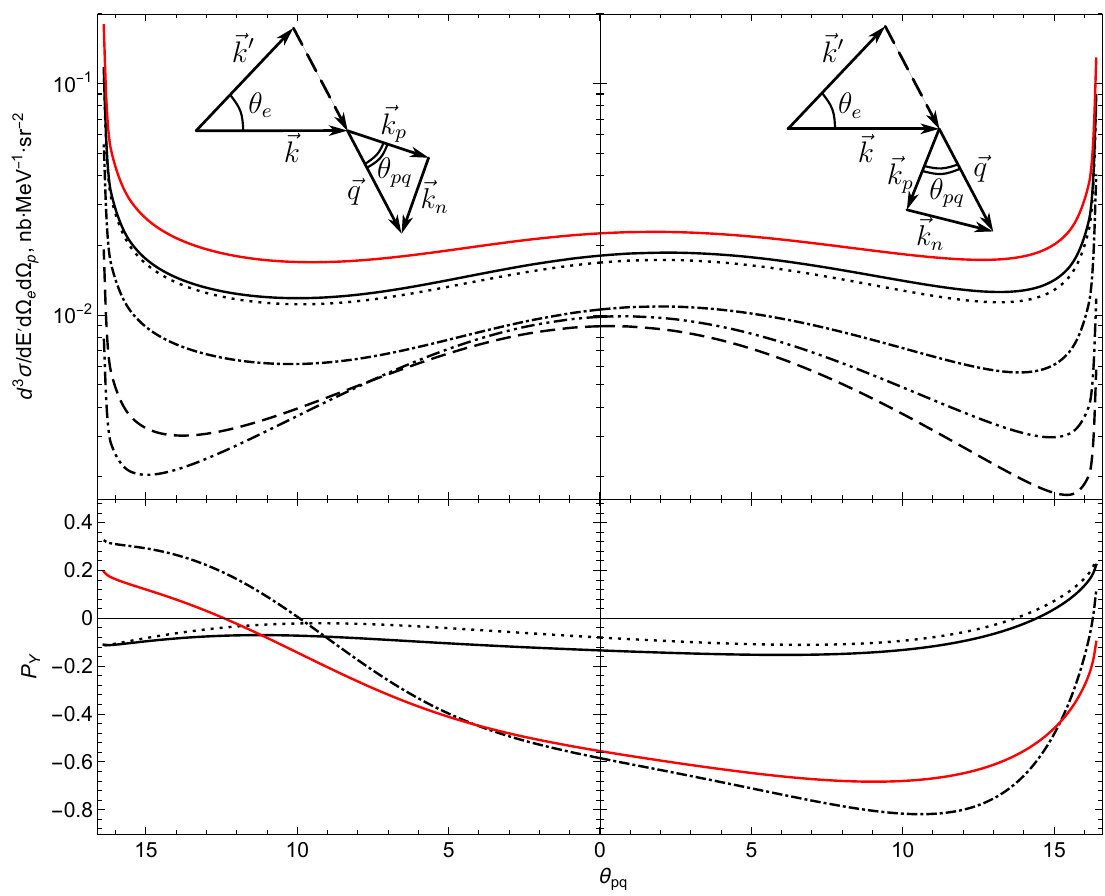}
  \caption{
	The same as in Fig.~\ref{fig:07} but versus the proton emission angle $\theta_{pq}$ for the kinematic conditions in the left wing of the QFP ($E=1~GeV$, $E'=827~MeV$, $\theta_e=50^\circ$) from \cite{KorMelShe90}.
	The left part of each plot corresponds to the azimuth angle (reaction plane angle) $\varphi=0^\circ$ and the right part -- to $\varphi=180^\circ$.
  }
  \label{fig:08}
\end{figure}
\begin{figure}[t]
  \centering
  \includegraphics[width=\linewidth]{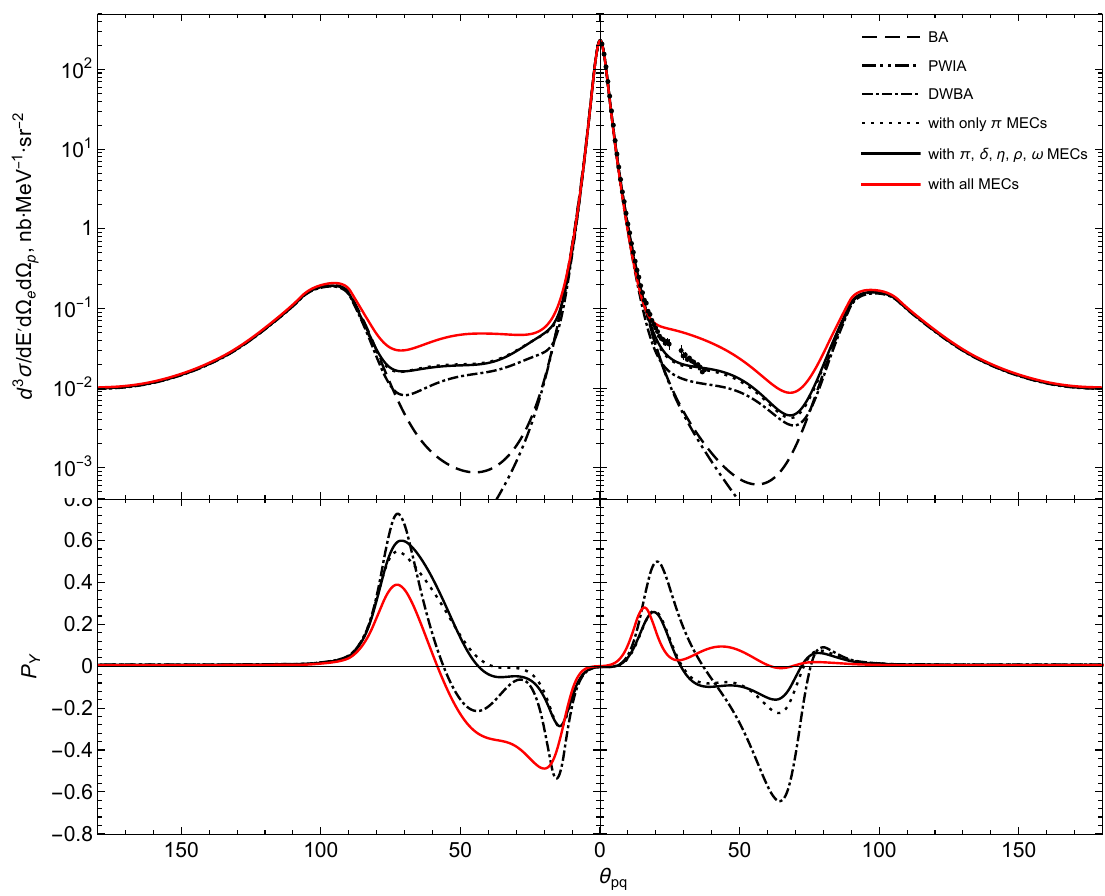}
  \caption{
	The same as in Fig.~\ref{fig:08} but for the kinematic conditions in the right wing of the QFP \cite{Ulmer2002}.
  }
  \label{fig:082}
\end{figure}
\begin{figure}[t]
  \centering
  \includegraphics[width=\linewidth]{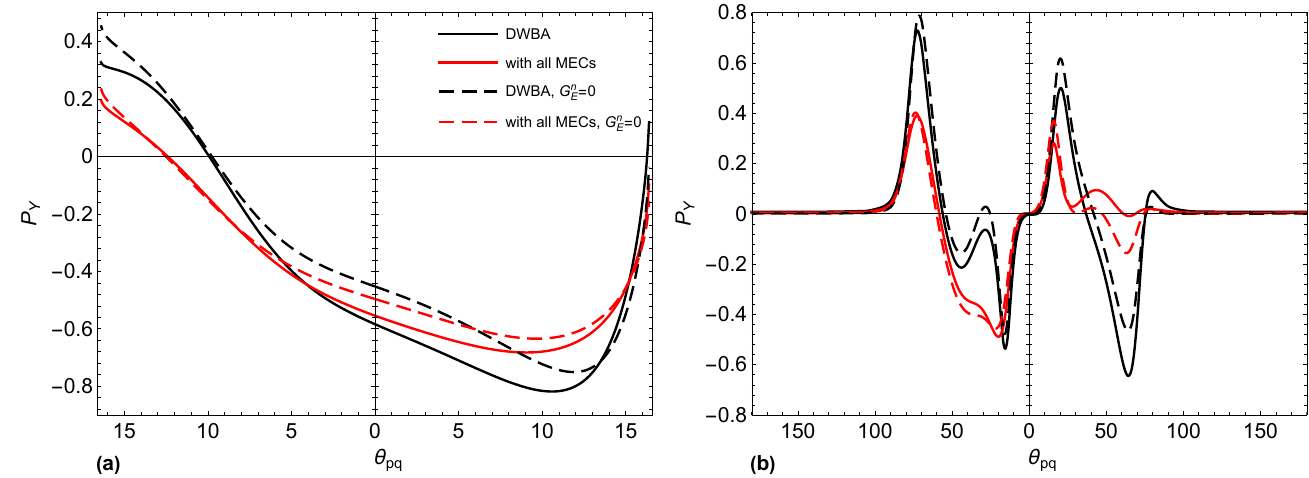}
  \caption{
	Dependence of the induced proton polarization on the choice of the $G_E^n$ model in the left wing of the QFP \cite{KorMelShe90} (a) and in the right wing of the QFP \cite{Ulmer2002} (b).
	The dashed curves were calculated with $G_E^n=0$; solid curves with $G_E^n$ from Ref.~\cite{Barut1968}. The black curves for DWBA results and the red ones for computations with all MECs contributions.
  }
  \label{fig:09}
\end{figure}
\begin{figure}[t]
  \centering
  \includegraphics[width=\linewidth]{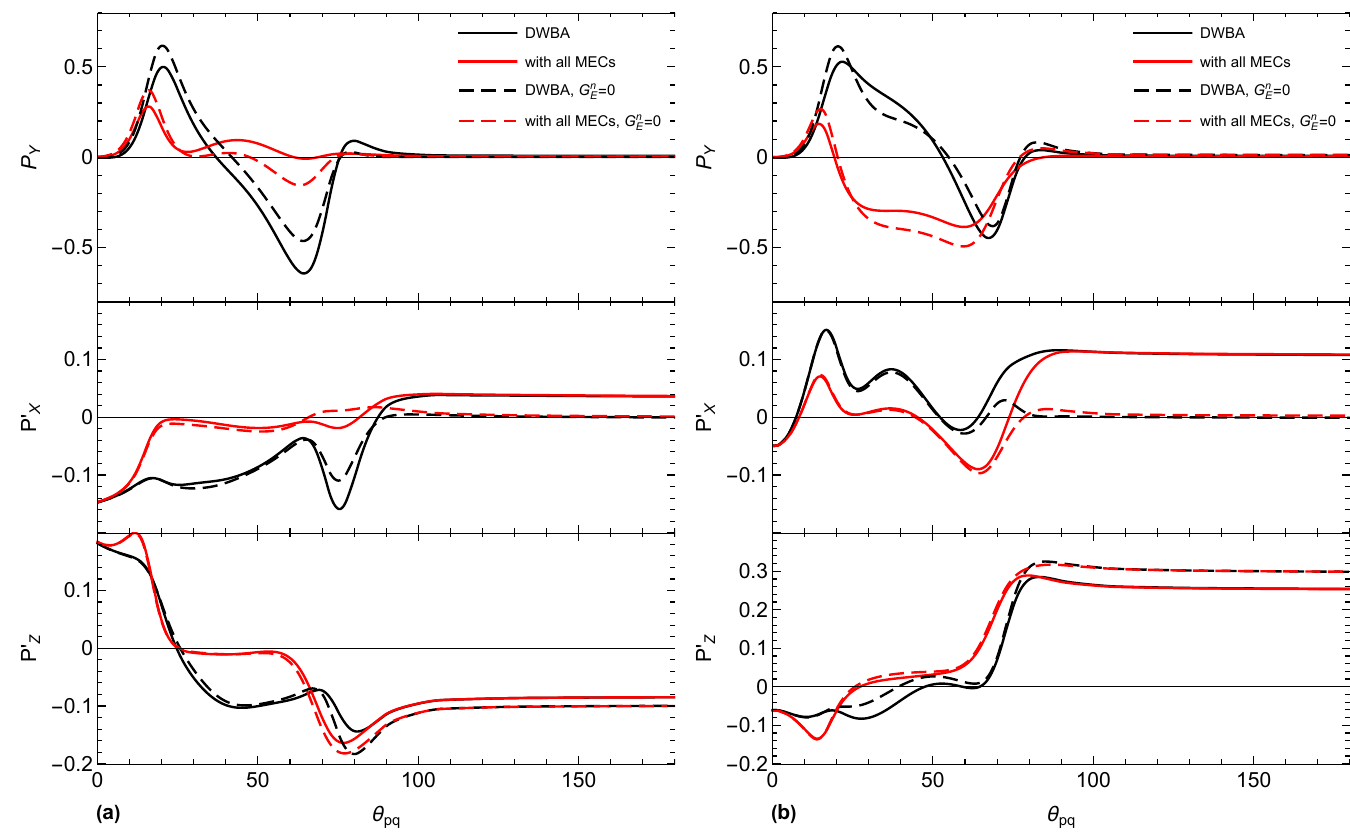}
  \caption{
	Proton (a) and neutron (b) polarization components versus the proton emission angle $\theta_{pq}$ for the kinematics of Ref.~\cite{Ulmer2002}.
	Curves have the same meaning as in Fig.~\ref{fig:09}.
  }
  \label{fig:10}
\end{figure}
    
\section{Results and Discussion}
\label{sec:results_and_discussion}

This research extends the results \cite{KorMelShe90, MelShe92} obtained in the 1990s. Therefore, let us compare our new calculations with those pioneering experimental data and theoretical predictions (see Figs.~\ref{fig:03}--\ref{fig:04}). In both calculations, the nucleon form factors are taken from Ref.~\cite{Barut1968}. The FSI effects were included by computing $n$-$p$ partial-wave scattering states with $J \leq J_{\textrm{max}} = 3$ while in Ref.~\cite{KorMelShe90} $J_{\textrm{max}} = 2$. The authors of Ref.~\cite{KorMelShe90} used the one-body current of Ref.~\cite{McVoy1962} and pion-induced MECs by Maize and Kim \cite{maize1984} with the Paris potential and deuteron WF \cite{ParisPotential}. Despite the non-relativistic nature of these ingredients, qualitatively these results have much in common with the relativistic calculations of this paper.
{\clb
The experimental data used for comparison are taken from Refs.\cite{Saclay1981} and \cite{Saclay1984}, corresponding to the kinematical ranges $155~MeV \leq k_n \leq 335~MeV$ with $Q^2=0.10~GeV^2$ and Bjorken variable $x_B=Q^2/2m\omega =0.36$, and $294~MeV \leq k_n \leq 500~MeV$ with $Q^2=0.038~GeV^2$ and $x_B=0.10$, respectively.
}
As the neutron momentum increases, the Born approximation (BA) and plane-wave impulse approximation (PWIA) curves diverge because the recoil mechanism becomes more important. For these kinematics, including FSI generally reduces the cross section over most of the neutron-momentum range. However, in our new calculations with an extended momentum range, we find that this trend does not hold at the highest momenta: in the far-right region, the distorted-wave Born approximation (DWBA) cross section exceeds the BA result.  
{\clb 
This behavior has been discussed extensively
(see, e.g., ~\cite{BOEGLIN2024})
and is related to the underlying structure of the respective amplitudes. In this kinematic region, the BA amplitude is dominated by the PWIA, which is directly proportional to the deuteron wave function evaluated at the specific value of the missing momentum, which drops off sharply as $k_n$ increases. In contrast, the FSI contribution involves an integral over momentum space. This integral is dominated by contributions from lower internal momenta where the wave function is significantly larger, thereby sustaining the cross section at higher values of $k_n$.
}

{\clb It is important to note that even in these kinematics, relativistic effects can be significant. This becomes evident already at the level of the one-body current.} In Figs.~\ref{fig:03}, \ref{fig:04} we see that the non-relativistic curves with only one-body current contribution are higher than the respective relativistic ones. 
That is demonstrated clearly in the Fig.~\ref{fig:05}(a), where three BA calculations with different one-body currents are shown: the relativistic current \eqref{one-body_current} (purple solid curve), its non-relativistic reduction
\begin{equation}
\begin{split}
	& F^0(\vect{p}'\mu',\vect{p}\mu)
	=F_1[(\vect{p}'-\vect{p})^2] \delta_{\mu'\mu},
	\\
	& \vect{F}(\vect{p}'\mu',\vect{p}\mu)
	=F_1[(\vect{p}'-\vect{p})^2] \frac{\vect{p}'+\vect{p}}{2m} \delta_{\mu'\mu}
	+ (F_1[(\vect{p}'-\vect{p})^2] + F_2[(\vect{p}'-\vect{p})^2]) 
	\frac{i \vect{\sigma}_{\mu'\mu} \times (\vect{p}'-\vect{p})}{2m}
\end{split}
\end{equation}
(purple dashed curve) and the current by McVoy and Van Hove \cite{McVoy1962} (orange dashed curve). 
In Fig.~\ref{fig:03}(a) we see that the DWBA result is already in good agreement with the data and the contribution of the MECs is not of great importance, while the relativistic DWBA calculation in Fig.~\ref{fig:03}(b) underestimates the cross section for $k_n \gtrsim 200~MeV$. The role of MECs is more important in the relativistic calculation, and only after their inclusion a good agreement with the data is obtained.
The kinematic regime of Fig.~\ref{fig:04} lies far to the right of the quasi-free peak (QFP) and therefore the role of FSI becomes even more significant. And again we see the destructive interference between MEC and FSI contributions to the cross section that confirms the conclusion drawn in Ref.~\cite{Arenhovel1982}. 

The cross section is also sensitive to the choice of the deuteron WF. Fig.~\ref{fig:05}(b) demonstrates the role of the deuteron WF in the BA calculations with the relativistic one-body current. For recoil momenta $k_n \lesssim 350~MeV$, these curves are almost indistinguishable, but for higher momenta they begin to differ significantly. 

Now let us discuss the effects of Fermi motion, i.e., the relative motion of the nucleons in the nucleus. Note that, after substituting \eqref{F_one-body} into Eq.~\eqref{MEFinal}, the nucleon FFs depend not only on the transferred momentum $q$, but also on the relative nucleon momentum in the deuteron $\vect{p}$. This dependence is integrated out in the matrix elements \eqref{MEFinal} but as shown in Fig.~\ref{fig:05}(c) its effect is still present in the resulting calculations. Models in which the form factors depend only on $Q^2$ are widely used.
The figure shows that the calculation with such "frozen" FFs (blue dotted curve) produces bigger cross section values for the entire range of recoil momenta. In such a model Fermi motion effects are still present, because the bispinors and the prefactor of $F_2$ in \eqref{F_one-body} depend on the relative momentum $\vect{p}$. If we "freeze" these dependencies as well, for example, putting $\vect{p}'=\vect{q}/2$ and $\vect{p}=-\vect{q}/2$ in \eqref{F_one-body} (orange dash-dotted curve) then the cross section drops by considerable amount for all $k_n$ except for the lowest and highest values. This clearly demonstrates the importance of the Fermi motion effects in the deuteron electrodisintegration.

Other relativistic ingredients are the boosted momenta $L\vect{p}'$, $L\vect{p}'_-$
and $D$-functions (Wigner rotations) that appear in the matrix elements \eqref{MEFinal}. To demonstrate the role of these effects, we performed the following DWBA calculations of the cross section and induced polarization for the Scalay kinematics \cite{Saclay1981} (see Fig.~\ref{fig:06}):
with all relativistic effects;
with a non-relativistic approximation to the Lorentz boosts, i.e., with the replacement $(Lp)^i \to (\vect{p} + \vect{q}/2)^i$;
with the $D$-functions replaced by the Kronecker deltas $D^{[1/2]}_{\bar\mu \mu}(W(L(q),p)) \to \delta_{\bar\mu\mu}$.
It turns out that the role of the $D$-functions is negligible for this and other kinematics we have considered (so we do not include the corresponding curve in Fig.~\ref{fig:06}), but the Lorentz boosts have a substantial impact. It is noteworthy that the influence of the latter is most noticeable at smaller values of recoil momentum ($k_n \lesssim 240~MeV$) for the cross section, but for the induced polarization, this influence is most pronounced at larger momenta ($k_n \gtrsim 300~MeV$). The relativistic treatment of the boosts is crucial for the correct description of the data in Fig.~\ref{fig:06}(a), and the non-relativistic approximation leads to a significant overestimation of the cross section.
The FSI and MEC contributions do not interfere destructively under all kinematic conditions. In Fig.~\ref{fig:07} we present the results of calculations for the kinematic of Ref.~\cite{Ulmer2002} ($E=3.11~GeV$, $E'=2.74~GeV$, $\theta_e=16.1^\circ$), where both FSI and MEC contributions lead to an increase in the cross section. This behavior was also observed for the Saclay kinematics in Figs.~\ref{fig:03}--\ref{fig:04} but for the highest values of neutron momenta in both cases. Since the kinematic used in Fig.~\ref{fig:07} allows much higher $k_n$ up to $550~MeV$ this behavior becomes more obvious. 
{\clb Moreover, the contribution from the FSI effects becomes dominant in comparison with the plane-wave contribution for $400~ MeV \lesssim k_n \lesssim 800~MeV$.}
Calculations with MECs induced by the $\pi$, $\delta$, $\eta$, $\rho$, $\omega$ mesons and only the $\pi$ meson show only minor differences. Contributions from different meson exchanges are not negligible, but they interfere destructively, so the final result with $\pi$, $\delta$, $\eta$, $\rho$, $\omega$ meson exchanges could be well described with only pion-induced MECs. The contribution of the $\sigma$ mesons is the largest amongst all MECs, and it leads to a significant increase of the cross section. This may appear unexpected because the role of the $\delta$ meson (which is also a scalar meson) is not that important, but the coupling constant for the $\sigma$ (T=1) meson (see UCT GS parameters from Table 1 in Ref.~\cite{ArsEtAl2021}) is much higher than for the $\delta$ meson ($g_\sigma^2/g_\delta^2\approx 5.5$), so the contribution of the former is more significant. Recall that the introduction of sigma mesons aims to effectively account for two-pion exchanges. In this context, their role in electromagnetic processes with nuclei requires further investigation. 

{\clb In general, our results are in good agreement with the experimental data. For values above $k_n\leq 335~MeV$ MECs begin to play a significant role, the expected behavior at the modest $Q^2$, where these experiments were performed. The cross-section calculations are quite sensitive to the corresponding two-body current model, for instance, the inclusion of sigma mesons leads to noticeable changes and overestimates the cross-section. An improved description can be achieved by introducing the $\Delta$-isobar and dissociation currents. In this regard, the implicit inclusion of two-body currents via the generalized Siegert theorem from Sec. 3.6 may prove useful. We will address these issues in our future works.

Figure 5(b) displays a linear plot for the low missing momentum data. In this region, the cross section is well described by PWIA. A reasonably accurate description of the experimental data cannot be achieved using a non-relativistic one-body current and only a fully relativistic form yields good agreement. Furthermore, the results are quite sensitive to the choice of form factor model. In addition to the calculations using form factors from \cite{Barut1968}, Fig.~\ref{fig:07}(b) includes results obtained with MMD form factors \cite{MMD1996}. The latter tends to overestimate the cross section at low $k_n$ , a result consistent with findings in \cite{Jeschonnek2008} where these FFs were also employed. As $k_n$ increases, both curves start getting closer to each other.}

As seen in Figs.~\ref{fig:08}--\ref{fig:082}, the $e'p$ coincidence cross section and proton polarization exhibit azimuthal (left-right) asymmetry relative to the geometry with $\theta_{pq}=0^{\circ}$ (parallel kinematics). This behavior is observed for both kinematic settings in the left (Fig.~\ref{fig:08}) and right (Fig.~\ref{fig:082}) wings of the QFP. As the momentum $\mathbf{k}_p$ lies on the right (left) of the momentum $\mathbf{q}$, the azimuthal angle $\varphi$ equals $180^{\circ}\left(0^{\circ}\right)$. 
This asymmetry can be explained by using the transformation properties
\begin{align}
R_z(180^{\circ}) J_{x, y}(\mathbf{q}) R_z^{-1}(180^{\circ}) & =-J_{x, y}(\mathbf{q}) 
\label{Jxy_180_rotation}
\\
R_z(180^{\circ}) \sigma_y R_z^{-1}(180^{\circ}) & =-\sigma_y
\label{sigma_180_rotation}
\end{align}
with respect to the frame rotation $R_z(180^{\circ})$ by $180^{\circ}$ about the vector $\hat{\mathbf{n}}_z$ ($\vect{q}$ direction). It follows from Eqs.~\eqref{Jxy_180_rotation} and \eqref{sigma_180_rotation} that for a given $\theta_{pq}$ value
\begin{equation}
\begin{array}{ll}
W_{C, T, S}(\varphi=0)=W_{C, T, S}(\varphi=\pi), 
& W_I(\varphi=0)=-W_I(\varphi=\pi), 
\\
\Sigma_{C, T, S}(\varphi=0)=-\Sigma_{C, T, S}(\varphi=\pi), 
& \Sigma_I(\varphi=0)=\Sigma_I(\varphi=\pi) .
\end{array}
\end{equation}

The deuteron electrodisintegration on the left wing of the QFP (Figs.~\ref{fig:08} and \ref{fig:09}(a)) is characterized by a limiting proton emission angle in the laboratory frame,
\begin{equation}
	\cos \theta_{\lim}=\frac{1}{2 q m_p}\sqrt{\left[\omega+m_d\right]^2-\left[\left(\omega+m_d\right)^2+m_p^2-m_n^2-\mathbf{q}^2\right]^2}.
\end{equation}
As $\theta_{pq} \to \theta_{\lim}$ (for the present case $\theta_{\lim} \simeq 16.6^{\circ}$), the differential cross section increases sharply due to the kinematic factor $R$, which diverges in this limit. Nevertheless, this singular behavior is mild, and for any finite angular acceptance, the cross section remains finite.

As seen in Fig.~\ref{fig:09} and noted in Ref.~\cite{KorMelShe90}, the neutron electric form factor $G_E^n$ plays a more pronounced role in the induced proton polarization on the left wing of the QFP than on the right wing. For parallel kinematics ($\theta_{pq}=0^\circ$), the MEC contribution nearly vanishes when $G_E^n$ is included, whereas setting $G_E^n=0$ leads to a noticeable reduction of the polarization due to MEC effects. Thus, measurements of the induced polarization may help discriminate between different models of $G_E^n$, although one should keep in mind that $P_Y$ in parallel kinematics remains sensitive to MEC contributions.

Measurements of the polarization transfer in $d(\vec{e},e'\vec{p})n$ and $d(\vec{e},e'\vec{n})p$ are a great source of information on the nucleon form factors, especially for the neutron. Fig.~\ref{fig:10} shows the results of calculations for the proton and neutron polarization components in the kinematics of Ref.~\cite{Ulmer2002}. For parallel kinematics ($\theta_{pq}=0^\circ$), MEC effects are negligible, and the same is true for the influence of $G_E^n$. However, for $\theta_{pq} \gtrsim 90^\circ$, the neutron electric form factor is of great importance and the MEC contributions remain negligible. This makes this range of the proton emission angles very promising for the extraction of $G_E^n$ from the polarization transfer measurements.

For the kinematics with $\vect{k}_n \parallel \vect{q}$ ($\theta_{pq} \simeq \theta_{\text{lim}}$) the following expressions can be derived in the BA under the assumption $u_{0,2}(q)/u_0(0) \ll 1$:
\begin{equation}
\begin{split}
	&I_0 P'_X \approx - \frac{1}{3} \xi \sqrt{\eta} \frac{q}{E_q} G_E^n G_M^n,
	\\
	&I_0 P'_Z \approx - \frac{1}{3} \sqrt{\eta(\eta+\xi)} \frac{q^2}{E_q(E_q+m)} [G_M^n]^2,
\end{split}
\end{equation}
for proton polarization transfer and
\begin{equation}
\begin{split}
	&I_0 P'_X \approx - \xi \sqrt{\eta} \frac{q}{E_q} G_E^n G_M^n,
	\\
	&I_0 P'_Z \approx \sqrt{\eta(\eta+\xi)} \frac{q^2}{E_q(E_q+m)} [G_M^n]^2,
\end{split}
\end{equation}
for neutron polarization transfer, with
\begin{equation}
	I_0 = \frac{\sigma_0}{\sigma_M R} \left( \frac{e^2}{4\pi} u_0^2(0) \right)^{-1}.
\end{equation}
Assumption $u_{0,2}(q)/u_0(0) \ll 1$ is justified for high $q$ values. The above formulae show that the ratio of the transverse and longitudinal polarization transfer components is directly proportional to the ratio of the neutron electric and magnetic form factors, i.e., $P'_X/P'_Z \propto G_E^n/G_M^n$. This is a well-known result for the quasi-elastic scattering on the nucleon, but it can also be obtained for the deuteron electrodisintegration in the parallel kinematics. It is important to note that these polarization components do not depend on the deuteron WF in this case.

\section{Conclusions}
To conclude:

(i) A fully relativistic 
description of the $d(e,e'p)n$ reaction has been developed within the clothed particle representation. The nucleon–nucleon interaction (the Kharkiv potential), electromagnetic current operators (including one– and two–body terms), boost operators and other Poincaré group generators are constructed on a common field–theoretical footing.

(ii) Comparison with the Saclay and Jefferson Lab data shows that relativistic effects are essential. In particular, the relativistic one–body current and the consistent treatment of Lorentz boosts significantly influence both the differential cross section and polarization observables. 

(iii) The role of FSI and MEC contributions depends on the kinematics. Contrary to the results in the considered Saclay kinematic regimes, for the higher-energy kinematics the FSI and MEC contributions may act constructively and increase the cross section at large recoil momenta. The contributions of different meson exchanges partially compensate each other, while the effective $\sigma$-meson mechanism gives a sizable effect that deserves further investigation.

(iv) The observables exhibit a pronounced left–right asymmetry with respect to the reaction plane. In the left wing of the QFP, the induced proton polarization is especially sensitive to the neutron electric form factor $G_E^n$, in agreement with earlier findings. At the same time, for parallel kinematics the polarization transfer components satisfy simple analytical relations and become directly proportional to the ratio $G_E^n/G_M^n$, while being independent of the deuteron WF, even within the relativistic approach. This makes the corresponding kinematic region particularly promising for the extraction of neutron form factors.

The present study demonstrates that a consistent field–theoretical construction of interactions and currents within the CPR framework provides a viable alternative to conventional non-relativistic models. Further developments may include a fully gauge-independent description, more detailed analysis of sub-nucleonic degrees of freedom and applications to other electroweak processes involving few-nucleon systems.

\bigskip

\section*{Acknowledgments}
This research was supported by the National Academy of Sciences (USA) and the Office of Naval Research Global (USA) in assistance of the Science and Technology Center in Ukraine (Grant No. 7134). 
{\clb
The work was also supported by the National Science Centre, Poland under Grant IMPRESS-U 2024/06/Y/ST2/00135; in part by the Excellence Initiative – Research University Program at the Jagiellonian University in Krak\'{o}w; by National Science Foundation Eager Award No. 2427848; and in part by the Japan Society for the Promotion of Science (JSPS) under Grant No.~JP25K07301. Some numerical calculations were performed on the supercomputers at the JSC, J\"{u}lich, Germany. 

We are grateful to 
H. Arenh\"{o}vel, S. Jeschonnek, J.-O. Hansen, and M. Jones
for providing us with the experimental dataset corresponding to Fig.~1 from Ref.~\cite{Ulmer2002}.
}

\setcounter{equation}{0}
\renewcommand{\theequation}{A.\arabic{equation}}

\section*{Appendix A: Links between the clothed one- and two-particle states and \textit{in}(\textit{out}) states}

In what follows we will explicitly derive the links (\ref{oneparticlerelation})--(\ref{twoparticlerelation_2}) between \textit{in}(\textit{out}) states and the clothed particle states. The derivation will be done for the complex scalar field $\varphi(x)$, but it can be easily generalized to other cases. The scalar field $\varphi(x)$  satisfies the following Klein-Gordon (KG) equation
\be
(\square_x + \mu^2)\varphi(x) = j(x),
\label{KGeq}
\ee
where $\mu$ is the physical mass and current $j(x)$ describes a possible interaction. The field $\varphi(x)$ is in the Heisenberg picture
\be
\varphi(x) = \varphi(\mbf{x},t) \equiv e^{iHt}\varphi(\mbf{x})e^{-iHt},
\label{phiH}
\ee
where $H$ is the corresponding total Hamiltonian operator. Further, one introduce the quantities
\be
A(\mbf{k},t) = (f^*_k,\varphi), \;\; B^\dagger(\mbf{k},t) = (\varphi,f_k)=-(f_k,\varphi),
\label{ABdef}
\ee
with respect to the product
\be
(F_1,F_2)=i\int d\mbf{x}[F_1(\mbf{x},t)\partial_tF_2(\mbf{x},t)-F_2(\mbf{x},t)\partial_tF_1(\mbf{x},t)].
\label{product}
\ee
Here $f_k(x)$ denotes the plane waves 
\be
f_k(x)=f(\mbf{x},t)=[(2\pi)^32k_0]^{-1/2}\exp(-ikx),
\label{planeWave}
\ee
with the four-vector $k=(k^0,\mbf{k})$ is determined on the mass shell $k^0=\sqrt{\mbf{k}^2+\mu^2}$, which enter in the expansion
\be
\varphi(\mbf{x})=\int d\mbf{k}[A(\mbf{k})f_k(\mbf{x},0)+B^\dagger(\mbf{k})f_k^*(\mbf{x},0)],
\label{phiS}
\ee
where the destruction and creation operators $A(\mbf{k})$, $B^\dagger(\mbf{k})$ correspond to the representation of the bare particles with physical masses. The clothed particle counterparts for these operators introduced via the unitary transformation $W$
\be
A(\mbf{k})=WA_c(\mbf{k})W^\dagger,\;\;B^\dagger(\mbf{k})=WB^\dagger_c(\mbf{k})W^\dagger, 
\label{AlinkAc}
\ee
which allows to express the operators (\ref{ABdef}) through the clothed particle operators. 

On the other hand, we shall employ the LSZ asymptotic condition,
\be
\lim_{t\rightarrow-\infty}\langle \Phi|A^\dagger(\mbf{k},t)|\Psi\rangle=\langle \Phi|A^\dagger_{in}(\mbf{k})|\Psi\rangle,
\label{Ain}
\ee
that determines the \textit{in}-operator $A_{in}(\mbf{k})$ for any normalizable states $\Phi$ and $\Psi$. 

Eqs. (\ref{ABdef}), (\ref{AlinkAc}) and (\ref{Ain}) are our working formulae. In the context, the one-particle \textit{in}-state is given by
\be
|\mbf{k};in\rangle = A^\dagger_{in}(\mbf{k})|\Omega\rangle=\lim_{t\rightarrow-\infty}A^\dagger(\mbf{k},t)|\Omega\rangle.
\ee
Explicit calculations give
\be
A(\mbf{k},t)= \,i\int d\mbf{x}\{f_k^*(\mbf{x},t)e^{iHt}[H,\varphi(\mbf{x},0)]e^{-iHt}-ik^0\varphi(\mbf{x},t)f_k^*(\mbf{k},t)\}\equiv \bar{A}(\mbf{k},t)+A_I(\mbf{k},t),
\ee
where
\be
\bar{A}(\mbf{k},t)=-e^{iHt}\int d\mbf{x}f_k^*(\mbf{x},t)([H_0,\varphi(\mbf{x})]-k^0\varphi(\mbf{x}))e^{-iHt},
\label{Abar}
\ee
and
\be
A_I(\mbf{k},t)=-e^{iHt}\int d\mbf{x}f_k^*(\mbf{x},t)[H_I,\varphi(\mbf{x})]e^{-iHt}.
\label{AI}
\ee
Here we assume the standard division of the total Hamiltonian operator into free and interaction parts $H=H_0(a)+H_I(a)$, where both terms are written in terms of the BPR operators $a=\{A(\mbf{k}),B(\mbf{k}),A^\dagger(\mbf{k}),B^\dagger(\mbf{k})\}$. In particular, the free part $H_0(a)$ is given by a usual expression
\be
H_0(a) = \int d\mbf{k}\sqrt{\mbf{k}^2+\m^2}\,[A^\dagger(\mbf{k})A(\mbf{k})+B^\dagger(\mbf{k})B(\mbf{k})].
\ee
Taking into account the commutation relations $[H_0,A(\mbf{k})]=-k^0A(\mbf{k})$ and $[H_0,B^\dagger(\mbf{k})]=k^0B^\dagger(\mbf{k})$ Eq.(\ref{Abar}) can be put in the form
\be
\bar{A}(\mbf{k},t)=e^{iHt}A(\mbf{k})e^{-iHt}e^{ik^0t}.
\ee
Similarly, after integration over $d\mbf{x}$ in Eq.(\ref{AI}) one gets
\be
A_I(\mbf{k},t)=-\frac{e^{iHt}}{2k_0}\left\{[H,A(\mbf{k})+B^\dagger(-\mbf{k})]+k_0(A(\mbf{k})-B^\dagger(-\mbf{k}))\right\}e^{ik^0t}e^{-iHt}.
\ee
Now, with help of Eq.(\ref{AlinkAc}) one can express  $\bar{A}(\mbf{k},t)$ in terms of the clothed particle operators $a_c=\{A_c,B_c,A^\dagger_c,B^\dagger_c\}$
\be
\bar{A}(\mbf{k},t)=e^{iHt}WA_c(\mbf{k})W^\dagger e^{-Ht}e^{ik_0t}=e^{iHt}e^{-iK_F(a_c)t}W_D(t)A_c(\mbf{k})W^\dagger_D(t)e^{iK_F(a_c)t} e^{-iHt},
\label{barCPR}
\ee
where we used that $e^{iK_F(\alpha_c)t}A_c(\mbf{k})e^{-iK_F(\alpha_c)t}=A_c(\mbf{k})e^{ik^0t}$ with a new
decomposition of the original Hamiltonian $H$ into  free and interaction parts $H=K_F(\a_c)+K_I(\a_c)$, where the free part is now given by
\be
K_F(a_c) = H_0(a_c)=\int d\mbf{k}\sqrt{\mbf{k}^2+\m^2}\,[A^\dagger_c(\mbf{k})A_c(\mbf{k})+B^\dagger_c(\mbf{k})B_c(\mbf{k})].
\ee
Under asymptotic condition (\ref{chpt2:equiv_condition}) one can get step by step the limits of interest
\be
\begin{split}
	&\lim_{t\rightarrow-\infty}\bar{A}^\dagger(\mbf{k},t)|\Omega\rangle=\lim_{t\rightarrow-\infty}e^{iHt}e^{-iH_F(\alpha_c)t}W_D(t)A^\dagger_c(\mbf{k})W^\dagger_D(t)e^{iH_F(\alpha_c)t} e^{-iHt}|\Omega\rangle\\
	&=\lim_{t\rightarrow-\infty}e^{iHt}e^{-iH_F(\alpha_c)t}A^\dagger_c(\mbf{k})|\Omega\rangle=\lim_{t\rightarrow-\infty}e^{ik_0t}e^{-ik_0t}A^\dagger_c(\mbf{k})|\Omega\rangle=A^\dagger_c(\mbf{k})|\Omega\rangle,
\end{split}
\label{AppA::barAacts}
\ee
where along with the asymptotic condition we have used the following properties of the clothed particle states and the physical vacuum
\be
K_F(\alpha_c)|\Omega\rangle = H|\Omega\rangle=0
\label{condOmega}
\ee
and
\be
K_F(\alpha_c)A^\dagger_c(\mbf{k})|\Omega\rangle=HA^\dagger_c(\mbf{k})|\Omega\rangle = k^0A^\dagger_c(\mbf{k})|\Omega\rangle.
\label{cond1particle}
\ee
In the same way one gets
\begin{small}
	\be
	\begin{split}
		&\lim_{t\rightarrow-\infty}A^\dagger_I(\mbf{k},t)|\Omega\rangle=\lim_{t\rightarrow-\infty} e^{iHt}\left(\frac{H-k_0}{2k_0}WA^\dagger_c(\mbf{k})W^\dagger + \frac{H+k_0}{2k_0}WB_c(-\mbf{k})W^\dagger\right)e^{-ik_0t}e^{-iHt}|\Omega\rangle\\
		&=\lim_{t\rightarrow-\infty} \frac{e^{iHt}}{2k_0}\bigg((H-k_0)e^{-iH_F(\a_c)t}W_D(t)A^\dagger_c(\mbf{k})W^\dagger_D(t) + (H+k_0)e^{-iH_F(\a_c)t}e^{-2ik_0t}W_D(t)B_c(-\mbf{k})\\
		&\times W^\dagger_D(t)\bigg)|\Omega\rangle=\lim_{t\rightarrow-\infty} e^{iHt}\frac{H-k_0}{2k_0}e^{-iH_F(\a_c)t}A^\dagger_c(\mbf{k})|\Omega\rangle 
		=\lim_{t\rightarrow-\infty} e^{iHt}e^{-ik_0t}\frac{k_0-k_0}{2k_0}A^\dagger_c(\mbf{k})|\Omega\rangle =0,
	\end{split}
	\label{AppA::AIacts}
	\ee
\end{small}
where we used important condition that clothed particle destruction operators destroys the physical vacuum $B_c(-\mbf{k})|\Omega\rangle = 0$. Therefore, $\lim A_I(\mbf{k},t)|\Omega\rangle=0$ and
\be
|\mbf{k};in\rangle = A^\dagger_{in}(\mbf{k})|\Omega\rangle=\lim_{t\rightarrow-\infty}\bar{A}^\dagger(\mbf{k},t)|\Omega\rangle =A^\dagger_c(\mbf{k})|\Omega\rangle = |\mbf{k};c\rangle.
\label{AppA::link1}
\ee
Analogously, one can obtain links between \textit{out}-states $A^\dagger_{out}(\mbf{k})|\Omega\rangle$, \textit{in}(\textit{out})-states $B^\dagger_{in(out)}(\mbf{k})|\Omega\rangle$ and the corresponding one-particle clothed states.

Now, lets consider the two-particle \textit{in}-state
\be
|\mbf{k}_1\mbf{k}_2;in\rangle = A^\dagger_{in}(\mbf{k}_1)A^\dagger_{in}(\mbf{k}_2)|\Omega\rangle.
\ee
First, using the link between one-particle states $A^\dagger_{in}(\mbf{k})|\Omega\rangle=A^\dagger_{c}(\mbf{k})|\Omega\rangle$ one can write
\be
|\mbf{k}_1\mbf{k}_2;in\rangle =A^\dagger_{in}(\mbf{k}_1)A^\dagger_c(\mbf{k}_2)|\Omega\rangle=\lim_{t\rightarrow-\infty}A^\dagger(\mbf{k}_1,t)A^\dagger_c(\mbf{k}_2)|\Omega\rangle.
\ee
Similarly to Eqs.(\ref{AppA::barAacts}) and (\ref{AppA::AIacts}) we obtain that
\be
\lim_{t\rightarrow-\infty}\bar{A}^\dagger(\mbf{k}_1,t)A^\dagger_c(\mbf{k}_2)|\Omega\rangle=\lim_{t\rightarrow-\infty}e^{i(H-k^0_1-k^0_2)t}A_c^\dagger(\mbf{k}_1)A_c^\dagger(\mbf{k}_2)|\Omega\rangle
\label{AppA::barAacts1State}
\ee
and
\be
\lim_{t\rightarrow-\infty}A^\dagger_I(\mbf{k}_1,t)A^\dagger_c(\mbf{k}_2)|\Omega\rangle=\frac{H-k_1^0-k_2^0}{2k^0_1}\lim_{t\rightarrow-\infty}e^{i(H-k_1^0-k_2^0)t}A_c^\dagger(\mbf{k}_1)A_c^\dagger(\mbf{k}_2)|\Omega\rangle.
\label{AppA::AIacts1State}
\ee
According to the definition (\ref{Moller_def}) of the M{\o}ller operators $\Omega_c^{(\pm)}$ in the CPR, limits in Eqs.(\ref{AppA::barAacts1State}) and (\ref{AppA::AIacts1State}) can be put in the form
\be
\lim_{t\rightarrow-\infty}e^{i(H-k^0_1-k^0_2)t}A_c^\dagger(\mbf{k}_1)A_c^\dagger(\mbf{k}_2)|\Omega\rangle=\lim_{t\rightarrow-\infty}e^{iKt}e^{-iK_Ft}A_c^\dagger(\mbf{k}_1)A_c^\dagger(\mbf{k}_2)|\Omega\rangle=\Omega_c^{(-)}A_c^\dagger(\mbf{k}_1)A_c^\dagger(\mbf{k}_2)|\Omega\rangle.
\ee
Further, using the property $H\Omega_c^{(\pm)}=\Omega^{(\pm)}_cK_F$ we obtain that Eq.(\ref{AppA::AIacts1State}) vanishes and we get the final result
\be
|\mbf{k}_1\mbf{k}_2;in\rangle =\lim_{t\rightarrow-\infty}e^{i(H-k^0_1-k^0_2)t}A_c^\dagger(\mbf{k}_1)A_c^\dagger(\mbf{k}_2)|\Omega\rangle=\Omega_c^{(-)}A_c^\dagger(\mbf{k}_1)A_c^\dagger(\mbf{k}_2)|\Omega\rangle=\Omega_c^{(-)}|\mbf{k}_1\mbf{k}_2;c\rangle.
\label{AppA::link2}
\ee
Its significant consequence is
\be
|\mbf{k}_1\mbf{k}_2;in\rangle \neq A^\dagger_c(\mbf{k}_1)A^\dagger_c(\mbf{k}_2)|\Omega\rangle.
\ee
Similar relations with $t\rightarrow+\infty$ can be derived for the \textit{out}-states.

\setcounter{equation}{0}
\renewcommand{\theequation}{B.\arabic{equation}}

\section*{Appendix B: Expressions for MEC operators}

The two-nucleon meson exchange current consists of the two parts
\begin{equation}
      J_{2N} =  J_{\mathcal{M}CC} + J_{\mathcal{M}NN}
\end{equation}
with the so-called mesonic $J_{\mathcal{M}CC}$ and seagull $J_{\mathcal{M}NN}$ currents. 
Both of them have the same operator structure
\begin{equation}\label{chpt3:J_MNN_through_R}
      J_{\mathcal{M}CC,\mathcal{M}NN}^{\alpha} =
      \sumint d1' d2' d1 d2\,
      F_{\mathcal{M}CC,\mathcal{M}NN}^{\alpha}(1',2',1,2)
      b^\dagger_{c}(1')b^\dagger_{c}(2')b_{c}(1)b_{c}(2),
      ~~~(\mathcal{M}= \delta, \sigma, \pi, \eta, \rho, \omega),
\end{equation}
{hereafter we denote $\sumintinline d1'\equiv \sum_{\mu'_1\eta'_1}\int{d \mathbf{p}'_1}$, $b_c^\dagger(1') \equiv b_c^\dagger(\vect{p}'_1\mu'_1\eta'_1)$ and so on, with spin indices $\mu$ and isospin indices $\eta$.}
The matrix elements of these operators sandwiched between the two-nucleon states
\begin{equation}
      | 1, 2 \rangle = b_c^\dagger(1) b_c^\dagger(2) | \Omega \rangle
\end{equation}
can be easily found
\begin{equation}\label{chpt3:vacuum_average}
      \langle 1',2'|J ^{\alpha}|1,2 \rangle 
      =
      - F ^{\alpha} (1',2',1,2)
      + F ^{\alpha} (1',2',2,1) 
      - F ^{\alpha} (2',1',2,1) 
      + F ^{\alpha} (2',1',1,2).
\end{equation}
Here we have used the commutation relation 
\begin{equation}\label{bb_commutator}
      \{b_c(\vect{p}\mu\eta),b_c^\dagger(\vect{p}'\mu'\eta')\}=\delta_{\mu'\mu}\delta_{\eta'\eta}\delta(\mathbf{p}'-\mathbf{p}).
\end{equation}

Before we proceed to the explicit expressions for the MECs, we would like to establish some notations.
For the sake of brevity, we denote the following isospin factors
\begin{equation}\label{isospin_factors}
	T_1 = \vect\tau_{\eta'_1\eta_1} \cdot \vect\tau_{\eta'_2\eta_2} + \tau_{\eta'_2\eta_2}^z,
	\quad
	T_2 =  1 + \tau^z_{\eta'_1\eta_1},
	\quad
	T_3 =  i[\vect\tau_{\eta'_1\eta_1} \times \vect\tau_{\eta'_2\eta_2}]^z,
\end{equation}
with $\vect\tau$ being the Pauli matrices in the isospin space.
Following Ref.~\cite{SheDub2010} we introduce cutoff functions that are defined as follows
\begin{equation}\label{cutoff_def}
	g_{\mathcal{M}}(n', n) = \left[\frac{\Lambda_{\mathcal{M}}^2-m_{\mathcal{M}}^2}{\Lambda_{\mathcal{M}}^2-
	\left(E'_n-E_n\right)^2 + \left(\vect{p}'_n-\vect{p}_n\right)^2
	}\right]^{n_{\mathcal{M}}},
\end{equation}
where $n_\mathcal{M}=1$ except for $n_\rho=n_\omega=2$ and all the parameters can be found in Table 1 from Ref.~\cite{ArsEtAl2021}. 
The cutoff factors are supposed to account for finite-size effects (see details in Refs.~\cite{SheDub2010, FroShe2012}). In Ref.~\cite{SheDub2012} these cutoffs have been used in the evaluation of the radial deuteron functions $u_{0,2}(|\vect{p}|)$.
They are present in the definition of the generator $R$ of the unitary clothing transformation and therefore enter the expressions for the MECs that stem from the commutator $\frac12[R,[R,J_c^\mu(0)]]$.

As discussed in Appendix C of Ref.~\cite{SheDub2010} we actually have the possibility to introduce two independent cutoff functions for each meson, one corresponds to the nucleon-meson-nucleon vertex \eqref{cutoff_def} and another one corresponds to the nucleon-meson-antinucleon vertex. But at the time authors of the paper considered only nucleon-nucleon interaction in the CPR that in the first non-vanishing approximation embodies only the first type of vertex. In contrast, the seagull currents we are considering here may contain this second type of cutoff factor.
Let us denote it analogously to \eqref{cutoff_def}:
\begin{equation}
	{\bar{g}_{\mathcal{M}}(n', n)} = \left[\frac{\bar{\Lambda}_{\mathcal{M}}^2-m_{\mathcal{M}}^2}{\bar{\Lambda}_{\mathcal{M}}^2-
	\left(E'_n-E_n\right)^2 + \left(\vect{p}'_n-\vect{p}_n\right)^2
	}\right]^{n_{\mathcal{M}}},
\end{equation}
with new parameters $\bar{\Lambda}_{\mathcal{M}}$ instead of ${\Lambda}_{\mathcal{M}}$. In our calculations presented in Sec.~\ref{sec:results_and_discussion} we have used the same values for $\bar{\Lambda}_{\mathcal{M}}$ as for ${\Lambda}_{\mathcal{M}}$, but in principle they can be different and should be fitted to the experimental data for $N$-$\bar{N}$ scattering.

Now we present the explicit expressions for $F^\alpha_{\mathcal{M} CC}$ and $F^\alpha_{\mathcal{M} NN}$ functions that determine the MEC operators and have been derived by computing the $\frac12[R,[R,J_c^\mu(0)]]$ commutator from the series \eqref{chpt3:J1N_J2N_series}. 
Mesonic MECs are given by the following expressions for\\
$\pi$-mesons (pseudoscalar with isospin $T=1$):
\be
\begin{split}\label{FpiCC}
	F^\alpha_{\pi CC}(1',2',1,2)=-\frac{eg_{\pi}^2m^2}{2(2\pi)^6}T_3\frac{g_\pi(1',1)g_\pi(2',2)}{\sqrt{E'_1E'_2E_1E_2}}(p'_1-p_1+p_2-&p'_2)^\alpha\frac{\bar{u}(\mathbf{p}'_1\mu'_1)\gamma_5u(\mathbf{p}_1\mu_1)}{(p'_1-p_1)^2-m^2_\pi}\\
	&\times\frac{\bar{u}(\mathbf{p}'_2\mu'_2)\gamma_5 u(\mathbf{p}_2\mu_2)}{(p'_2-p_2)^2-m^2_\pi},
\end{split}
\ee
$\delta$-mesons (scalar with isospin $T=1$):
\be
\begin{split}\label{explicit_mesonic_delta}
	F^\alpha_{\delta CC}(1',2',1,2)=\frac{eg_{\delta}^2m^2}{2(2\pi)^6}T_3\frac{g_\delta(1',1)g_\delta(2',2)}{\sqrt{E'_1E'_2E_1E_2}}(p'_1-p_1+p_2-&p'_2)^\alpha\frac{\bar{u}(\mathbf{p}'_1\mu'_1)u(\mathbf{p}_1\mu_1)}{(p'_1-p_1)^2-m^2_\delta}\\
	&\times\frac{\bar{u}(\mathbf{p}'_2\mu'_2) u(\mathbf{p}_2\mu_2)}{(p'_2-p_2)^2-m^2_\delta},
\end{split}
\ee
$\rho$-mesons (vector with isospin $T=1$):
\begin{multline}
      F_{\rho CC}^{\alpha}(1',2',1,2)
      =
      \frac{em^2}{8(2\pi)^6} 
      \frac{T_3}{\omega_{1} \omega_{2}} 
      \frac{g_{\rho}(1', 1) g_{\rho}(2', 2)}{\sqrt{E'_1E'_2E_1E_2}}
      \\\times
      \bigg\{
            \bigg[
                  g^{\alpha\delta} \left(
                        k_1^\sigma-k_{2-}^\sigma \frac{(k_1\cdot k_{2-})}{m_\rho^2}
                  \right) 
                  -k^{\alpha}_1 \left(
                        g^{\delta\sigma} - \frac{k_{2-}^\delta k_{2-}^\sigma}{m_{\rho}^2}
                  \right)
                  +
                  g^{\alpha\sigma} \left(
                        k_{2-}^\delta-k_1^\delta \frac{(k_1\cdot k_{2-})}{m_\rho^2}
                  \right) 
				  \\
                  -k^{\alpha}_{2-} \left(
                        g^{\delta\sigma} - \frac{k_1^\delta k_1^\sigma}{m_{\rho}^2}
                  \right)
            \bigg]
            \frac{
                  \bar{u}(1')
                  \left[
                        g_\rho  \gamma_\delta-\frac{f_\rho}{2m}ik_{1}^{\nu}\sigma_{\nu\delta}
                  \right]
                  u(1)
            }
            {
                  E'_1 - E_1 -\omega_1
            }
            \frac{
                  \bar{u}(2')
                  \left[
                        g_\rho  \gamma_\sigma+\frac{f_\rho}{2m}ik_{2-}^{\nu}\sigma_{\nu\sigma}
                  \right]
                  u(2)
            }
            {
            -E'_2 + E_2 -\omega_2 
            }
            \\
            - 
            \bigg[
                  g^{\alpha\delta} \left(
                        k^\sigma_{1-}-k_2^\sigma \frac{(k_{1-}\cdot k_2)}{m_\rho^2}
                  \right) 
                  -k^{\alpha}_{1-} \left(
                        g^{\delta\sigma} - \frac{k_2^\delta k_2^\sigma}{m_{\rho}^2}
                  \right)
                  +
                  g^{\alpha\sigma} \left(
                        k_2^\delta-k^\delta_{1-} \frac{(k_{1-}\cdot k_2)}{m_\rho^2}
                  \right) 
				  \\
                  -k^{\alpha}_2 \left(
                        g^{\delta\sigma} - \frac{k_{1-}^\delta k_{1-}^\sigma}{m_{\rho}^2}
                  \right)
            \bigg]
            \frac{
                  \bar{u}(1')
                  \left[
                        g_\rho  \gamma_\delta+\frac{f_\rho}{2m}ik_{1-}^{\nu}\sigma_{\nu\delta}
                  \right]
                  u(1)
            }
            {
            -E'_1 + E_1 -\omega_1
            }
            \frac{
                  \bar{u}(2')
                  \left[
                        g_\rho  \gamma_\sigma-\frac{f_\rho}{2m}ik_{2}^{\nu}\sigma_{\nu\sigma}
                  \right]
                  u(2)
            }
            {
                  E'_2 - E_2 -\omega_2
            }
            \\
            + 
            \bigg[
                  g^{\alpha\delta} \left(
                        k^\sigma_1-k_2^\sigma \frac{(k_1\cdot k_2)}{m_\rho^2}
                  \right) 
                  -k^{\alpha}_1 \left(
                        g^{\delta\sigma} - \frac{k_2^\delta k_2^\sigma}{m_{\rho}^2}
                  \right)
                  -
                  g^{\alpha\sigma} \left(
                        k_2^\delta-k^\delta_1 \frac{(k_1\cdot k_2)}{m_\rho^2}
                  \right) 
				  \\
                  +k^{\alpha}_2 \left(
                        g^{\delta\sigma} - \frac{k_1^\delta k_1^\sigma}{m_{\rho}^2}
                  \right)
            \bigg]
            \frac{
                  \bar{u}(1')
                  \left[
                        g_\rho  \gamma_\delta-\frac{f_\rho}{2m}ik_{1}^{\nu}\sigma_{\nu\delta}
                  \right]
                  u(1)
            }
            {
                  E'_1 - E_1 -\omega_1
            }
            \frac{
                  \bar{u}(2')
                  \left[
                        g_\rho  \gamma_\sigma-\frac{f_\rho}{2m}ik_{2}^{\nu}\sigma_{\nu\sigma}
                  \right]
                  u(2)
            }
            {
                  E'_2 - E_2 -\omega_2
            }
            \\
            - 
            \bigg[
                  g^{\alpha\delta} \left(
                        k^\sigma_{1-}-k_{2-}^\sigma \frac{(k_{1}\cdot k_{2})}{m_\rho^2}
                  \right) 
                  -k^{\alpha}_{1-} \left(
                        g^{\delta\sigma} - \frac{k_{2-}^\delta k_{2-}^\sigma}{m_{\rho}^2}
                  \right)
                  -
                  g^{\alpha\sigma} \left(
                        k_{2-}^\delta-k^\delta_{1-} \frac{(k_{1}\cdot k_{2})}{m_\rho^2}
                  \right) 
				  \\
                  +k^{\alpha}_{2-} \left(
                        g^{\delta\sigma} - \frac{k_{1-}^\delta k_{1-}^\sigma}{m_{\rho}^2}
                  \right)
            \bigg]
            \frac{
                  \bar{u}(1')
                  \left[
                        g_\rho  \gamma_\delta+\frac{f_\rho}{2m}ik_{1-}^{\nu}\sigma_{\nu\delta}
                  \right]
                  u(1)
            }
            {
            -E'_1 + E_1 -\omega_1
            }
            \frac{
                  \bar{u}(2')
                  \left[
                        g_\rho  \gamma_\sigma+\frac{f_\rho}{2m}ik_{2-}^{\nu}\sigma_{\nu\sigma}
                  \right]
                  u(2)
            }
            {
            -E'_2 + E_2 -\omega_2 
            }
      \bigg\}.
\end{multline}
Here
\begin{equation}
\begin{split}
      &\vect{k}_1=\vect{p}'_1-\vect{p}_1,
      ~~
      \vect{k}_2=\vect{p}'_2-\vect{p}_2,
      ~~
      \vect{p}_3= \vect{p}_1 + \vect{p}_2 - \vect{p}'_2,
      ~~
      \vect{p}'_3= \vect{p}'_1 + \vect{p}'_2 - \vect{p}_2,
      \\
      &\omega_1=\sqrt{m_{\rho}^2+\vect{k}_1^2},
      ~~
      \omega_2=\sqrt{m_{\rho}^2+\vect{k}_2^2},
      ~~
      E_n=\sqrt{m+\vect{p}_n^2}
      ~~
      E'_n=\sqrt{m+\vect{p}'^2_n} ~~ \textrm{for} ~~ n=1,2,3,
      \\
      &k_{1}=(\omega_1,\vect{k}_1),
      ~~
      k_{2}=(\omega_2,\vect{k}_2),
      ~~
      k_{1-}=(\omega_1,-\vect{k}_1),
      ~~
      k_{2-}=(\omega_2,-\vect{k}_2),
      \\
      &p'_{3-}=(E'_3,-\vect{p}'_3), 
      ~~
      p_{3-}=(E_3,-\vect{p}_3),
      ~~
      p'_n=(E'_n,\vect{p}'_n), 
      ~~
      p_n=(E_n,\vect{p}_n) ~~ \textrm{for} ~~ n=1,2,3,
      \\
      &\cross{p}_3=\gamma_\mu p^\mu_3,
      ~~
      \cross{p}_{3-}=\gamma_\mu p^\mu_{3-},
      ~~ 
      \cross{p}'_3=\gamma_\mu p'^\mu_3,
      ~~ 
      \cross{p}'_{3-}=\gamma_\mu p'^\mu_{3-},
      ~~
      g^{\delta\sigma}=\textrm{diag}(1,-1,-1,-1).
\end{split}
\end{equation}

Seagull MECs are given by the following expressions for\\
$\pi$-mesons:
\be\label{FpiNN}
\begin{split}
	&F^\alpha_{\pi NN}(1',2',1,2)=\frac{eg^2_{\pi}m^2}{2(2\pi)^6}\frac{{g_\pi(2',2)}}{\sqrt{E'_1E'_2E_1E_2}}\frac{\bar{u}(2')\gamma_5 u(2)}{(p'_2-p_2)^2-m^2_\pi}\\
	&\times\bigg[\frac{T_1-T_3}{2} 
      \bar{u}(1')\gamma^\alpha A(1',2',1,2)\gamma_5u(1) + \frac{T_1+T_3}{2}
      \bar{u}(1')\gamma_5 A(1,2,1',2')\gamma^\alpha u(1)\bigg],
\end{split}
\ee
$\delta$-mesons:
\be
\begin{split}\label{FdeltaNN}
	&F^\alpha_{\delta NN}(1',2',1,2)=-\frac{eg^2_{\delta}m^2}{2(2\pi)^6}\frac{{g_{\delta}(2',2)}}{\sqrt{E'_1E'_2E_1E_2}}\frac{\bar{u}(2')u(2)}{(p'_2-p_2)^2-m^2_\delta}\\
	&\times\bigg[\frac{T_1-T_3}{2}
      \bar{u}(1')\gamma^\alpha A(1',2',1,2)u(1) + \frac{T_1+T_3}{2}
      \bar{u}(1') A(1,2,1',2')\gamma^\alpha u(1)\bigg],
\end{split}
\ee
with
\begin{multline}
A(1',2',1,2) = \frac{1}{2E_3}\bigg[
      {g_{\pi,\delta}(1,3)}(\slashed{p}_3+m)\frac{E_1-E_2+E'_2-E_3}{(p_1-p_3)^2-m^2_\mathcal{M}}
      \\
      +{\bar{g}_{\pi,\delta}(1,3)} (\slashed{p}_{3-}-m)\frac{E_1-E_2+E'_2+E_3}{(p_1+p_{3-})^2-m^2_\mathcal{M}}\bigg],
\end{multline}
and for $\rho$-mesons:
\begin{multline}\label{FrhoNN}
      F_{\rho NN}^{\alpha}(1',2',1,2)
      =
      \frac{em^2}{16(2\pi)^6} \frac{1}{\omega_{2}}  \frac{{g_{\rho}(2', 2)}}{\sqrt{E'_1E'_2E_1E_2}}
      \\\times
      \bigg\{
            \frac{T_1 - T_3}{E_{3}} 
            \bigg(
                  \left[
                        -g^{\delta\sigma} + \frac{k_{2-}^\delta k_{2-}^\sigma}{m_{\rho}^2}
                  \right] 
                  \frac{
                        \bar{u}(2')
                        \left[
                              g_\rho  \gamma_\delta+\frac{f_\rho}{2m}ik_{2-}^{\nu}\sigma_{\nu\delta}
                        \right]
                        u(2)
                  }
                  {
                  -E'_2 + E_2 -\omega_2 
                  }
                  \\\times
                  \bar{u}(1')
                        \gamma^\alpha 
                  \left[
                        {g_{\rho}(3, 1)}
                  \frac{
                        (\cross{p}_3 + m)
                  }
                  {
                        E_3 - E_1 -\omega_2
                  }
                  +
                  {\bar{g}_{\rho}(3, 1)}
                  \frac{
                        (\cross{p}_{3-} - m)
                  }
                  {
                        -E_3 - E_1 -\omega_2
                  }
                  \right]
                  \left[
                        g_\rho  \gamma_\sigma-\frac{f_\rho}{2m}ik_{2-}^{\nu}\sigma_{\nu\sigma}
                  \right]
                  u(1)
                  \\- 
                  \left[
                        -g^{\delta\sigma} + \frac{k_2^\delta k_2^\sigma}{m_{\rho}^2}
                  \right] 
                  \frac{
                        \bar{u}(2')
                        \left[
                              g_\rho  \gamma_\delta-\frac{f_\rho}{2m}ik_{2}^{\nu}\sigma_{\nu\delta}
                        \right]
                        u(2)
                  }
                  {
                        E'_2 - E_2 -\omega_2
                  } 
                  \\\times
                  \bar{u}(1') \gamma^\alpha 
                  \left[
                        {g_{\rho}(3, 1)}
                  \frac{
                        (\cross{p}_3 + m)
                  }
                  {
                  -E_3 + E_1 -\omega_2
                  }
                  +
                  {\bar{g}_{\rho}(3, 1)}
                  \frac{
                        (\cross{p}_{3-} - m)
                  }
                  {
                  E_3 + E_1 -\omega_2
                  }
                  \right]
                  \left[
                        g_\rho  \gamma_\sigma+\frac{f_\rho}{2m}ik_{2}^{\nu}\sigma_{\nu\sigma}
                  \right]
                  u(1)
            \bigg)
            \\+
            \frac{T_1 + T_3}{E'_{3}} 
            \bigg(
                  \left[
                        -g^{\delta\sigma} + \frac{k_2^\delta k_2^\sigma}{m_{\rho}^2}
                  \right]
                  \frac{
                        \bar{u}(2')
                        \left[
                              g_\rho  \gamma_\delta-\frac{f_\rho}{2m}ik_{2}^{\nu}\sigma_{\nu\delta}
                        \right]
                        u(2)
                  }
                  {
                        E'_2 - E_2 -\omega_2
                  } 
                  \\\times
                  \bar{u}(1')
                        \left[
                              g_\rho  \gamma_\sigma+\frac{f_\rho}{2m}ik_{2}^{\nu}\sigma_{\nu\sigma}
                        \right]
                  \left[
                        {g_{\rho}(1', 3')} 
                  \frac{
                        (\cross{p}'_3 + m)
                  }
                  {
                  -E'_1 + E'_3 -\omega_2
                  } 
                  +
                  {\bar{g}_{\rho}(1', 3')} 
                  \frac{
                        (\cross{p}'_{3-} - m)
                  }
                  {
                        - E'_1 - E'_3 -\omega_2
                  }
                  \right]
                  \gamma^\alpha u(1)
                  \\- 
                  \left[
                        -g^{\delta\sigma} + \frac{k_{2-}^\delta k_{2-}^\sigma}{m_{\rho}^2}
                  \right] 
                  \frac{
                        \bar{u}(2')
                        \left[
                              g_\rho  \gamma_\delta+\frac{f_\rho}{2m}ik_{2-}^{\nu}\sigma_{\nu\delta}
                        \right]
                        u(2)
                  }
                  {
                  -E'_2 + E_2 -\omega_2 
                  }
                  \\\times
                  \bar{u}(1')
                        \left[
                              g_\rho  \gamma_\sigma-\frac{f_\rho}{2m}ik_{2-}^{\nu}\sigma_{\nu\sigma}
                        \right]
                  \left[
                        {g_{\rho}(1', 3')} 
                  \frac{
                        (\cross{p}'_3 + m)
                  }
                  {
                        E'_1 - E'_3 -\omega_2
                  }
                  +
                  {\bar{g}_{\rho}(1', 3')} 
                  \frac{
                        (\cross{p}'_{3-} - m)
                  }
                  {
                  E'_1 + E'_3 -\omega_2  
                  } 
                  \right]
                  \gamma^\alpha u(1)
            \bigg)
      \bigg\}.
\end{multline}

The $\sigma$ mesons (scalar) with isospin $T=1$ produces the same currents as the $\delta$ mesons, but with different coupling constant, cutoff parameter and mass. 
In order to get the expressions for the MECs induced by the mesons with isospin $T=0$: $\sigma(0)$ (scalar), $\eta$ (pseudoscalar) and $\omega$ (vector), one needs to replace $T_1 \to T_2$, $T_3 \to 0$ and substitude the corresponding parameters (coupling constants, cutoff parameters and masses) in the respective expressions above. 
It is clear that the isoscalar mesons that has zero charge do not produce mesonic MECs, but they do produce seagull ones.

Although the expressions \eqref{FpiCC}--\eqref{FrhoNN} for MECs may appear rather complex, they can be efficiently calculated via the relations from Ref.~\cite{LevShe1995}:
\begin{align}
      &\bar{u}(\vect{p}'\mu') {\Gamma} u(\vect{p}\mu) =
      A(\vect{p}',\vect{p}) \delta_{\mu'\mu} +  \vect{B}(\vect{p}',\vect{p})\cdot \vect{\sigma}_{\mu'\mu},
      \label{Eq18}
      \\
      &A(\vect{p}',\vect{p})={\textrm{Tr}\left[(\cross{p}'+m)\Gamma(\cross{p}+m)(1+\gamma_0)\right]} / {8m\sqrt{(E_{\vect{p}'}+m)(E_{\vect{p}}+m)}},
      \label{A}
      \\
      &\vect{B}(\vect{p}',\vect{p})={\textrm{Tr}\left[(\cross{p}'+m)\Gamma(\cross{p}+m)(1+\gamma_0)\vect{\gamma}\gamma_5\right]} / {8m\sqrt{(E_{\vect{p}'}+m)(E_{\vect{p}}+m)}},
      \label{B}
\end{align}
with $\cross{p}=\gamma_\alpha p^\alpha$, 
$p'=(E_{\vect{p}'},\vect{p}')$, $p=(E_{\vect{p}},\vect{p})$ and $E_{\vect{p}}=\sqrt{\vect{p}^2+m^2}$, for any combination of the $\gamma$ matrices $\Gamma$.
We have the same definitions for $\gamma$ matrices and bispinors as in \cite{BjorkenDrell}
\begin{equation}
      {u}(\vect{p}\mu) =
      \sqrt{\frac{E_{\vect{p}}+m}{2m}} 
      \begin{pmatrix}
            | \mu \rangle \\
            \frac{\vect{\sigma}\cdot\vect{p}}{E_{\vect{p}}+m}| \mu \rangle
      \end{pmatrix}
      ,
      ~~~
      {v}(\vect{p}\mu) =
      \sqrt{\frac{E_{\vect{p}}+m}{2m}} 
      \begin{pmatrix}
            \frac{\vect{\sigma}\cdot\vect{p}}{E_{\vect{p}}+m} | \mu \rangle^c
            \\
            | \mu \rangle^c
      \end{pmatrix}
     ,
\end{equation}
\begin{equation}
      | \tfrac12 \rangle= 
      \begin{pmatrix}
            1
            \\
            0
      \end{pmatrix},
      ~~~
      | -\tfrac12 \rangle= 
      \begin{pmatrix}
            0
            \\
            1
      \end{pmatrix},
      ~~~
      | \mu \rangle^c = (-1)^{\tfrac{1}{2}-\mu} | -\mu \rangle.
\end{equation}

\setcounter{equation}{0}
\renewcommand{\theequation}{C.\arabic{equation}}

\section*{\clb Appendix C: Additional test of \textit{np}-pair partial wave calculations}

In Sec.~\ref{subsec:deuteron_and_np_states} we have described our method of calculating the $np$-pair partial wave functions $\Psi^{\a(-)}_{p_0ll'}(p')$ in the momentum space. 
The matrix inversion method allows us to find the representation \eqref{chpt5:phiMIM} for the WFs $\varphi^{\a}_{pll'}(p')$ that have the asymptotics of standing waves. 
The links \eqref{chpt5::coupledLink} give us an analogous representation for the $np$-pair partial WFs
\begin{equation}\label{inegral_rep_np_partial_WF}
      \Psi^{\a(\pm)}_{p_0ll'}(p')=\sum_{j=1}^{N+1}\Omega^{\a(\pm)}_{p_0ll'}(j)\frac{\d(p'-p_j)}{p_j^2},
\end{equation}
\begin{equation}
      \Omega^{\a(\pm)}_{p_0ll'}(j)\equiv \sum_{l''}{O}^{\a(\pm)}_{ll''}(p_0)B^{\a}_{p_0l''l'}(j),
\end{equation}
that should be meant in an integral sense.
Here $p_{N+1}=p_0$ is the relative momentum of the $np$-pair and for the uncoupled partial waves the matrix ${O}{\a(\pm)}_{ll'}(p_0)$ reduces to the phase factor
\begin{equation}
      {O}^{\a(\pm)}_{ll}(p_0)=\exp(\pm i\d^\a_l)\cos\d^\a_l,
      ~~~
      (l'=l).
\end{equation}

\begin{figure}[t]
  \centering
  \includegraphics[width=\linewidth]{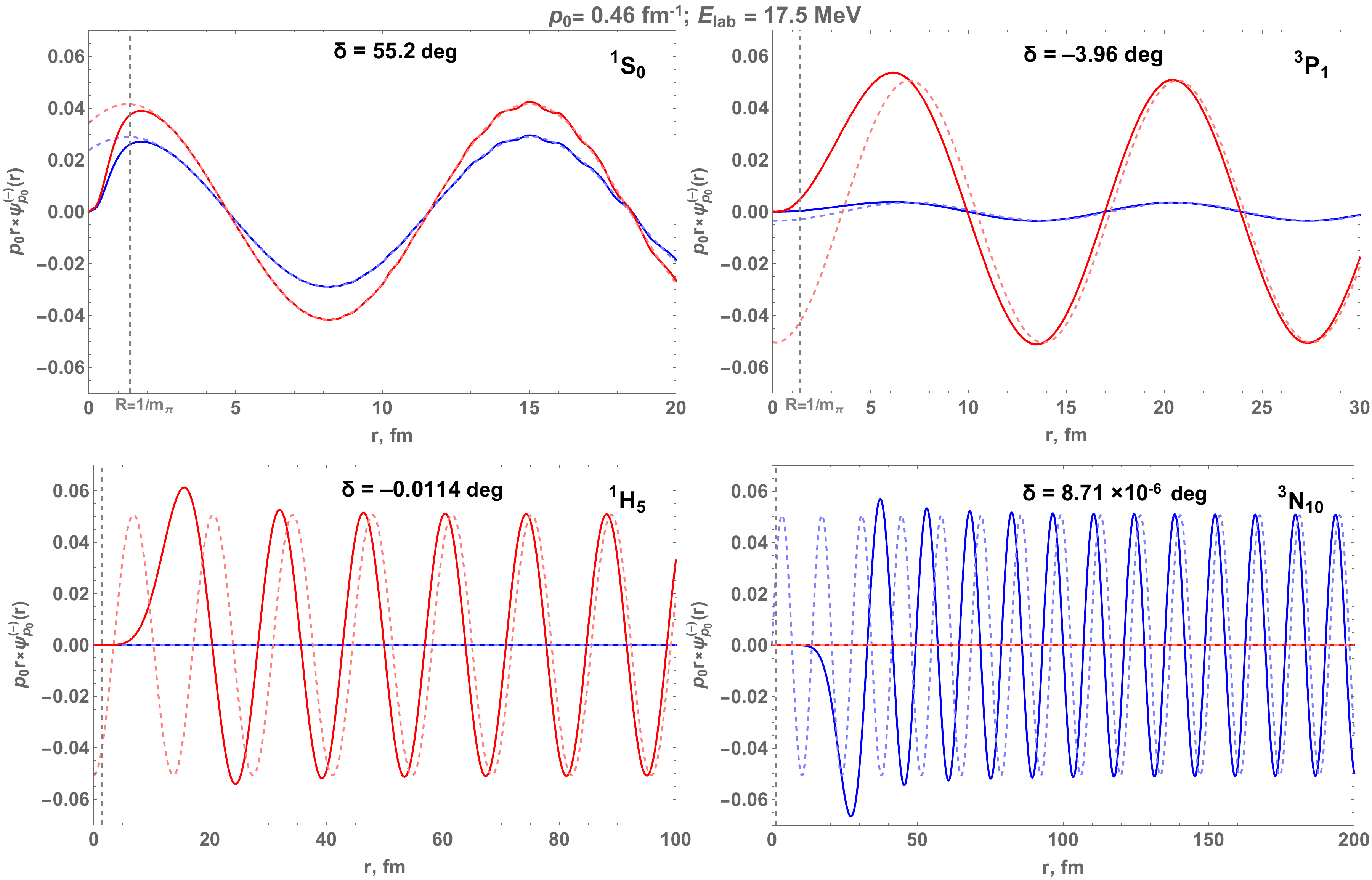}
  \caption{
      Verification of the asymptotic behavior \eqref{approx_np_WF} for the $\,^1S_0$, $\,^3P_1$, $\,^1H_5$ and $\,^3N_{10}$ uncoupled partial waves as examples for $np$-pair lab energy $E_{\text{lab}}=17.5~MeV$ ($p_0=0.46~ fm^{-1}$).
      Blue solid lines: real part of $p_0r \Psi^{\a(-)}_{p_0ll}(r)$.
      Red solid lines: imaginary part of $p_0r \Psi^{\a(-)}_{p_0ll'}(r)$.
      Dotted lines are the respective real (blue) and imaginary (red) parts of the asymptotic functions \eqref{approx_np_WF} multiplied by $p_0 r$.
      The characteristic radius of the nuclear forces (vertical dotted line) $R=1/m_\pi=1.4~ fm$ is shown too.
  }
  \label{fig:12}
\end{figure}

In fact, Eq.~\eqref{inegral_rep_np_partial_WF} allows us to find analytical expressions for the $np$-pair partial WFs in the coordinate space and verify their asymptotic behavior.
From Eq.~\eqref{typical_integral_final_form} it follows that the radial partial wave functions
\begin{equation}
      \Psi^{\a(\pm)}_{p_0ll'}(r)=
      \frac{1}{2\pi^2} i^{l'} \int_0^\infty p'^2dp'\, \Psi^{\a(\pm)}_{p_0ll'}(p') j_{l'}(p'r)
\end{equation}
can be written as
\begin{equation}\label{np_partial_WF_coordinate_space}
      \Psi^{\a(\pm)}_{p_0ll'}(r)
      = \frac{1}{2\pi^2} i^{l'} \sum_{j=1}^{N+1}\Omega^{\a(\pm)}_{p_0ll'}(j) j_{l'}(p_j r)
      .
\end{equation}

Recall the known result of the scattering theory (see, e.g., Sec.~6.2 and Sec.~7.1 in Ref.~\cite{GoldWat}) for the asymptotic behavior of the partial wave functions of the continuous spectrum at large $p_0r$
\begin{equation}\label{approx_np_WF}
\begin{split}
      &\textrm{uncoupled waves:}~~~
      \Psi^{\a(\pm)}_{p_0ll}(r)
      \asymp 
      \frac{1}{2\pi^2} i^{l} e^{\pm i\delta^\a_{l}} \frac{\sin\left(p_0 r - \frac{l\pi}{2} + \delta^\a_{l}\right)}{p_0 r}
      ,
      ~~~
      p_0r \gg 1
      ,
      \\
      &\textrm{coupled waves:}~~~
      \Psi^{\a(\pm)}_{p_0ll'}(r)
      \asymp 
      \frac{1}{2\pi^2} \frac{1}{p_0 r}\begin{pmatrix}
      \cos \varepsilon_\alpha & -\sin \varepsilon_\alpha
      \\
      \sin \varepsilon_\alpha & \cos \varepsilon_\alpha
      \end{pmatrix}
      \\
      &\hspace{1cm}
      \times 
      \begin{pmatrix}
      i^{J-1} e^{\pm i\delta^\a_{-}} \sin\left(p_0 r - \frac{(J-1)\pi}{2} + \delta^\a_{-}\right)
      &
      0\\0
      &
      i^{J+1} e^{\pm i\delta^\a_{+}} \sin\left(p_0 r - \frac{(J+1)\pi}{2} + \delta^\a_{+}\right)
      \end{pmatrix}\\
      &\hspace{9cm}
      \times\begin{pmatrix}
      \cos \varepsilon_\alpha & \sin \varepsilon_\alpha
      \\
      -\sin \varepsilon_\alpha & \cos \varepsilon_\alpha
      \end{pmatrix}
      ,
      ~~~
      p_0r \gg 1.
\end{split}
\end{equation}
That gives us a possibility to verify once again the correctness of our calculations.

\begin{figure}[t]
  \centering
  \includegraphics[width=0.92\linewidth]{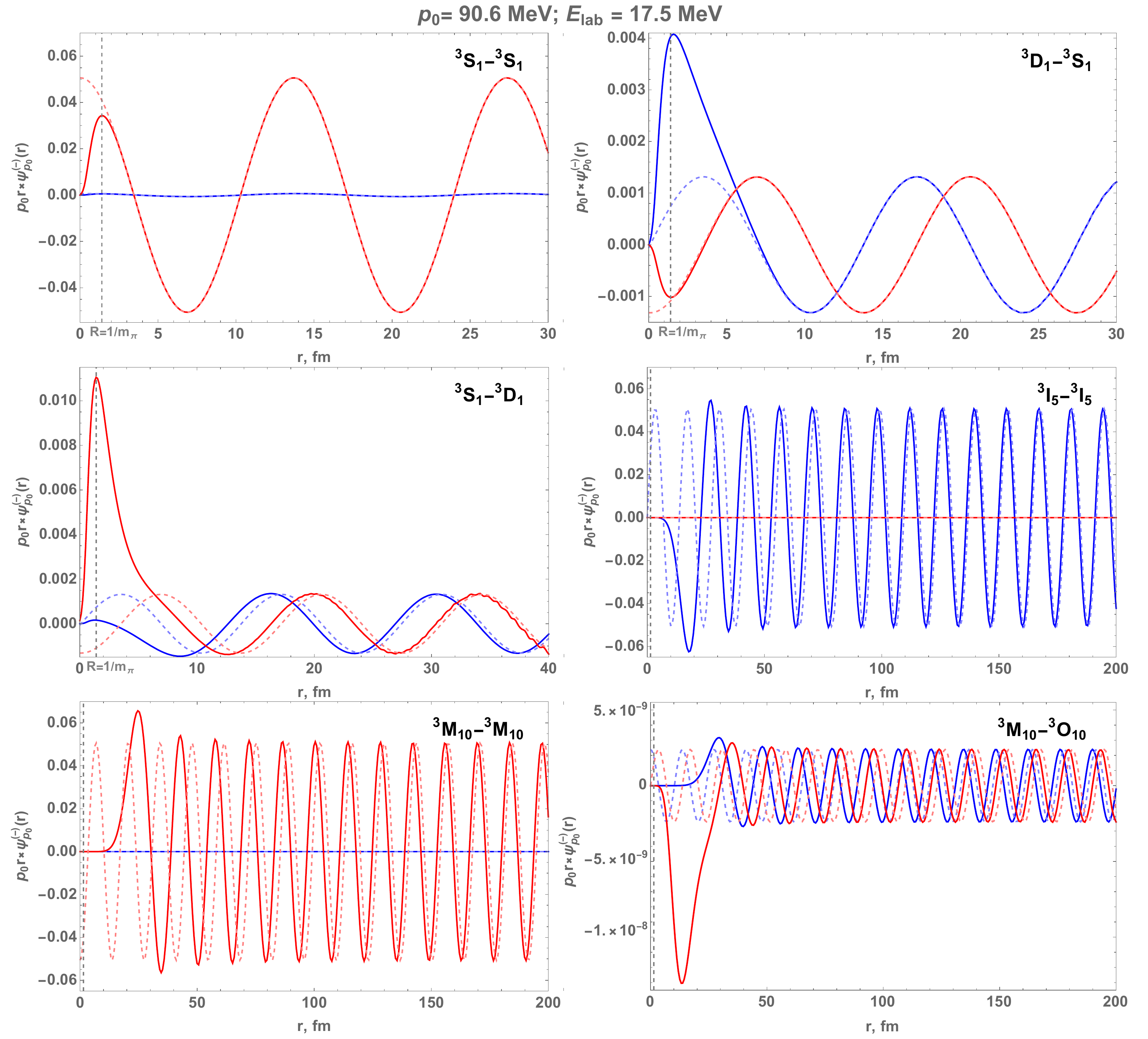}
  \caption{
  The same as in Fig.~\ref{fig:12} but for the coupled partial waves:
  $\,^3S_1$--$\,^3S_1$, $\,^3D_1$--$\,^3S_1$,
  $\,^3S_1$--$\,^3D_1$, $\,^3I_5$--$\,^3I_5$,
  $\,^3M_{10}$--$\,^3M_{10}$ and $\,^3M_{10}$--$\,^3O_{10}$.
  }
  \label{fig:13}
\end{figure}

In Figs.~\ref{fig:12}--\ref{fig:13} we show that the functions $p_0 r \Psi^{\a(-)}_{p_0ll'}(r)$ indeed have the expected asymptotic behavior.
The numerical result for the state $\,^1S_0$ starts to follow the asymptotic one as soon as $r$ exceeds the characteristic radius of nuclear forces $R=1/m_\pi=1.4~ fm$, while for the states with higher orbital momenta the asymptotic behavior is observed at larger $r$.

\FloatBarrier

\providecommand{\href}[2]{#2}\begingroup\raggedright\endgroup

\end{document}